\documentclass[%
 reprint,
 amsmath,amssymb,
 aps,
 prab,
floatfix,
]{revtex4-2}

\usepackage{textcomp}
\usepackage[T1]{fontenc}
\usepackage{lmodern}
\usepackage{tgtermes}
\usepackage{newtxmath}
\usepackage{siunitx}
\usepackage{graphicx}
\usepackage{dcolumn}
\usepackage{bm}

\newcommand\En{E_{\text{total}}}
\newcommand\betar{\beta_{\text{r}}}
\newcommand\Trev{T_\text{rev}}
\newcommand\ITibs{T^{-1}_{\text{IBS},u}}
\newcommand\Tibs{T_{\text{IBS},u}}

\begin{document}

\preprint{arXiv:2308.02196}

\title{Interplay of space charge, intrabeam scattering and synchrotron radiation\\in the Compact Linear Collider damping rings}

\author{M.~Zampetakis}
\email{michail.zampetakis@cern.ch}
\affiliation{European Organization for Nuclear Research (CERN), CH-1211 Geneva 23, Switzerland;}
\affiliation{Department of Physics, University of Crete, P.O. Box 2208, GR-71003 Heraklion, Greece}
\author{F.~Antoniou}
\author{F.~Asvesta}
\author{H.~Bartosik}
\author{Y.~Papaphilippou}
\affiliation{European Organization for Nuclear Research (CERN), CH-1211 Geneva 23, Switzerland;}

\begin{abstract}
    Future ultra-low emittance rings for $e^-/e^+$ colliders require extremely high beam brightness and can thus be limited by collective effects. In this paper, the interplay of effects such as synchrotron radiation, intra-beam scattering (IBS) and space charge in the vicinity of excited betatron resonances is assessed. In this respect, two algorithms were developed to simulate IBS and synchrotron radiation effects and integrated in the PyORBIT tracking code, to be combined with its widely used space charge module. The impact of these effects on the achievable beam parameters of the Compact Linear Collider (CLIC) Damping Rings was studied, showing that synchrotron radiation damping mitigates the adverse effects of IBS and space charge induced resonance crossing. The studies include also a full dynamic simulation of the CLIC damping ring cycle starting from the injection beam parameters. It is demonstrated that a careful working point choice is necessary, in order to accommodate the transition from a non-linear lattice induced detuning to a space-charge dominated one and thereby avoid excessive losses and emittance growth generated in the vicinity of strong resonances.
\end{abstract}


\maketitle
\section{\label{sec:1}Introduction}
In ultra-low emittance electron/positron storage rings, such as the Damping Rings (DRs) of the Compact Linear Collider (CLIC), magnet non-linearities from the strong chromatic sextupoles as well as errors from other magnets can excite resonances and induce particle losses. This, in turn, can compromise the achievable target parameters of the ring. Collective effects due to the high intensity and low target beam emittances in all 3 planes (i.e.~high brightness) can further limit the performance. While coherent effects result mostly in beam instabilities, incoherent effects from Intra-Beam Scattering (IBS) and Space Charge (SC) result mostly in losses and an increase of the achievable beam emittances.

Incoherent effects have been intensively studied in several accelerators. In particular, IBS plays a crucial role in the evolution of the beam emittances in ion machines and in proton storage rings or colliders where the beam is stored for many hours~\cite{MSc_Mertens,JWei_RHIC,Fischer_RHIC,Bruce:ibs_kick,Lebedev_1,Papadopoulou:2018uvl,Rogelio,Antoniou:SPS, Evans:1980}. It can also limit the performance of damping rings, light sources and linear accelerators~\cite{PhD_Papadopoulou,Lebedev_2,PhD_Antoniou,Antoniou:SLS,Bane_2,Bane_KEK,Kubo_KEK,Huang,Xiao_2010,DiMitri_2020,Dalena_2022,Etisken_2017}. Space charge effects have been studied in many low-energy hadron machines~\cite{Franchetti:2003aa,Metral:2006qw,Franchetti:2010zz,Franchetti:2017aa,Fermi_res,Fermi_IOTA,Fermi_boost,Fermi_highI,SC_RHIC,SC_eRHIC,Asvesta:ps,SaaHernandez:2018zqv}, and in damping rings~\cite{Venturini:2006rp, Venturini:2006ilc,Venturini:2007ler,Xiao:2007ilc}. The SC induced tune spread can push particles on driven resonances that, in turn, cause particle losses and emittance increase. On the other hand, the damping from Synchrotron Radiation (SR) counteracts the emittance growth.

The CLIC DRs need to deliver beam emittances with strict requirements~\cite{CLIC_CDR}, and therefore it is important to characterize any possible sources of beam degradation in the presence of all these incoherent effects. In the past, lattice non-linearities were studied in the CLIC DRs~\cite{Ghasem} and analytical estimations have been performed for the equilibrium parameters in the presence of IBS and SR. However, the effect of SC in the presence of lattice non-linearities has not been studied and the final steady state of the beam has not been verified with macro-particle tracking simulations in the presence of all the effects mentioned.

The goal of this work is to study the SC effect in the CLIC DRs, first in the absence of the strong radiation damping to avoid the change of emittances affecting the dynamics. Once the SC effect is studied, the interplay of SC and IBS in the vicinity of an excited vertical resonance will be investigated. Last, the question of whether the SR damping can help preserving the vertical emittance by compensating these two effects and their potential interplay will be examined and the equilibrium parameters for the full CLIC DR cycle will be assessed for the first time with macro-particle tracking simulations. To this end, two algorithms were implemented to simulate the IBS and SR effects. These algorithms were integrated in the PyORBIT macro-particle tracking code~\cite{pyorbit} to be combined with the 3D potentials of its widely used SC module, in a quasi self-consistent way. This allows the combined study of these effects and eventually identifying their impact on the full cycle of the CLIC DRs. 

\begin{figure*}[!htb]
    \centering
    \includegraphics[width=0.8\textwidth]{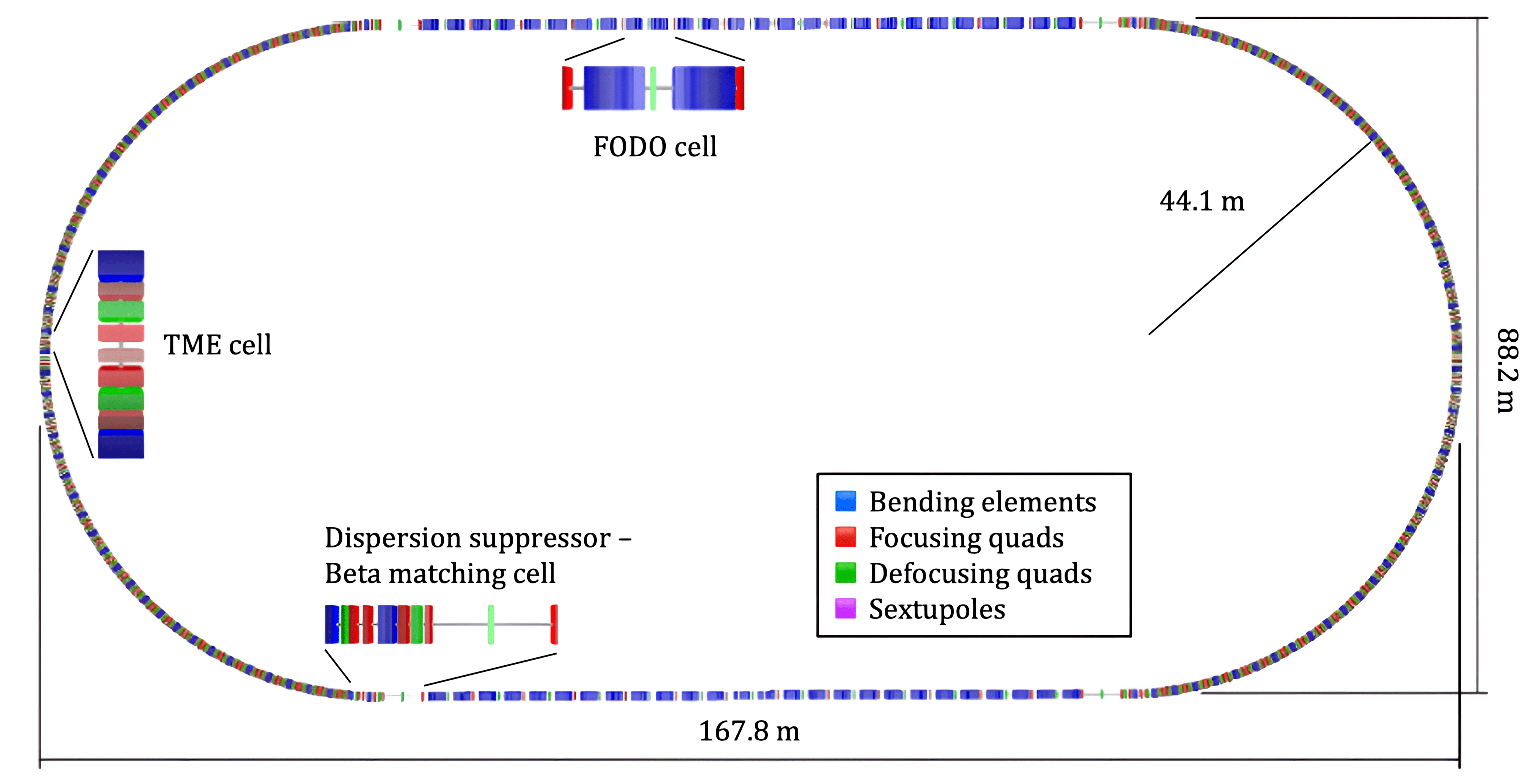}
    \caption{General layout of the CLIC DR.}
    \label{fig:dr_layout}
\end{figure*}

The paper is organized as follows. After this Introduction, the description of the CLIC DRs and beam parameters are discussed in Sec.~\ref{sec:2}. Section~\ref{sec:3} starts with a brief description of the PyORBIT tracking code and its available SC models. Then, the implementation of the two algorithms that describe the IBS and SR effects are presented. Both algorithms are benchmarked against analytical calculations. In Sec.~\ref{sec:4} the SC induced tune spread is discussed and it is shown that despite the high energy at which the CLIC DRs operate, the SC tune spread is significant due to the ultra-low emittances, and can lead to emittance blow-up or particle losses in the presence of excited resonances. In this respect, the configuration that is used to excite the $3Q_y=31$ skew resonance is described. In addition, the tune modulation that particles with synchrotron motion experience, is investigated. In Sec.~\ref{sec:5}, IBS is added in the simulations together with SC and their interplay is studied in the vicinity of the $3Q_y=31$ resonance. Section~\ref{sec:6} shows the results of similar simulations that include also the SR effect to investigate if the strong radiation damping can mitigate the beam degradation due to SC and IBS in the presence of the excited resonance. Simulations of the full cycle of the CLIC DRs in the presence of all three effects are also shown and the sensitivity to different skew sextupolar errors is investigated. The effect that the vertical dispersion has on the full cycle, in the absence of any other errors, is discussed in Sec.~\ref{sec:7}. The main conclusions of this work are presented in Sec.~\ref{sec:8}. Appendix~\ref{app:conv} contains convergence studies for the number of macro-particles and the number of the longitudinal profile slices that are used in the simulations.

\section{\label{sec:2}The CLIC Damping Rings}
The CLIC main DRs are part of the CLIC injector complex. Their purpose is to damp the large transverse emittance of the beam as received from the upstream accelerators by several orders of magnitude within the repetition rate of \SI{50}{Hz}. For this study, the CLIC DRs conceptual design~\cite{Papaphilippou:1464093} is used as reference with the main parameters summarized in Table~\ref{tab:params}.

\begin{figure}[!t]
    \centering
    \includegraphics*[width=1.\columnwidth]{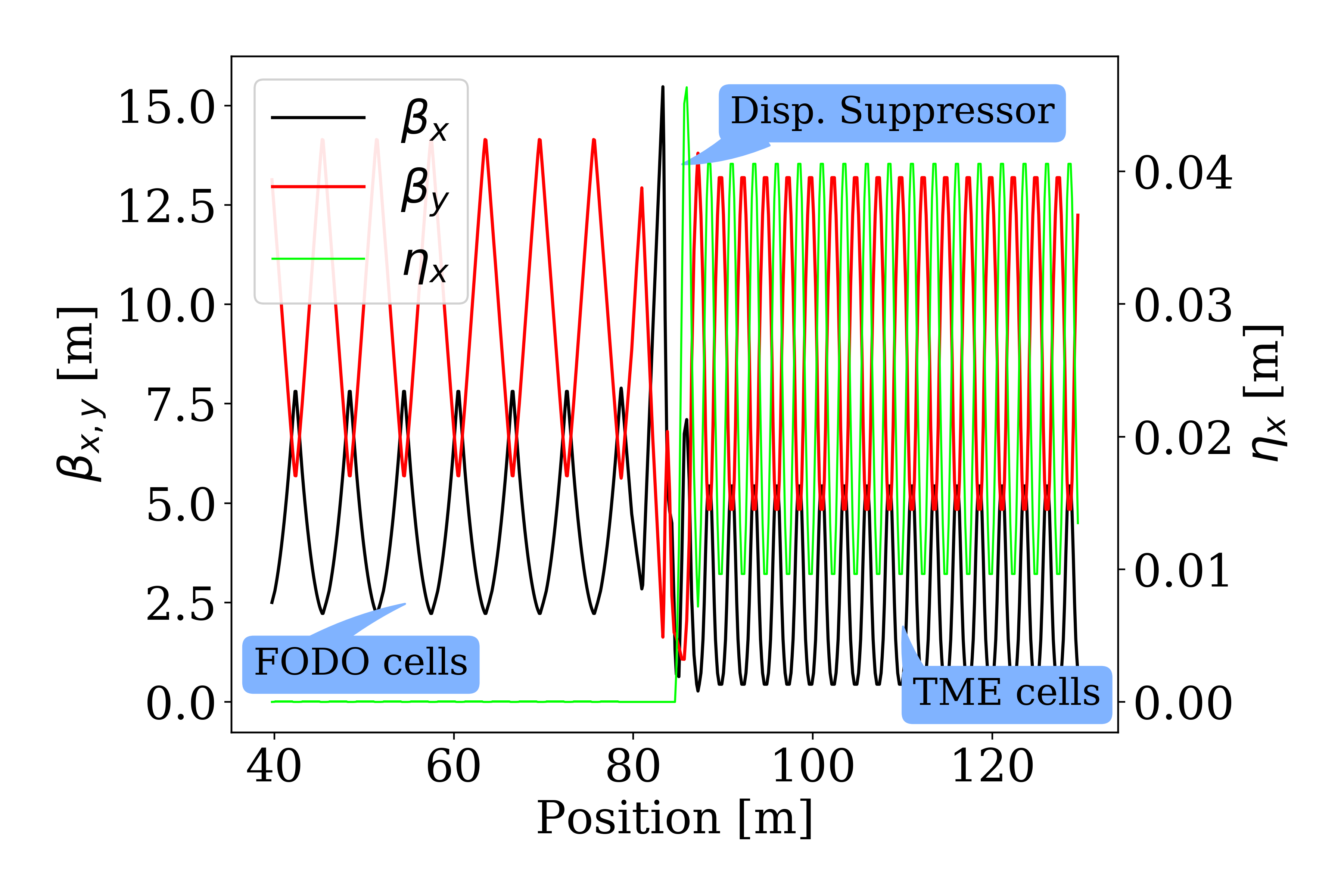}
    \caption{Optics functions along a quarter of the CLIC DR.}
    \label{fig:dr_optics}
\end{figure}

The CLIC DRs have a race-track layout as shown in Fig.~\ref{fig:dr_layout}. The two arcs consist of theoretical minimum emittance (TME) cells, while the two long straight sections are composed of FODO cells, incorporating damping wiggler magnets. Matching of the optics between the arcs and the zero dispersion straight sections is achieved by the dispersion suppressor sections. The optics functions along a quarter of the ring are shown in Fig.~\ref{fig:dr_optics}.

The natural equilibrium emittance at the output of the main DRs is determined by the effects of SR and Quantum Excitation (QE). The steady state emittance, on the other hand, is larger as the IBS effect is taken into account. The CLIC DRs were designed to mitigate the effect of IBS~\cite{PhD_Antoniou, Papadopoulou_2019}. The injection and extraction parameters such as the number of particles per bunch ($N_p$), the transverse geometrical emittances $\varepsilon^\text{geom}$, the longitudinal RMS bunch length ($\sigma_z$) and momentum spread ($\sigma_{\delta}$), for the aforementioned configuration, are summarized in Table~\ref{tab:ineqams}.

\renewcommand{\thefootnote}{\alph{footnote}}
\begin{table}[h]
   \centering
   \caption{CLIC DRs Parameters}
   \begin{tabular}{lc}
       \toprule
       \multicolumn{2}{c}{\textbf{Ring Parameters}}     \\
       \hline
        Circumference [m]              & 427.5          \\ 
        Energy [GeV]                   & 2.86           \\
        Energy loss [MeV/turn]         & 3.98           \\
        Relativistic gamma, $\gamma_r$ & 5597           \\
        Harmonic number, $h$           & 2852           \\ 
        RF voltage [MV]                & 4.5            \\ 
        Momentum compaction factor     & 1.3$\times10^{-4}$       \\
        Bare machine tunes $Q_x$, $Q_y$                   & 48.357, 10.387 \\
        $\delta Q_x$, $\delta Q_y$\footnotemark[1]     & -0.005, -0.17  \\
       \hline
   \end{tabular}
   \label{tab:params}
    \footnotetext[1]{Maximum tune-shift from SC for the 
                    on-momentum particles in the core of the beam}
\end{table}

\begin{table}[!hbt]
   \centering
   \caption{Beam Parameters for the CLIC DRs}
   \begin{tabular}{lccc}
       \toprule
        &\textbf{Injection} &\textbf{Output Values} & \textbf{Target}\\
        &                   &\textbf{without/with IBS}  &               \\
        \hline
        $N_p [10^{9}]$\rule{0pt}{2.5ex}                
                                    & 4.4   & 4.4   & 4.1 \\
        $\varepsilon_x^\text{geom}$[pm]\rule{0pt}{2.5ex} 
                                    & 9.63e3  & 55.7 / 103  & 89.3  \\
        $\varepsilon_y^\text{geom}$[pm]\rule{0pt}{2.5ex} 
                                    & 268     & 0.59 / 0.68 & 0.89  \\
        $\sigma_z$[mm]\rule{0pt}{2.5ex}                  
                                    & 3.2     & 1.46 / 1.56 &  1.81\footnotemark[2] \\
        $\sigma_{\delta}[10^{-3}]$\rule{0pt}{2.5ex}      
                                    & 1.04    & 1.09 / 1.16 & 1.16  \\
       \hline
   \end{tabular}
   \label{tab:ineqams}
   \footnotetext[2]{Evaluated from the momentum spread, considering 
                    the requirement of a normalized longitudinal emittance of$\varepsilon_s=$~6~keV$\cdot$m}
\end{table}

\section{\label{sec:3}Simulation set-up}
The simulations studies of the incoherent effects of SC, IBS and SR were performed using the Polymorphic Tracking Code (PTC)~\cite{PTC} within PyORBIT~\cite{pyorbit}. PyORBIT is a well-known simulation tool that has been intensively benchmarked and used for SC studies for many different accelerators and provides the possibility to use various models for the SC potential, e.g.~a fully self-consistent 2.5 D Particle-In-Cell (PIC) solver, a slice-by-slice SC solver and an analytical frozen potential solver. For the presented studies, the SC effect is included using the SC frozen potential model, where the SC kick is analytically calculated from the lattice functions, the beam intensity and the transverse beam sizes using the Bassetti-Erskine formula~\cite{Bassetti} and modulated by the line density at the longitudinal position of the particle being tracked.

To perform the studies of the interplay between the different incoherent effects mentioned above, two additional modules were implemented in PyORBIT to take into account IBS and SR. These two modules are implemented in Python and can be used, with the appropriate modifications, in any Pythonic code. For example, the IBS module can be used for any particle type and has already been used in codes such as Xsuite~\cite{xsuite}, BLonD~\cite{blond} and PyHEADTAIL~\cite{pyHT}. Their modularity also gives the possibility for parallelization and the use of GPUs for computationally demanding simulations. Both modules were successfully benchmarked against analytical calculations, as shown in the next sections.

\subsection{\label{sec:3sub1}Synchrotron Radiation}
In $e^-/e^+$ rings such as the CLIC DRs, the natural equilibrium emittances are defined by the equilibrium of SR damping and QE. The equilibrium emittances and the radiation damping times are given by the SR integrals~\cite{SRint} and depend only on the optical functions of the ring. The RMS emittance evolution of the beam under the influence of SR and QE is given by:
\begin{equation}
    \frac{d\varepsilon_u}{dt}=-\frac{2}{\tau_u}\varepsilon_u+\frac{2}{\tau_u}\varepsilon_u^{\text{EQ}},
    \qquad \text{$u=x,~y,~z$},
    \label{eq:sreq}
\end{equation}
where $\varepsilon_u^{\text{EQ}}$ are the equilibrium emittances and $\tau_u$ the radiation damping times.

This study is investigating the interplay between multiple incoherent effects that act on a particle distribution. To this end, a momentum kick was implemented into a python module included in PyORBIT, similar to the one found in PyHEADTAIL~\cite{pyHT} and is given by 
\begin{equation}
    p_i(t+dt)=-\frac{2p_i(t)}{\tau_i}+\sigma^{\text{EQ}}_{i}\sqrt{\frac{1}{\tau_i}} R,
    \qquad \text{$i=x,~y$},
    \label{eq:srTkick}
\end{equation}
\begin{equation}
    p_z(t+dt)=-\frac{2p_z(t)}{\tau_z}+\sigma^{\text{EQ}}_{\delta}\sqrt{\frac{1}{\tau_z}} R-\frac{U_0}{\betar^2 \En},
    \label{eq:srLkick}
\end{equation}
for the transverse and longitudinal planes, where $\sigma^{\text{EQ}}_{x,y,\delta}$ are the equilibrium transverse beam sizes and momentum spread, respectively, $R$ is a random number with normal distribution, zero mean, unit standard deviation and without any cutoff, $U_0$ is the energy losses over time $dt$, $\En$ the total energy and $\betar$ is the normalized velocity ($\betar=v/c$ with velocity $v$ and speed of light $c$). The first term in these equations describes the damping from SR due to the different momenta of each particle and thus, different energy loss, the second term the QE and the third term, that is found only in the longitudinal equation, accounts for the average energy loss per particle over time $dt$~\cite{Sands:SR}.

\begin{figure}[!h]
    \centering
    \includegraphics*[width=1.\columnwidth]{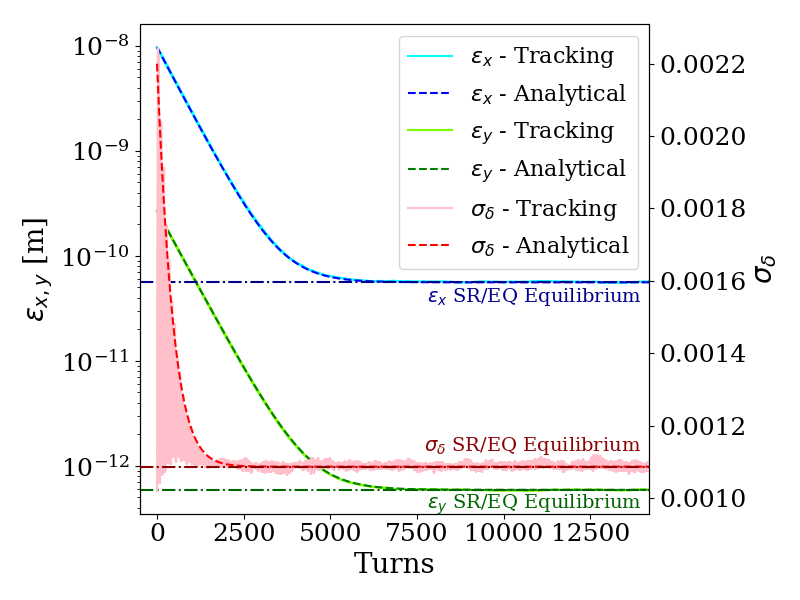}
    \caption{Comparison of the horizontal emittance (blue), vertical emittance
            (green) and momentum spread (red) between analytical calculations
            (dark dashed lines) and the RMS values of the particle distribution in 
            the tracking simulations with SR (light colors).}
    \label{fig:sr_bench}
\end{figure}

The comparison between the analytical evolution of the RMS emittances, as calculated from the solution of Eq.~\eqref{eq:sreq}, and the emittance evolution of a macro-particle distribution in the PyORBIT tracking code including the effect of SR is shown in Fig.~\ref{fig:sr_bench}. The beam distribution is initialized as Gaussian in all three planes using the injection parameters shown in Table~\ref{tab:ineqams}, and it consists of 5000 macro-particles, which is sufficient to perform such simulations as shown by the convergence study reported in Appendix~\ref{app:conv}. The simulated number of turns corresponds to about 20~ms storage time, which is the duration of the full cycle of the CLIC DRs. For the transverse emittances, perfect agreement is observed along the full cycle. For the momentum spread $\sigma_{\delta}$, a difference is observed at the injection of the macro-particle distribution. This is due to the slight longitudinal mismatch of the beam when extracted from the pre-damping rings and injected into the damping rings. On the other hand, for the analytical prediction, the beam is considered always perfectly matched into the bucket. Regardless of this difference of the starting conditions, the longitudinal oscillations and, in turn, the momentum spread are damped to the equilibrium values as predicted from the analytical evolution. Some minor fluctuations can still be observed due to the random nature of the QE and the statistical calculation of the momentum spread.

\subsection{\label{sec:3lsub2}Intra-Beam Scattering}
Intra-beam scattering plays a crucial role in the final steady state emittance of the beam in $e^-/e^+$ damping rings due to the ultra-low emittances that are reached. Intra-beam scattering is a statistical effect due to multiple, small-angle Coulomb scattering events that lead to phase space re-distribution~\cite{Piwinski:0}. Apart from the number of particles in the beam, the growth rates $1/T_{\text{IBS}}$ depend on the optics of the ring, the energy and the transverse and longitudinal emittances. Since the beam parameters evolve in time, the IBS growth rates need to be calculated iteratively over time, to have a self-consistent evaluation. Analytically, the emittance evolution can be expressed through the following differential equation:
\begin{equation}
    \frac{d\varepsilon_u}{dt}=\frac{2}{\Tibs}\varepsilon_u,
    \qquad \text{$u=x,~y,~z$}.
    \label{eq:ibseq}
\end{equation}

Traditionally, most of the IBS studies were performed with the use of this differential equation in combination with the well-known IBS analytical models like the classical model of Piwinski~\cite{Piwinski:0}, the quantum model of Bjorken and Mtingwa~\cite{Bjorken:0}, the high energy approximation of Bane~\cite{Bane_HE} and more. These analytical models are summarized also in~\cite{Martini:2016,Martini:2017} and are useful for calculating the RMS emittance evolution in time, under the assumption of Gaussian beam distributions in all 3 planes. Simulating the IBS effect in more realistic tracking scenarios, where particle distributions with variable tail populations are needed, is not trivial~\cite{PhD_Raubenheimer}. To this end, various macro-particle IBS tracking codes were developed in the past, e.g., the Software for IBS and Radiation Effects (SIRE)~\cite{vivoli:1} and the IBStrack~\cite{osti_1} implemented also in the collective effects simulation tool CMAD~\cite{Pivi:1,Sonnad:1}. Both of these algorithms are inspired by MOCAC (Monte CArlo Code) developed by Zenkevich et al~\cite{Zenkevich:0, Zenkevich:1}, which calculates the IBS effect for arbitrary particle distributions. Most of these codes  are not modular and it is thus impractical to incorporate other beam dynamics effects. The other codes require relatively long computation times. Performing combined simulations of IBS with other effects thus demand for additional simplifications, as discussed below.

When operating above transition, as in the case of the CLIC DRs, the transverse emittances and the momentum spread all have a positive growth rate when IBS is considered alone~\cite{Piwinski:0}, i.e. ignoring all other effects. In this case, one can simulate the IBS effect through a simplified effective kick as proposed by~\cite{Bruce:ibs_kick}, given by the following formula:

\begin{equation}\label{eq:ibs}
    \Delta p_u=\sigma_{p_u}\sqrt{2\ITibs\Trev\sigma_z\sqrt{\pi}\rho(z)} R,
\end{equation}
where $R$ is a random number with normal distribution, zero mean, unit standard deviation and without any cutoff, $\Trev$ is the revolution period, $\sigma_{p_u}$ the standard deviation of the momentum $p_u$ in plane $u$, $\sigma_z$ the bunch length, $\rho(z)$ the longitudinal line density and $\Tibs$ the IBS growth rates. Every turn, each particle receives a change of its momenta depending on the beam parameters and the particle's longitudinal position.

\begin{figure}[!b]
    \centering
    \includegraphics*[width=1.\columnwidth]{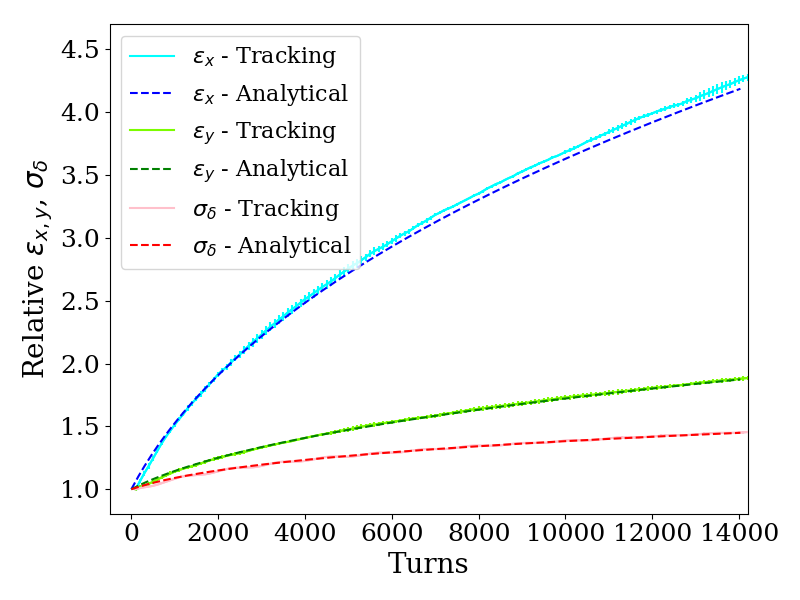}
    \caption{Comparison of the relative horizontal emittance (blue), relative
            vertical emittance (green) and relative momentum spread (red)
            between analytical calculations (dark dashed lines) and IBS tracking
            simulations (light colors). The output values with IBS of 
            Table~\ref{tab:ineqams} are used as initial parameters.}
    \label{fig:ibs_bench}
\end{figure}

This simplified kick has been implemented in the PyORBIT tracking code and has been benchmarked against analytical predictions using the Nagaitsev IBS approach~\cite{Nagaitsev}. This is based on the Bjorken and Mtingwa (BM) model~\cite{Bjorken:0} and the IBS growth rates are expressed through the complete elliptic integrals of the second kind, making the IBS kick calculations and the simulations significantly more efficient in terms of computation time. Both the BM model and Nagaitsev's approach assume Gaussian beam distributions in all three planes to derive the IBS growth rates, and since the kick is applied in a random manner of a normal distribution, it is more accurate for Gaussian distributions. Figure~\ref{fig:ibs_bench} presents a comparison between the analytical predictions (dark dashed lines) and the tracking simulations using 5000 macro-particles (solid lines), starting from the output values with IBS of Table~\ref{tab:ineqams}. An excellent agreement is demonstrated, with less than 2$\%$ difference in all three planes. The difference observed among multiple tracking simulation runs is negligible, and hence, the error-bars are too small to be shown. This excellent agreement verifies that the macro-particle distribution remains Gaussian throughout the simulation and, thus, it agrees with the analytical models that always assume Gaussian beams.

\section{\label{sec:4}Space Charge Studies}
Space charge induces an incoherent tune spread in the transverse planes. The maximum 
tune shift is given by Eq.~(\ref{eq:sc_ts}):
\begin{equation}\label{eq:sc_ts}
    \Delta Q_{x,y}= -\frac{N_p r_c}{(2\pi)^{3/2}\beta_r^2\gamma_r^3\sigma_z} \int_{0}^{C} \frac{\beta_{x,y}(s)}{\sigma_{x,y}(s)\big(\sigma_x(s)+\sigma_y(s)\big)} \,ds,
\end{equation}
where $N_p$ is the number of particles per bunch, $r_c$ is the particle's classical radius, $\sigma_z$ the longitudinal RMS bunch length, $C$ is the ring circumference and $\beta_{x,y}$ and $\sigma_{x,y}$ are the optical beta functions and the transverse beam sizes, respectively.

The SC induced tune spread scales with the normalised energy as $1/\gamma_r^3$ meaning that at high energies SC becomes weaker, and the spread smaller. It scales also with the transverse emittances, becoming stronger as the beam size decreases. The damping rings have ultra-low emittances, and thus SC effects are still relevant despite the high $\gamma_r$ of the particles. In fact, the SC induced maximum tune shifts, as reported in Tab.~\ref{tab:params}, can still be large enough, in particular in the vertical plane, so that particles can cross excited betatron resonances, leading to emittance blow-up or particle losses.

\begin{figure}[t]
    \centering
    \includegraphics*[trim=14 0 10 0, clip, width=1.\columnwidth]{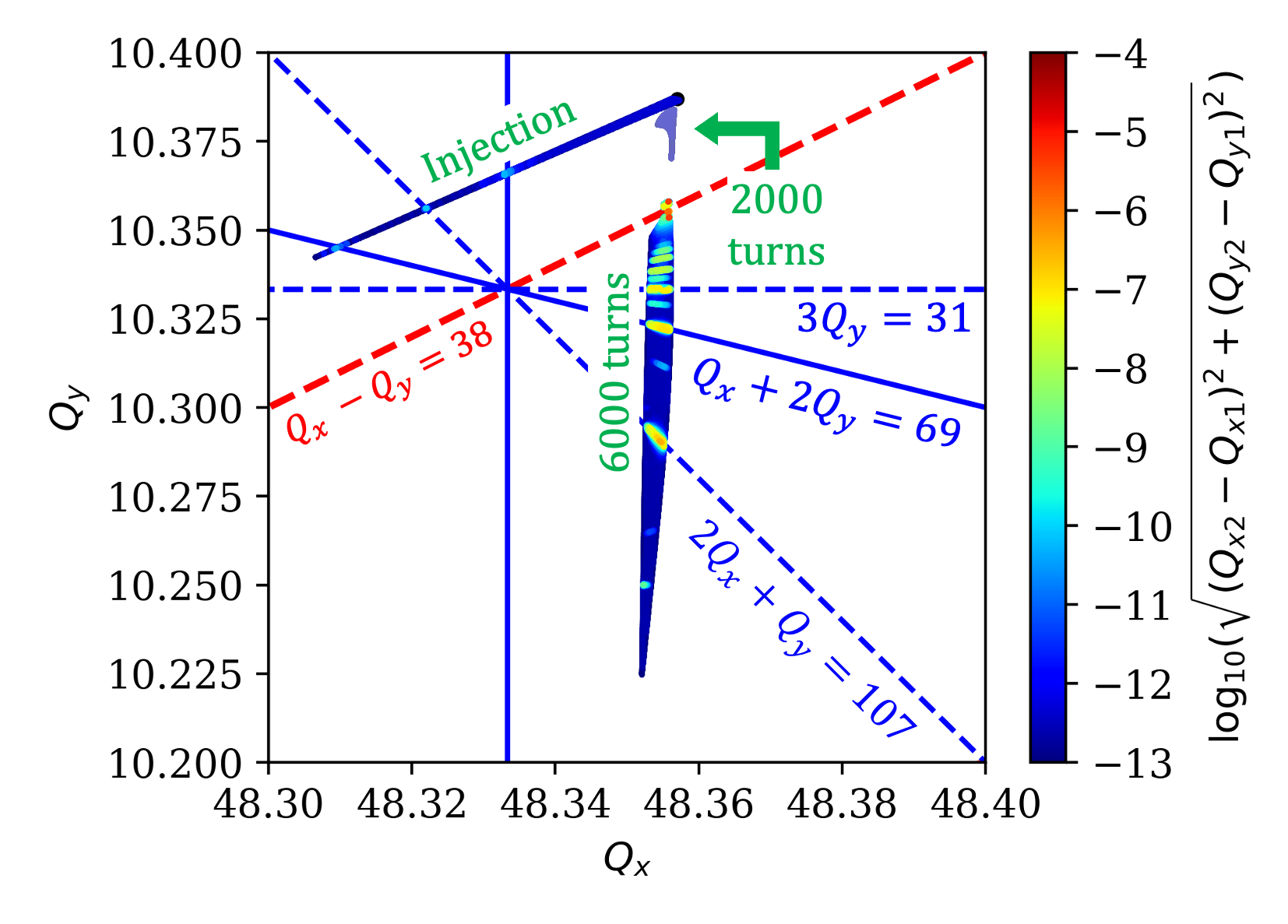}
    \caption{Evolution of the tune footprint during the CLIC DR cycle. Three cases are shown as indicated by the labels on the plot: 1) at injection, 2) after 2000 turns and 3) after 6000 turns. The low order resonances that are expected to affect the beam are annotated. The bare machine working point is $Q_x/Q_y=48.357/10.387$.}
   \label{fig:foot_evol}
\end{figure}

In order to investigate the impact of the SC effect along the CLIC DR cycle, tune footprints were calculated for on-momentum particles up to $3\upsigma$, as shown in Fig.~\ref{fig:foot_evol}. In these studies the frozen SC potential was used, i.e.~the SC potential was pre-computed according to the initial beam parameters and the corresponding beam sizes at each SC kick of the about 1200 SC nodes around the lattice, with a longitudinal line density profile of 200~slices, which was found to be sufficient, as shown in Appendix~\ref{app:conv}. The tunes of the test-particles were determined using the NAFF algorithm~\cite{Papafilippou:naff} and the tune diffusion is indicated by the color code. The tune diffusion is calculated as the difference of the tunes over two time spans (normally in multiples of the synchrotron period), indicated as $Q_{x,y1}$ and $Q_{x,y2}$, and thus given by $\log_{10}\left(\sqrt{(Q_{x2}-Q_{x1})^2+(Q_{y2}-Q_{y1})^2}\right)$. The red and blue lines correspond to systematic and non-systematic resonances accordingly, while the solid and dashed lines correspond to normal and skew resonances, respectively. 

Starting at injection with large transverse emittances, particles experience large amplitude detuning from the non-linearities introduced by the sextupoles and the tune footprint looks like a line, parallel to the coupling resonance, as shown in Fig.~\ref{fig:foot_evol}. As the emittances are damped down, the detuning from magnet non-linearities is decreased, but the SC induced tune spread becomes dominant reaching a maximum vertical tune shift of the order of $\Delta Q_y=-0.17$.

\begin{figure}[h]
    \centering
    \includegraphics*[trim=15 10 5 10, width=.9\columnwidth]{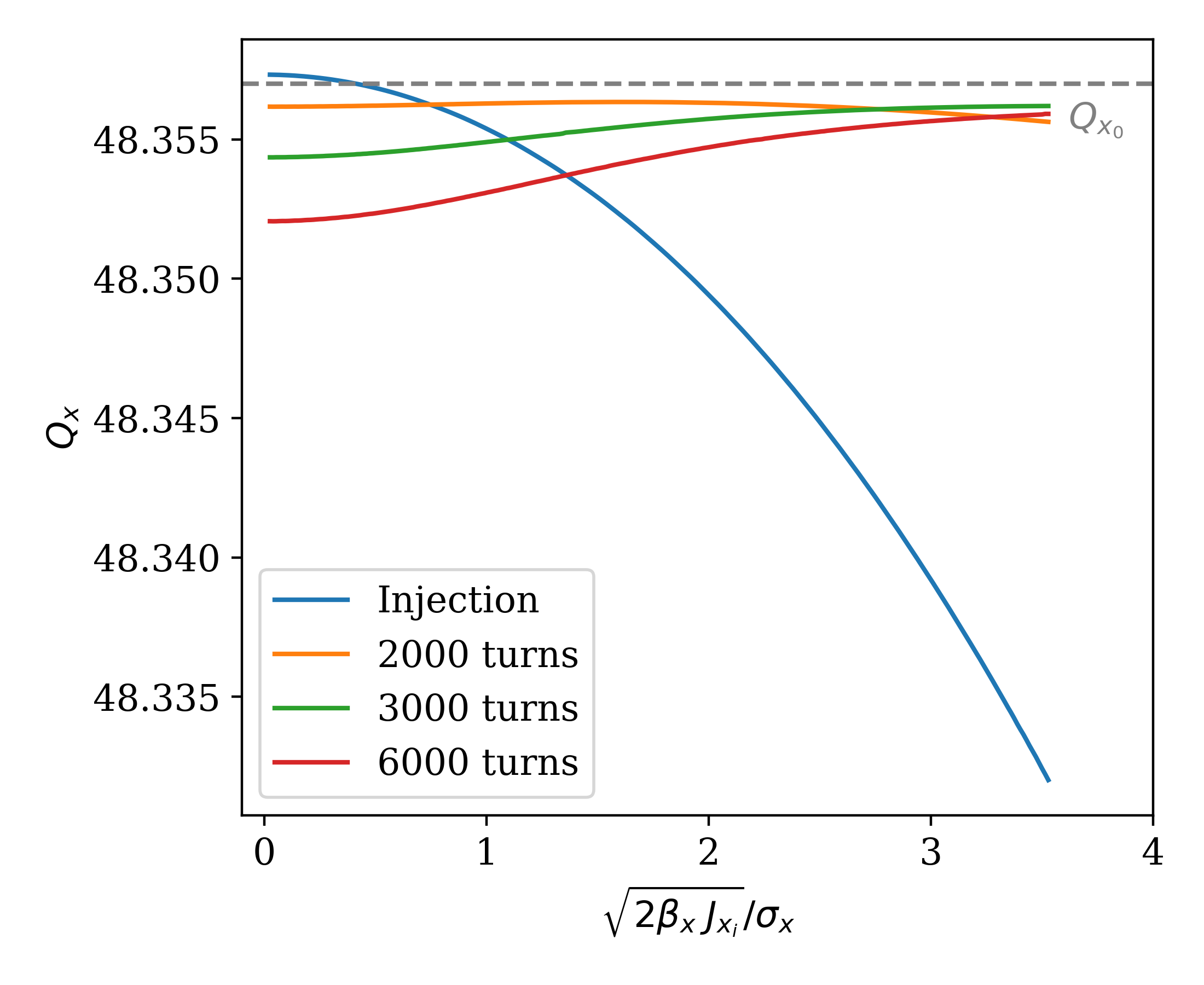}
    \includegraphics*[trim=15 10 5 10, width=.9\columnwidth]{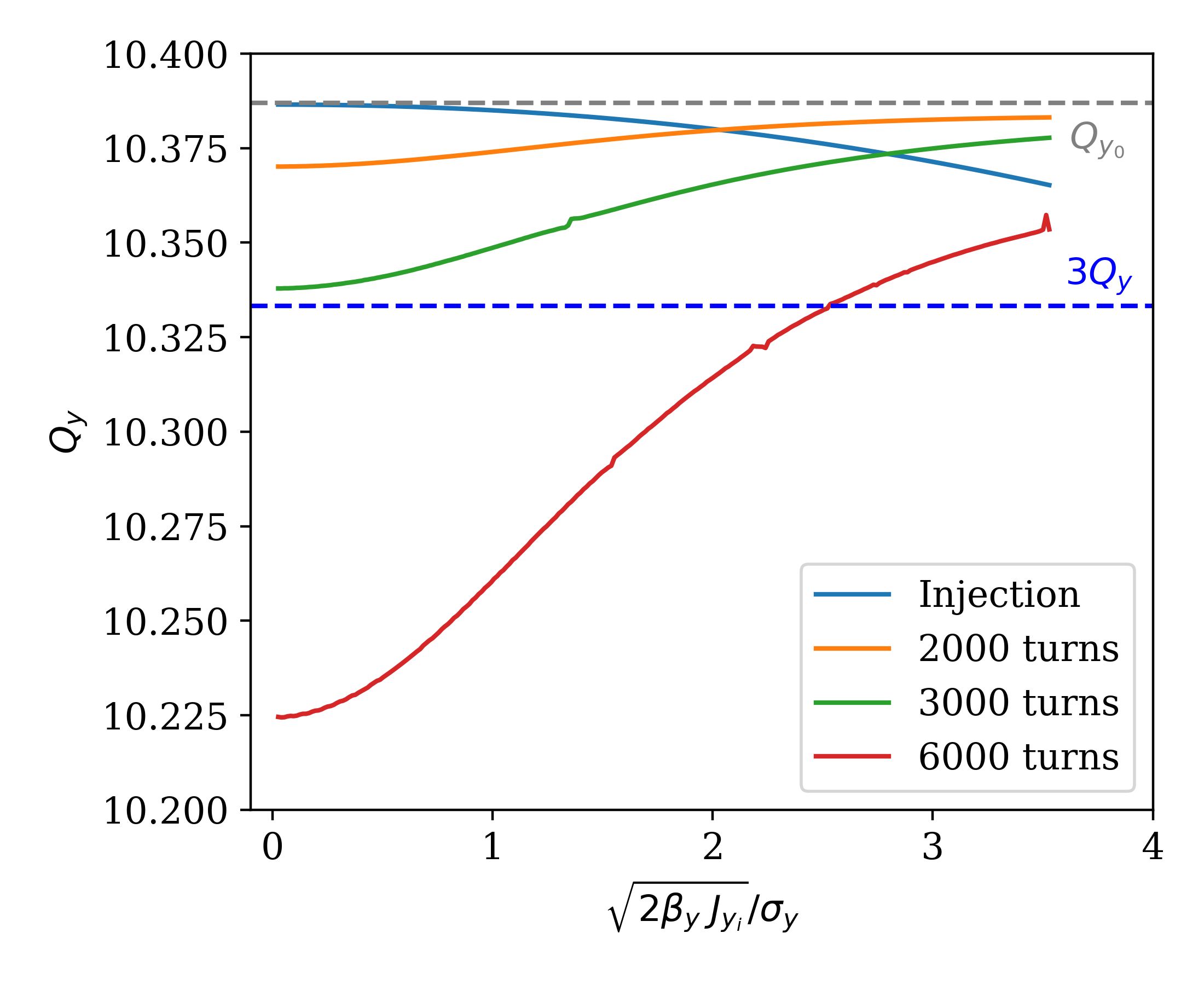}
    \caption{Horizontal~(top) and vertical~(bottom) tune as a function of the horizontal and vertical initial positions, respectively, of the particles in $\sigma$, for particles along the diagonal in the normalized configuration space, at injection~(blue), at 2000~(orange), 3000~(green) and 6000~(red) turns from injection as obtained from analyzing tracking data.}
   \label{fig:sigx_qx}
\end{figure}

This transition from the detuning from lattice non-linearities to the SC induced one can also be seen in Fig.~\ref{fig:sigx_qx} for the horizontal (top) and vertical (bottom) planes. At injection, particles with high action experience amplitude detuning while at approximately 2000~turns there is barely any tune shift. As the emittances are further damped down, SC becomes dominant leading to large tuneshift of particles with small action. Due to this behavior, the beam with the injection parameters can be affected by other resonances than the resonances that are encountered at the steady state emittance of the beam. These two regimes thus need to be studied separately.

The SC dominated incoherent tune footprint overlaps several low order resonances. The ones that seem more likely to affect the performance of the machine are the systematic skew resonance $Q_x-Q_y=38$ (linear coupling) or the higher order coupling resonances, the vertical skew resonance $3Q_y=31$, the non-linear coupling resonance $Q_x+2Q_y=69$, and the skew non-linear coupling resonance $2Q_x+Q_y=107$, as shown in Fig.~\ref{fig:foot_evol}. The linear coupling resonance needs to be well corrected in order to achieve the small vertical emittance target through SR damping, even in case SC and IBS would be of no concern. The work presented in this paper is focused on the $3Q_y=31$ resonance, as from the resonances mentioned above it is the closest to the nominal working point of the CLIC DRs. The aim is to study the acceptable resonance excitation without preventing the CLIC DRs from delivering the vertical emittance extraction requirements. This resonance can be excited principally by tilt errors in the strong chromatic sextupoles of the machine.

\subsection{The $3Q_y=31$ resonance}
\begin{figure}[ht]
    \centering
    \includegraphics*[trim=12 0 10 0, clip, width=1.\columnwidth]{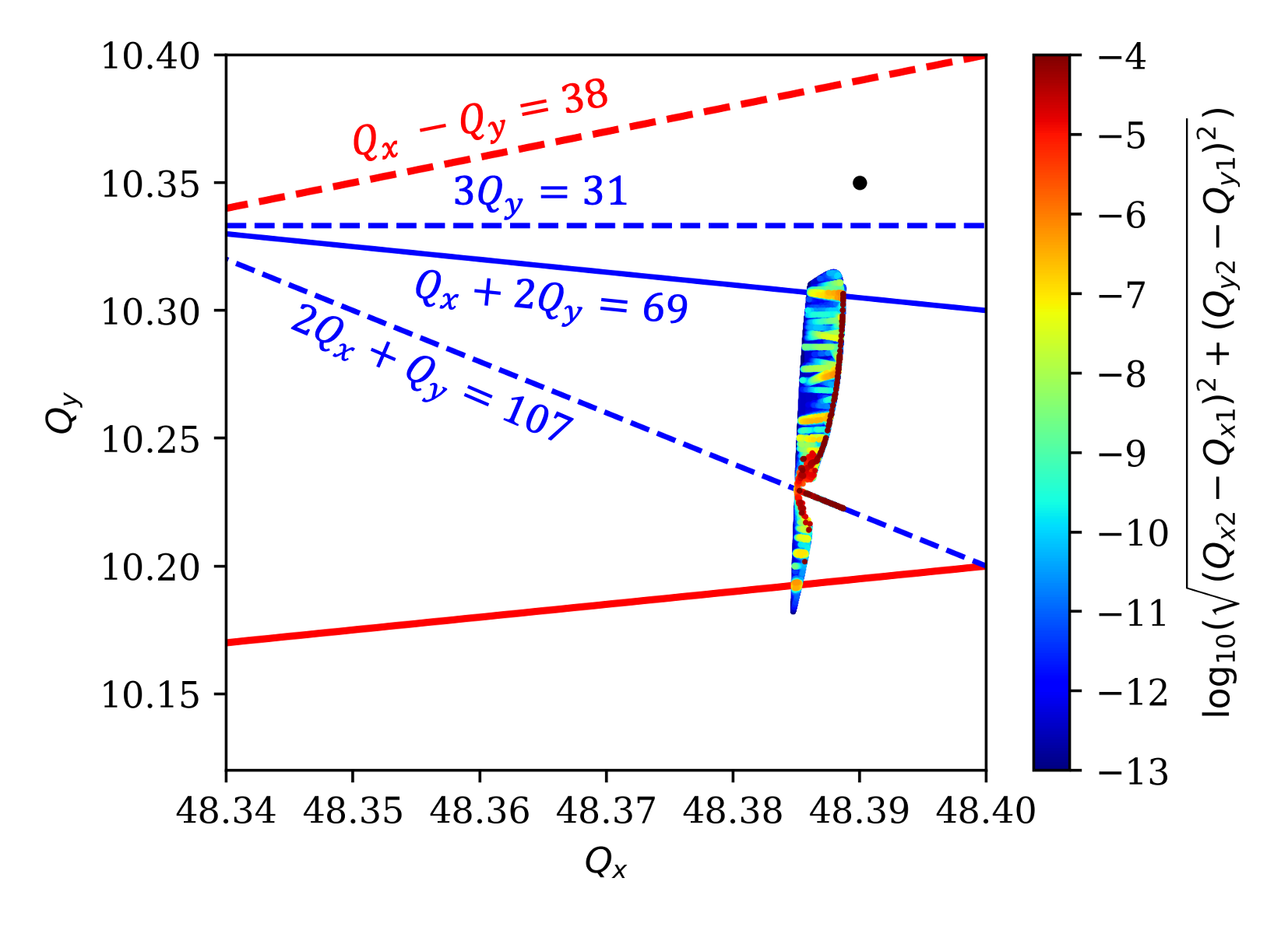}
    \caption{Frequency Map Analysis for the on-momentum particles with $Q_x=48.39,
            \;Q_y=10.35$ and one skew sextupole with normalized integrated
            strength $k_{\text{2s}L}=150~\text{m}^{-2}$ in presence of space charge. The resonances that are most likely to affect the beam are annotated.}
    \label{fig:foot_comp}
\end{figure}

To excite the $3Q_y=31$ resonance in this study, a skew sextupole error is included in the straight section of the lattice. This choice was made in order to avoid any effect from the dispersion that is present in the arcs, which would create chromatic coupling. Adding only one skew sextupole error in a non-dispersive region of the lattice will not only excite the $3Q_y=31$ resonance but also other (unwanted) skew sextupole resonances. For example, with a single skew sextupole error, the $2Q_x+Q_y=107$ resonance gets strongly excited, as revealed by the frequency map analysis for on-momentum particles shown in Fig.~\ref{fig:foot_comp}. Particles get trapped on the resonance as revealed from the large tune diffusion. This study is mainly focused on the $3Q_y=31$ resonance and any contribution from other resonances should be avoided.

\begin{figure}[b]
    \centering
    \includegraphics*[width=1.\columnwidth]{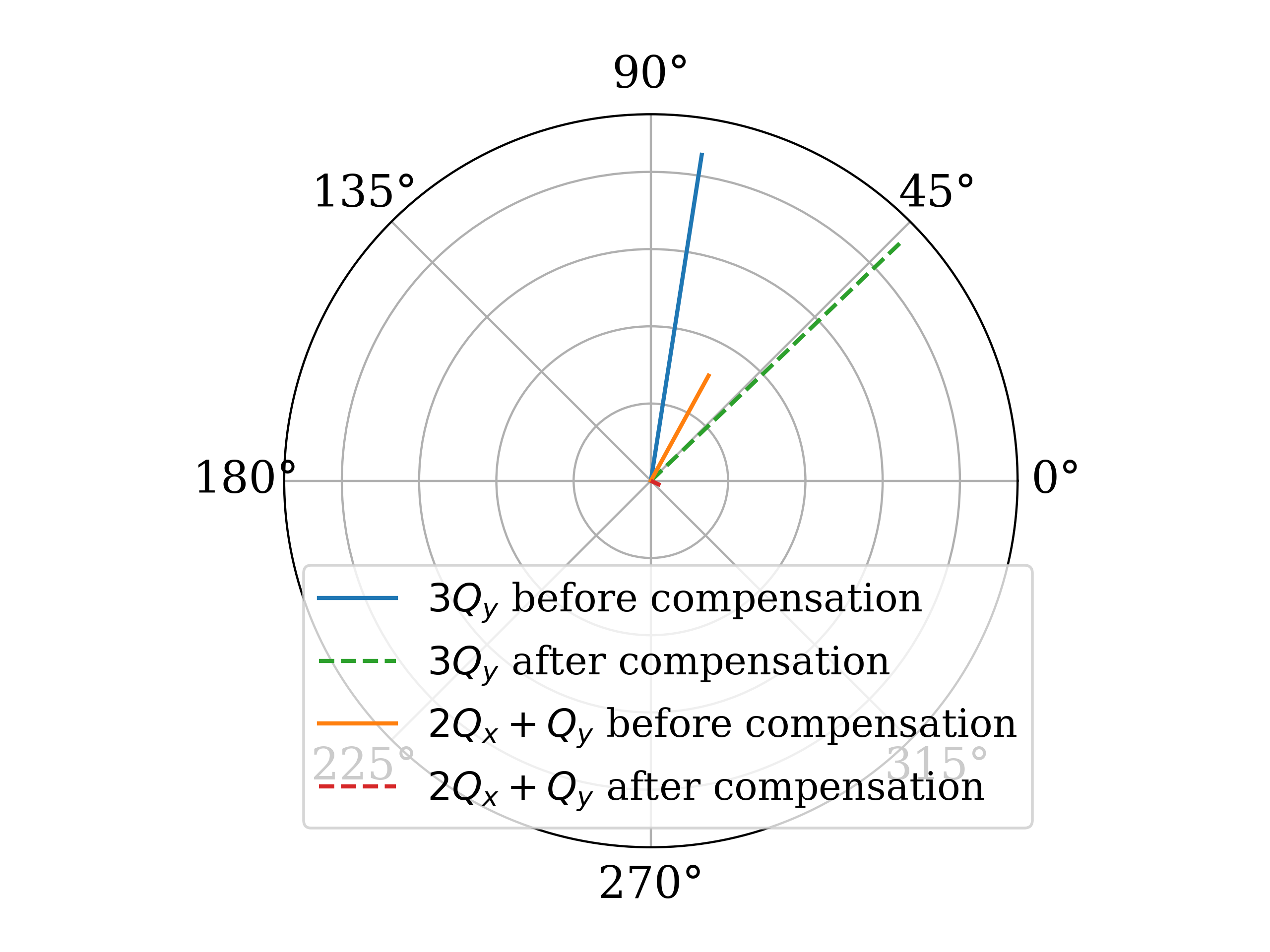}
    \caption{Illustration of the RDTs of the $3Q_y=31$ and $2Q_x+Q_y=107$ resonances using a single skew sextupole~(full lines) and two identical skew sextupoles in the same straight section to compensate the $2Q_x+Q_y=107$ resonance~(dashed lines). The strength  of two skew sextupoles had to be reduced by $33\%$ compared to the single skew sextupole to achieve the same amplitude of the $3Q_y=31$ RDT.}
    \label{fig:res_comp}
\end{figure}

In order to compensate the strongly excited $2Q_x+Q_y=107$ resonance the Resonance Driving Terms (RDTs) for the $3Q_y$ and the $2Q_x+Q_y$ resonances were computed using PTC in MAD-X~\cite{madx}. Based on this analysis, a second identical skew sextupole with the same normalized integrated strength was installed towards the end of the same straight section with such a phase advance that the $2Q_x+Q_y$ resonance is suppressed. To maintain the same level of excitation in terms of RDT amplitude for the $3Q_y$ resonance, the strength of the two skew sextupoles had to be reduced by 33\% compared to the case of using an individual magnet. The RDT calculations before and after these lattice manipulations are shown in Fig.~\ref{fig:res_comp}.

With the two skew sextupole configurations, a sensitivity study was performed in order to define a normalized integrated skew sextupole strength $k_{\text{2s}}L$ that allows studying the interplay between SC, IBS and SR. The initial beam parameters correspond to the steady state (indicated as SS) emittances and the initial particle coordinates are generated to follow a Gaussian distribution in all three planes, with 10000 macro-particles (this choice is based on convergence studies as shown in Appendix~\ref{app:conv}). The working point was set to $(Q_x, Q_y)~=~(48.39, 10.35)$, which is expected to be significantly affected by the resonances, as shown in Fig.~\ref{fig:foot_comp}. The simulation is performed for 14100 turns, which corresponds to the full cycle of the CLIC DRs. Treating the variations of the beam parameters in a self-consistent way can be expensive in terms of computational resources. Thus, in these studies, the frozen SC potential was used. In order to preserve an adiabatic self-consistency of the SC kick during the simulation, the SC potential was recomputed every $100$~turns according to the evolution of the transverse beam parameters and the longitudinal line density of the macro-particle distribution.

\begin{figure}[!h]
    \centering
    \includegraphics*[width=1.\columnwidth]{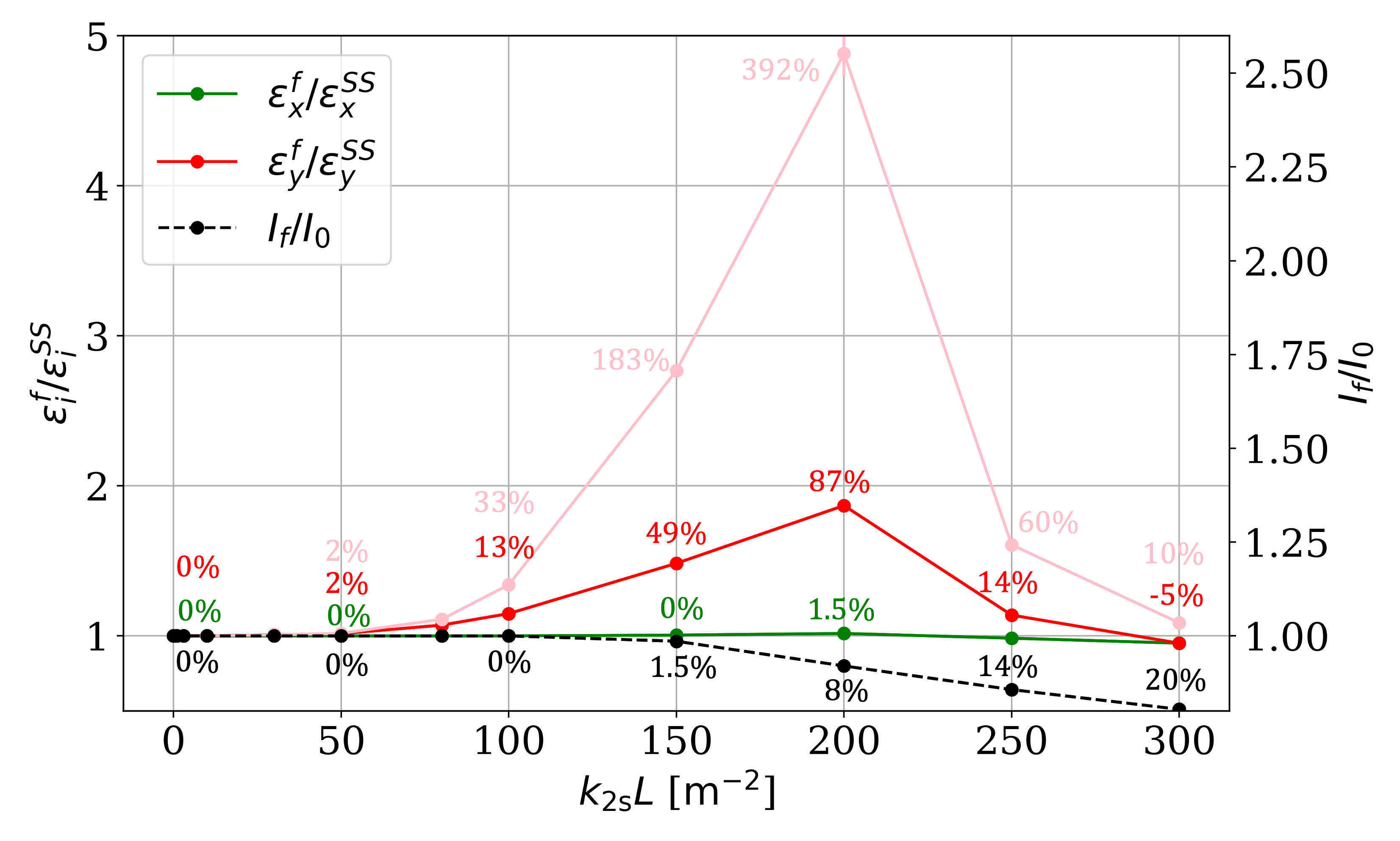}
    \caption{Ratios~(final over initial) of the horizontal~(green) and vertical~(red) emittances and intensity~(black) for different integrated normalized skew sextupole strengths for the WP $Q_x=48.39$, $Q_y=10.35$. The vertical emittance is evaluated for particles up to $7\sigma$ (red) and $18\sigma$~(pink). The resonance $2Q_x+Q_y=107$ is compensated.}
    \label{fig:k2sl_2sex_comp}
\end{figure}

The result of a scan of the skew sextupole strength is shown in Fig.~\ref{fig:k2sl_2sex_comp}, where the ratios of the final over the initial values of the horizontal~(green) and vertical~(red) emittances and intensity~(black) are presented as a function of the normalized integrated strength $k_{\text{2s}}L$ of the skew sextupoles. The horizontal and the vertical emittances are evaluated using the second-order moments of the particle distribution following a similar approach as described in~\cite{Franchetti:2017aa}. In particular, a Gaussian fit is initially applied to the beam profile, and the weighted RMS emittances are calculated using the profile up to the $7\sigma$ extent of the Gaussian fit. For comparison, the vertical emittance evaluated for up to $18\sigma$ of the Gaussian fitted distribution is shown in pink. Using the full profile for the computation of the emittance leads to considerable larger values. This is due to the fact that a limited number of outlier macro-particles move to very high amplitudes, without getting lost as the ring has a very large physical aperture, and they have a significant impact on the emittance evaluation.

Based on this sensitivity study, the integrated strength of the two skew sextupoles was set to $k_{\text{2s}}L=100~\text{m}^{-2}$, as it allows for a controllable beam behavior but sufficient blow-up to observe any possible interplay between IBS and SR. For reference, the magnitude of this error corresponds to the case that one of the normal sextupoles of the ring is tilted about the longitudinal axis by about $70^\circ$, which is clearly unrealistic. However, in a more realistic scenario, this error could arise from the combination from many of the 400 sextupoles of the CLIC DRs. In this configuration, a vertical emittance blow-up of $10\%$ is observed within a full cycle duration, without any induced particle losses. This very weak response of the beam to this resonance could be related to the fact that the CLIC DRs have a synchrotron motion with a period of about $T_s=100$ turns, which is quite typical for lepton storage rings, but is fast compared to typical hadron synchrotons. In fact, it was shown that fast tune modulation leads to the excitation of resonance sidebands~\cite{Satogata,Kostoglou} instead of the usual periodic resonance crossing and the associated particle scattering that is commonly observed in most low-energy hadron rings~\cite{Franchetti:2006aa}, where the synchrotron motion is multiple times slower.

\subsubsection{Resonance Sidebands}
\begin{figure}[h!]
    \centering
    \includegraphics[trim=20 0 5 0, width=1.\columnwidth]{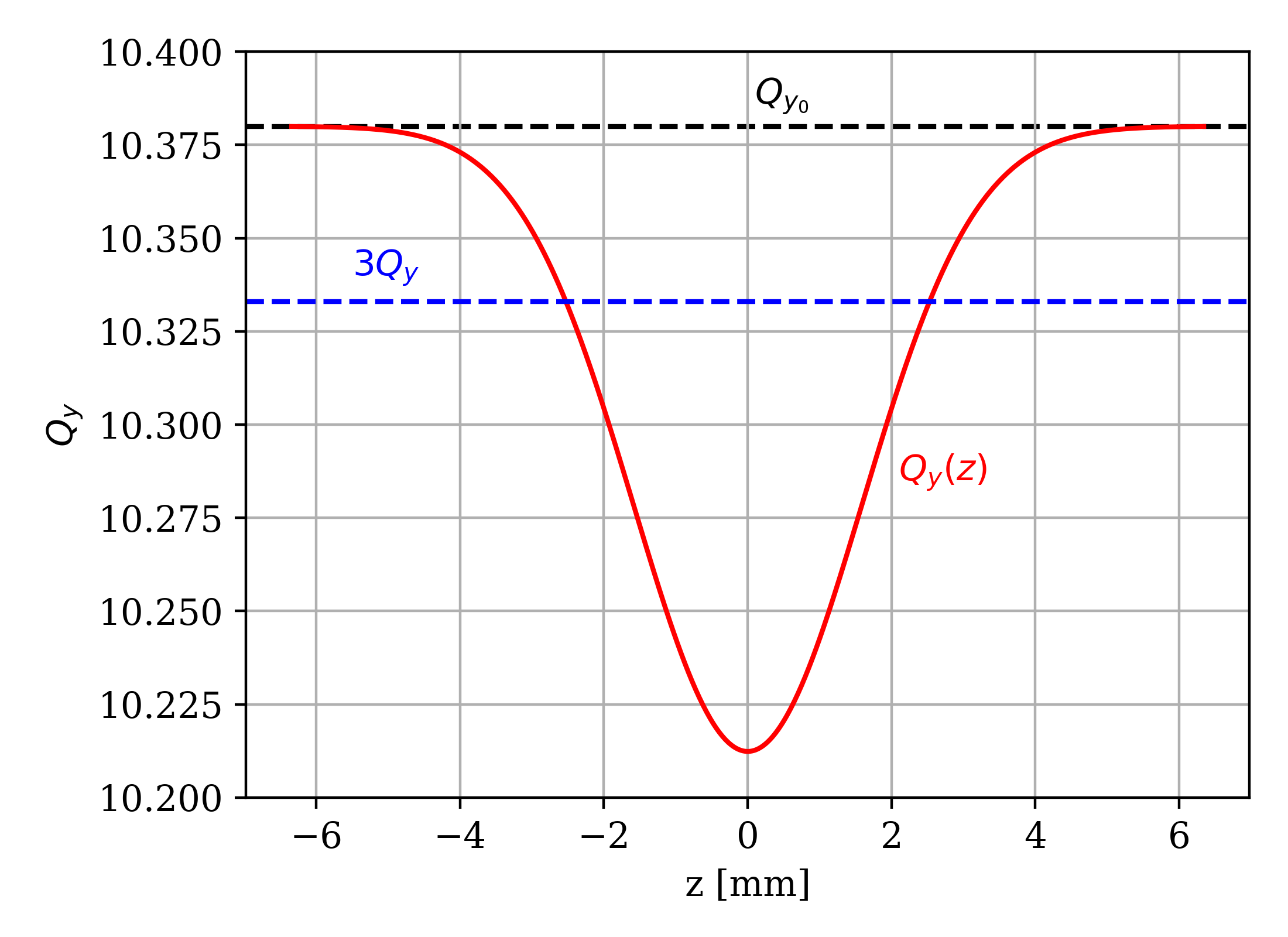}
    \caption{Maximum vertical tune shift due to SC as a function of the 
            longitudinal coordinate z with respect to the reference particle.
            The effect of chromaticity is not considered.}
    \label{fig:DQy_z}
\end{figure}

Longitudinal motion modulates the SC induced tune shift that a particle experiences depending on its location in the longitudinal bunch profile. Particles at the tails of the longitudinal bunch profile experience the minimum tune shift while particles located in the core, experience the maximum tune shift. An example of the maximum SC tune shift as a function of the longitudinal position $z$ for the vertical tune $Q_y=10.38$ is shown in Fig.~\ref{fig:DQy_z}. In this example, the tune modulation from chromaticity is not taken into account, and the calculation is done for a particle at the closed orbit.

\begin{figure}
    \centering
    \begin{tabular}{@{}c@{}}
        \includegraphics[width=1.\columnwidth]{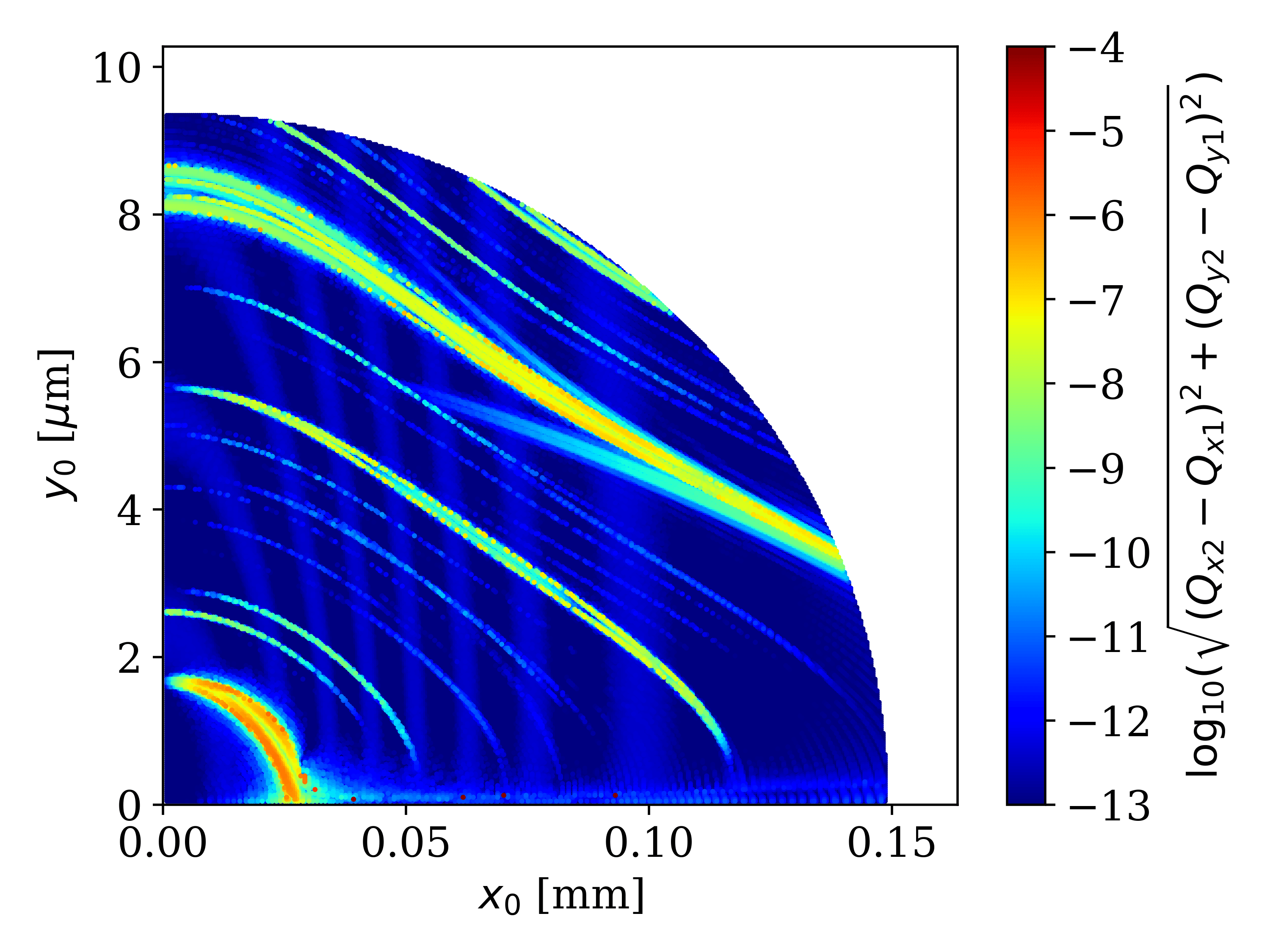} \\
        \small (a)
    \end{tabular}
    \begin{tabular}{@{}c@{}}
        \includegraphics[width=1.\columnwidth]{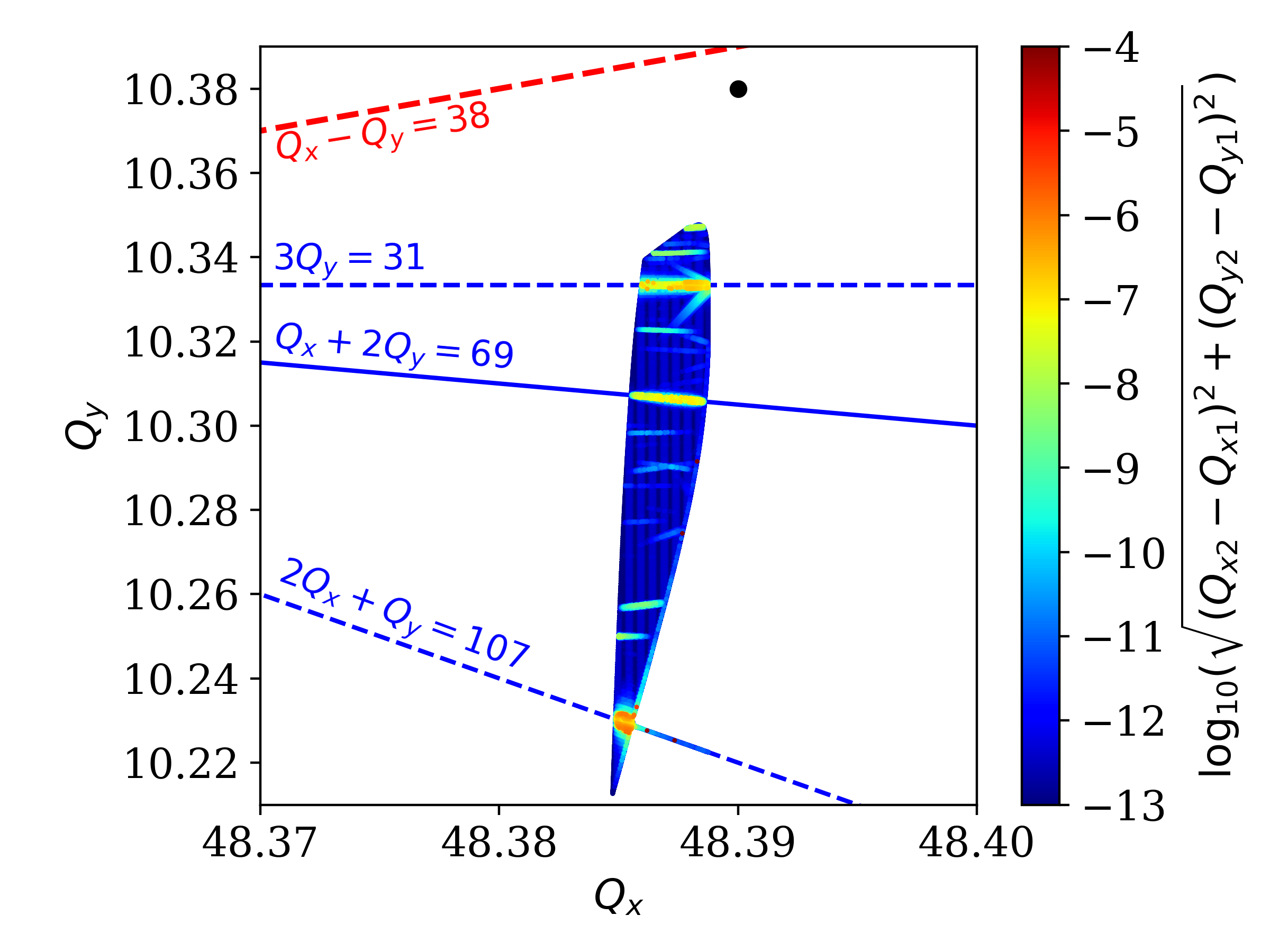} \\
        \small (b)
    \end{tabular}
    \caption{Initial distribution~(a) and Frequency map analysis of the 
            tune footprint~(b) of the on-momentum particles. Tune diffusion shown with
            the color map.} \label{fig:on-mom}
\end{figure}

On-momentum particles, located in the core of the longitudinal distribution, do not experience any tune modulation due to the lack of longitudinal motion. In this case, the particles sample the usual betatron resonances of the lattice as shown in Fig.~\ref{fig:on-mom}b. The initial transverse coordinates of these particles are shown in Fig.~\ref{fig:on-mom}a. On the other hand, the SC induced tune shift is modulated for off-momentum particles due to their synchrotron motion. As particles move closer to the core of the beam, their SC tune shift grows, while as they move to the edges of the longitudinal bunch profile their tune shift is minimized. Each particle passes through the maximum tune shift twice per synchrotron period and thus, the SC tune shift is modulated at even multiples of the synchrotron tune. 

\begin{figure}
    \centering
    \begin{tabular}{@{}c@{}}
        \includegraphics[width=1.\columnwidth]{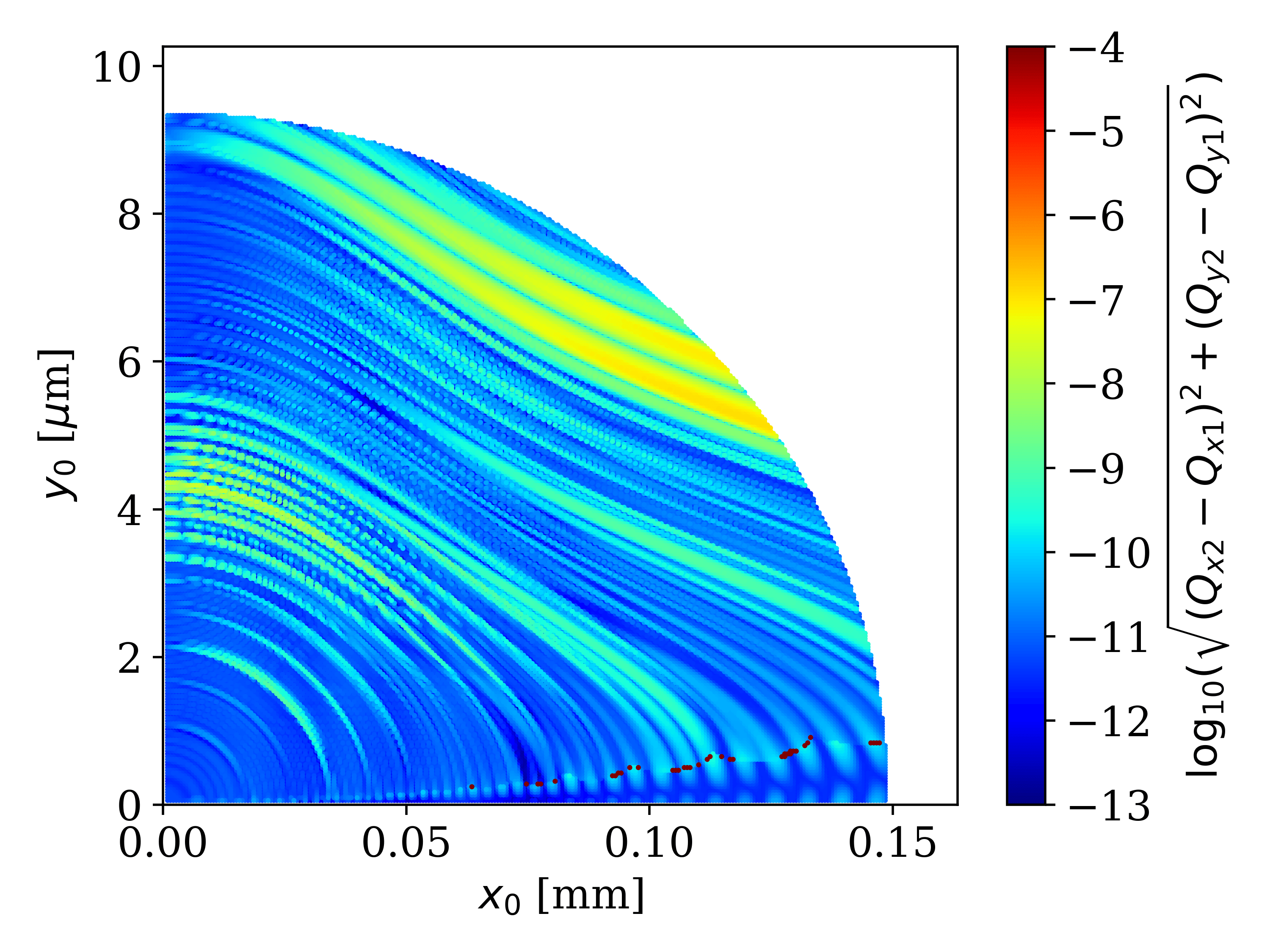} \\
        \small (a)
    \end{tabular}
    \begin{tabular}{@{}c@{}}
        \includegraphics[width=1.\columnwidth]{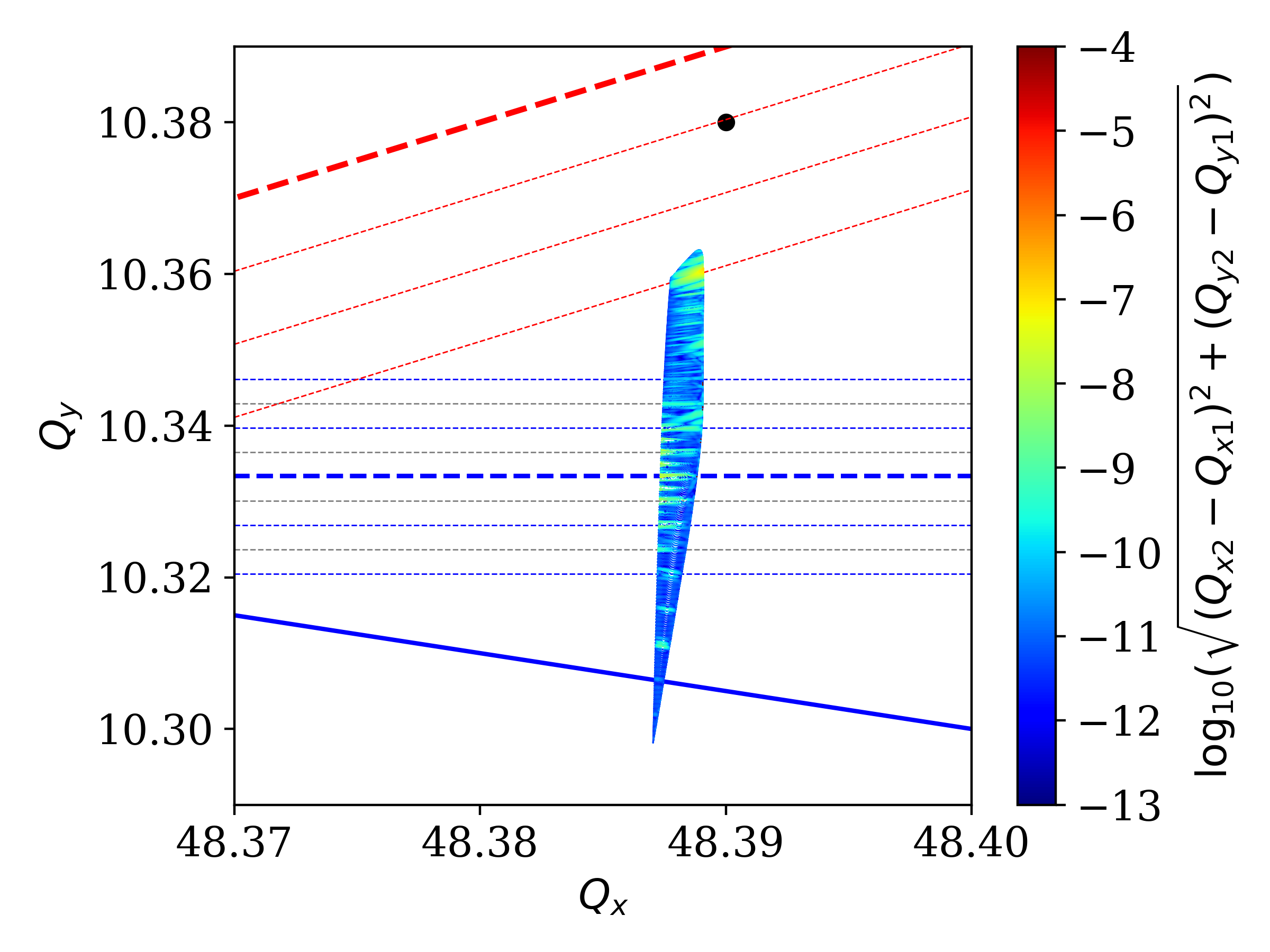} \\
        \small (b)
    \end{tabular}
    \caption{Initial distribution~(a) and Frequency map analysis of the 
            tune footprint~(b) of off momentum particles at $1\sigma_z$, 
            $Q_x=48.39$ and $Q_y=10.38$~(black dot). Tune diffusion is 
            observed around the sidebands of the $3Q_y=31$~(thin blue 
            dashed lines), $6Q_y=62$~(thin grey dashed lines) and
            $Q_x-Q_y=38$~(thin red dashed lines) or higher order resonances.} 
            \label{fig:ts100_1sp}
\end{figure}

\begin{figure}[!t]
    \centering
    \includegraphics*[width=1.\columnwidth]{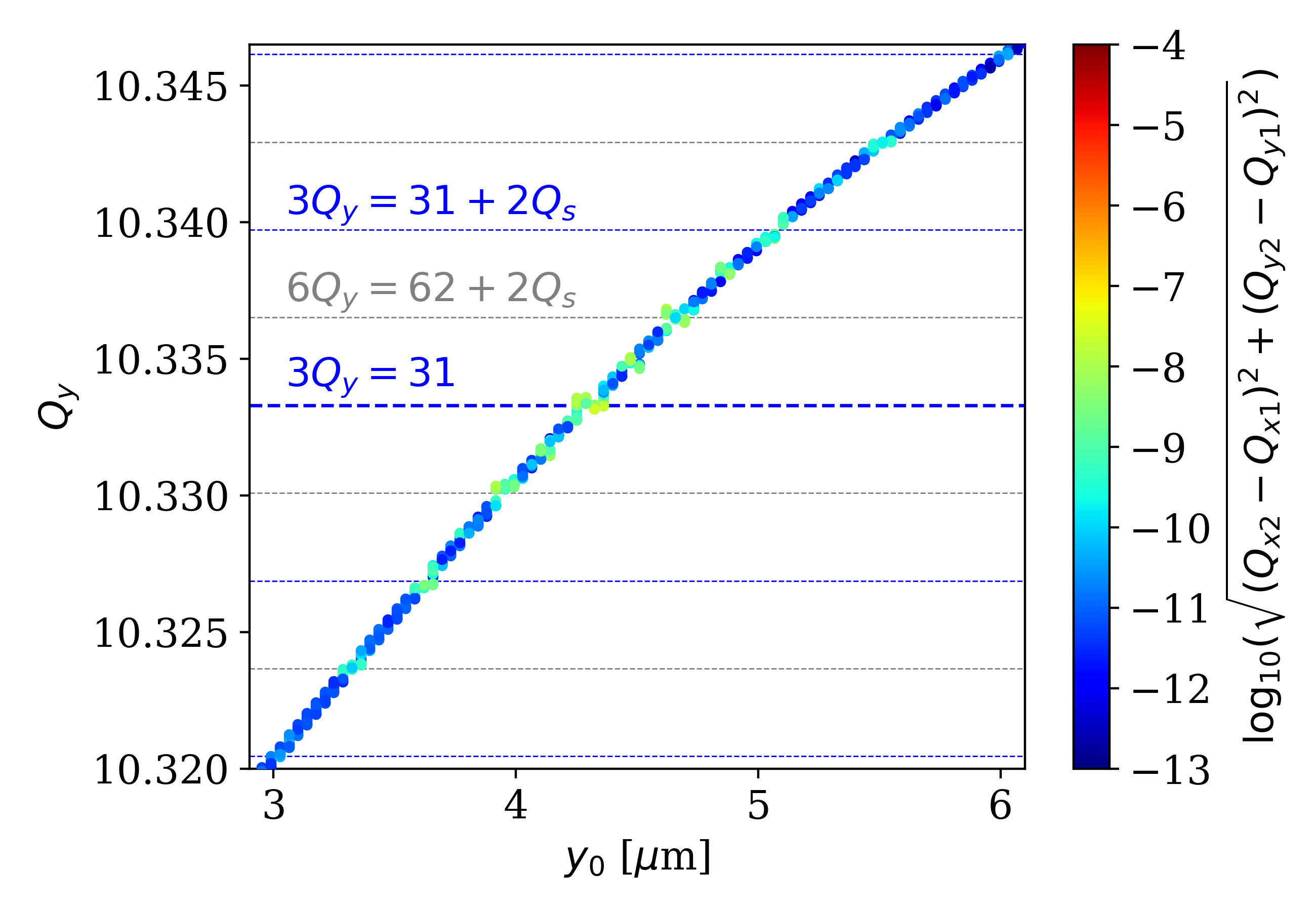}
    \caption{Tune diffusion for a line of particles at $x_0=0$. Diffusion is
            observed not only around of the sidebands of the $3Q_y=31$ resonance,
            $3Q_y=31\pm2nQ_s$, with $n=1,2,3...$, shown in blue dashed lines,
            but also around the sidebands of the $6^{\text{th}}$ order resonance,
            $6Q_y=62\pm2nQ_s$ shown in grey. Weaker higher order sidebands are
            also observed.}
    \label{fig:ts100_y0Q}
\end{figure}

The movement of the off-momentum particles in the tune space creates chaotic motion around resonances. The size of the chaotic phase space area depends on the strength of the resonance and on the frequency of the synchrotron oscillations. In the case of fast synchrotron motion, resonance sidebands start to arise at even multiples of the synchrotron tune since the SC tune shift is modulated with twice the synchrotron tune. In the case of the $3Q_y$ resonance the sideband condition will be $3Q_y=31\pm2nQ_s$ with $n$ an integer number. Figure~\ref{fig:ts100_1sp}a shows the position of the particles of the initial transverse distribution for off-momentum particles at $1\sigma_z$, while Fig.~\ref{fig:ts100_1sp}b depicts the tune footprint of the same distribution. It becomes clear that the sidebands of the $3Q_y$ resonance cause tune diffusion. Another representation of the same tracking study is illustrated in Fig.~\ref{fig:ts100_y0Q}, where the vertical tune as a function of the initial vertical position for particles at $x_0=0$ is shown. Resonance sidebands produce significantly weaker diffusion than the usual scattering on the resonance islands, as shown in more detail below.

For comparison, a study case was investigated with a different synchrotron period. To change the synchrotron period of the machine, the harmonic number $h$ and the RF voltage were modified, keeping the momentum spread constant. Since the properties of the RF bucket are changed, the same momentum spread corresponds to a different bunch length. Since SC strongly depends on the longitudinal line density, the beam intensity $N_p$ was also modified accordingly in order to to maintain the same SC tune shift as the nominal case. The full list of parameters, for the different synchrotron periods that are used for these studies are summarized in Table~\ref{tab:N_h_V}.

\begin{table}[!hbt]
   \centering
   \caption{Summary of parameters adapted to modify the synchrotron period $T_s$ for a given SC tune spread, i.e., harmonic number $h$, RF voltage $V_{RF}$ and beam intensity $N_p$.}
   \begin{tabular}{cccc}
       \toprule
        $\mathbf{T_s}$ \textbf{[Turns]}\rule{0pt}{2.5ex} & $\mathbf{h}$ & $\mathbf{N_p [10^{9}]}$ & $\mathbf{V_{RF}}$ \textbf{[MV]}\\[0.4ex]
        \hline
        $25$\rule{0pt}{2.5ex}    & 11839 & 1.06  & 19   \\
        $50$    & 5919  & 2.1   & 9.3  \\
        $100$   & 2852  & 4.4   & 4.5  \\
        $1000$  & 296   & 42.4  & 0.47  \\
       \hline
   \end{tabular}
   \label{tab:N_h_V}
\end{table}

\begin{figure}
    \centering
    \begin{tabular}{@{}c@{}}
        \includegraphics[width=1.\columnwidth]{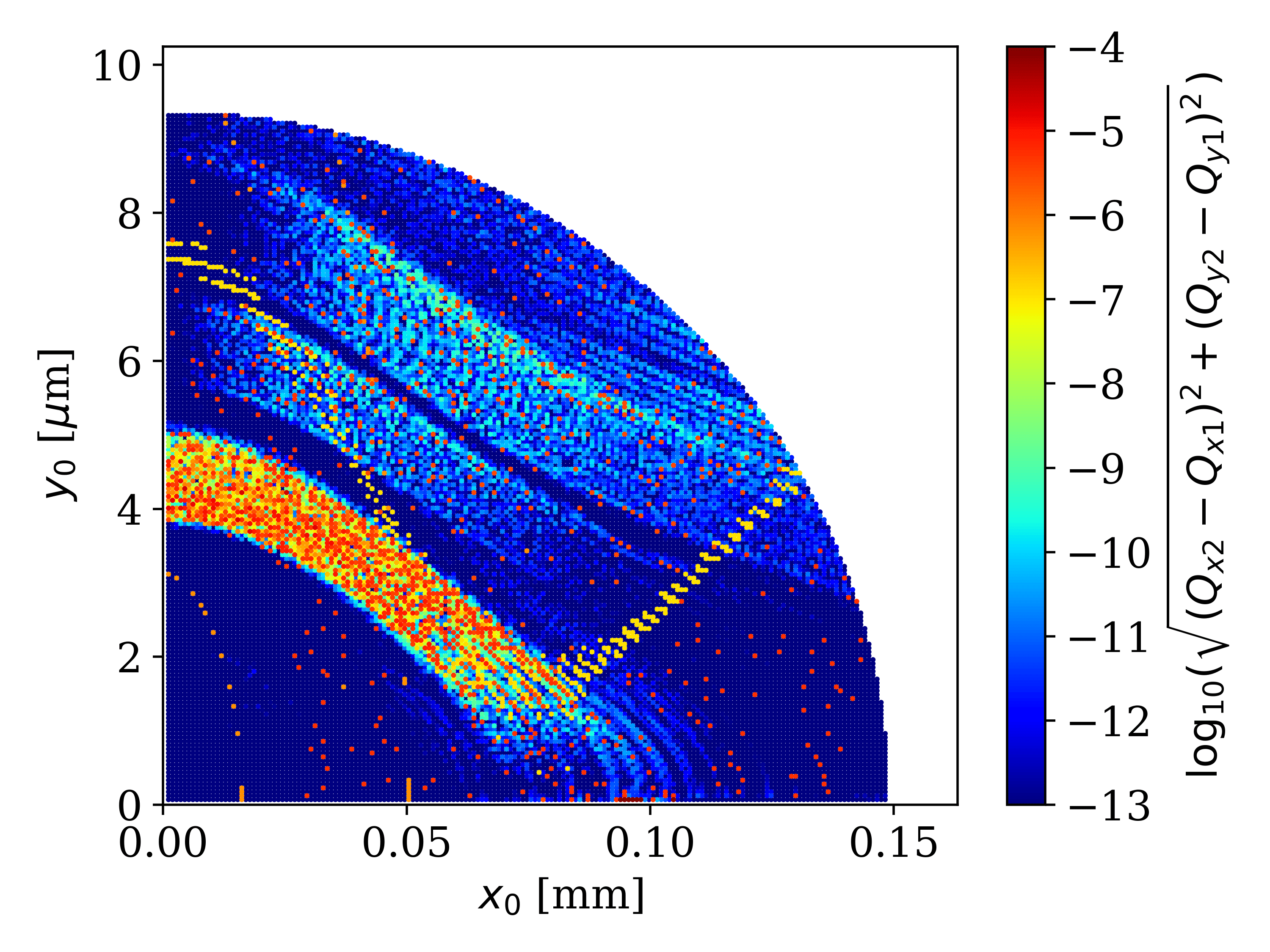} \\
        \small (a)
    \end{tabular}
    \begin{tabular}{@{}c@{}}
        \includegraphics[width=1.\columnwidth]{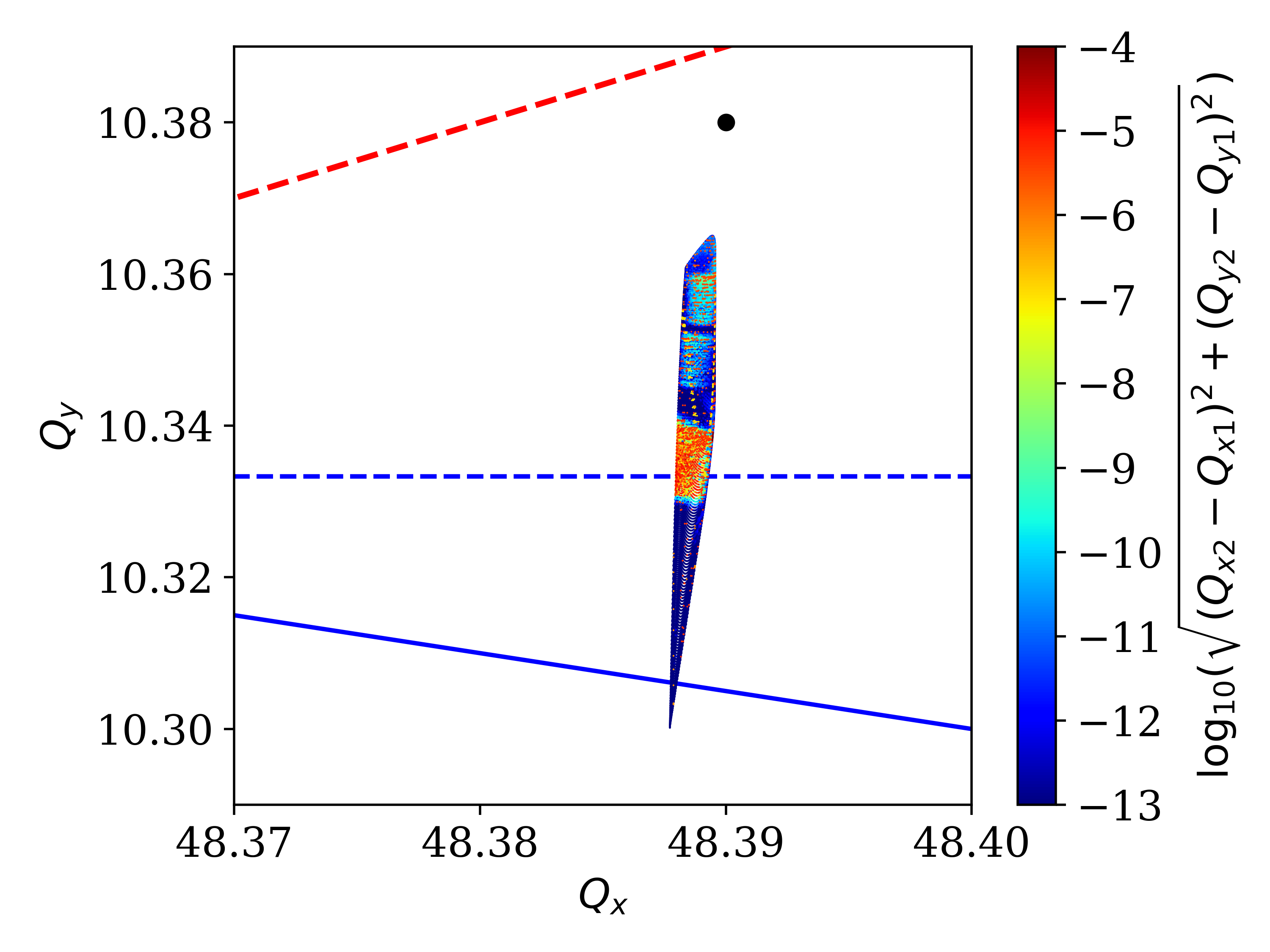} \\
        \small (b)
    \end{tabular}
    \caption{Initial distribution~(a) and Frequency map analysis of the tune footprint~(b) for off momentum particles at $1\sigma_z$, $Q_x=48.39$ and $Q_y=10.38$~(black dot) for the slow synchrotron motion case ($T_s=1000$~turns). Tune diffusion is observed around the resonances.} 
    \label{fig:ts1000_foot}
\end{figure}

For this comparison, the synchrotron period was modified to $T_s=1000$ turns, 10 times slower than the nominal case. This case is expected to be in the usual resonance trapping regime, with significantly stronger beam degradation due to resonances. The initial distribution and the tune footprint are shown in Fig.~\ref{fig:ts1000_foot}a and in Fig.~\ref{fig:ts1000_foot}b respectively, for off-momentum particles at $1\sigma_z$ and with the tune diffusion evaluated over 2 synchrotron periods. It is evident that in this case there are no resonance sidebands excited, but there is a chaotic area around the resonance due to the synchrotron motion of the particles that leads to variations of their SC tune shift and thus, to resonance crossing and associated scattering at the resonance islands~\cite{Franchetti:2006aa}.

\subsubsection{Sensitivity to the Synchrotron Period}
\begin{figure}[!t]
    \centering
    \includegraphics*[trim=15 0 13 0, clip, width=1.\columnwidth]{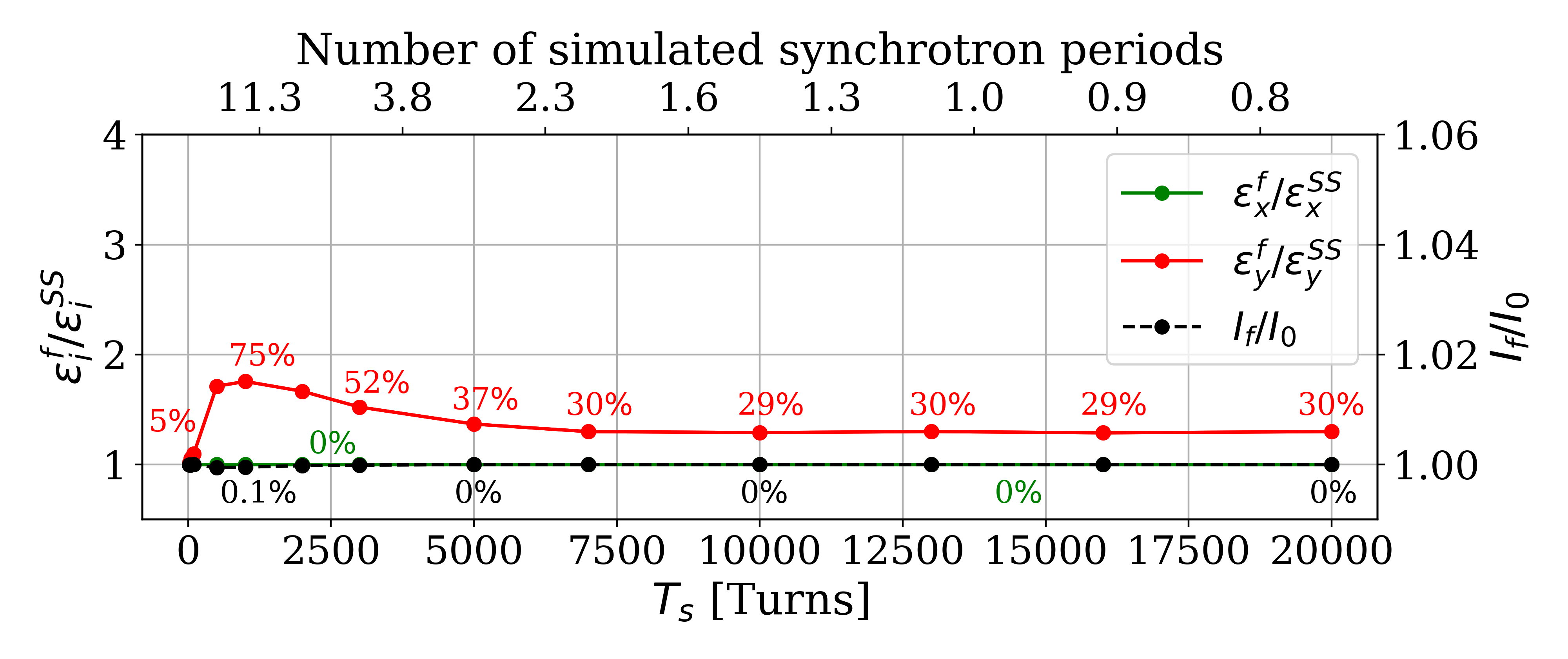}
    \caption{Ratios (final over initial) of the emittances and intensity for a scan of different synchrotron periods $T_s$, for the working point $Q_x=48.39$, $Q_y=10.35$, skew sextupoles with a normalized integrated strength $k_{\textit{2s}L}=100~\text{m}^{-2}$ and a duration of 14100~turns. The number of synchrotron periods that correspond to the 14100 turns, are shown on the top horizontal axis.}
    \label{fig:qs_scan_7s}
\end{figure}

To investigate the sensitivity to the $3Q_y$ resonance, a scan on a wide range of synchrotron periods was performed, for the working point $Q_x=48.39$, $Q_y=10.35$. The synchrotron period was modified as described in the previous paragraph. The initial beam parameters are, as before, the ones of the steady state. Even though it takes only about 7000 turns to reach the steady state, the duration of the simulations is twice as long (i.e.~14100 turns) to observe a clear evolution of the emittances. The skew sextupoles normalized integrated strength is set to $k_{\text{2s}}L=100~\text{m}^{-2}$. The ratios (final over initial) of the horizontal emittance (green), vertical emittance (red) and intensity (black) for the CLIC DRs operating with different synchrotron periods $T_s$ are summarized in Fig.~\ref{fig:qs_scan_7s}. For very small synchrotron periods $T_s$ (below 100~turns), there is no response of the beam to the resonance. In the range between 1000~turns and 2000~turns the response is maximized while for larger periods the response decreases due to the limited number of resonance crossings, as the synchrotron period becomes comparable to (or in some cases even larger than) the number of turns simulated (i.e.~14100 turns).

\begin{figure}
    \centering
    \begin{tabular}{@{}c@{}c@{}c@{}}
        \includegraphics[trim=20 10 30 0, clip, width=1.\columnwidth]{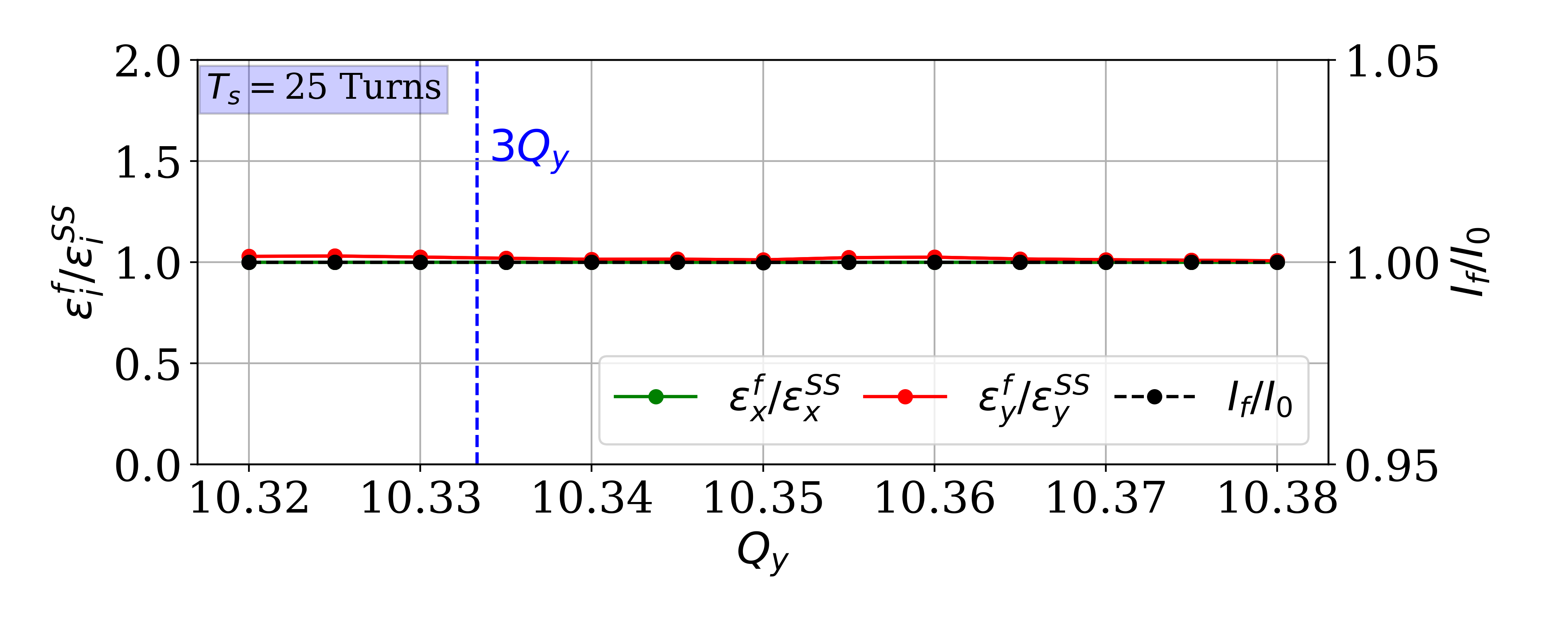} \\
        \small (a)
    \end{tabular}
    \begin{tabular}{@{}c@{}c@{}c@{}}
        \includegraphics[trim=20 10 30 0, clip, width=1.\columnwidth]{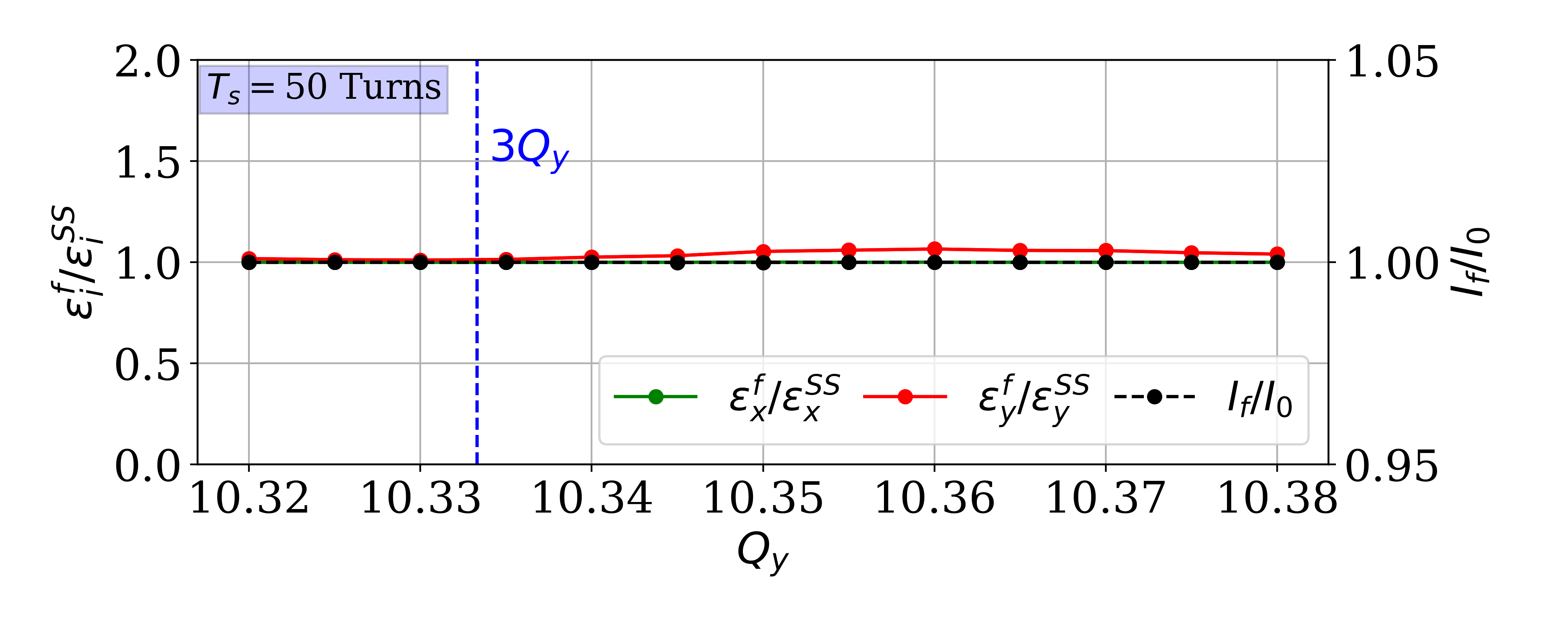} \\
        \small (b)
    \end{tabular}
    \begin{tabular}{@{}c@{}c@{}c@{}}
        \includegraphics[trim=20 10 30 0, clip, width=1.\columnwidth]{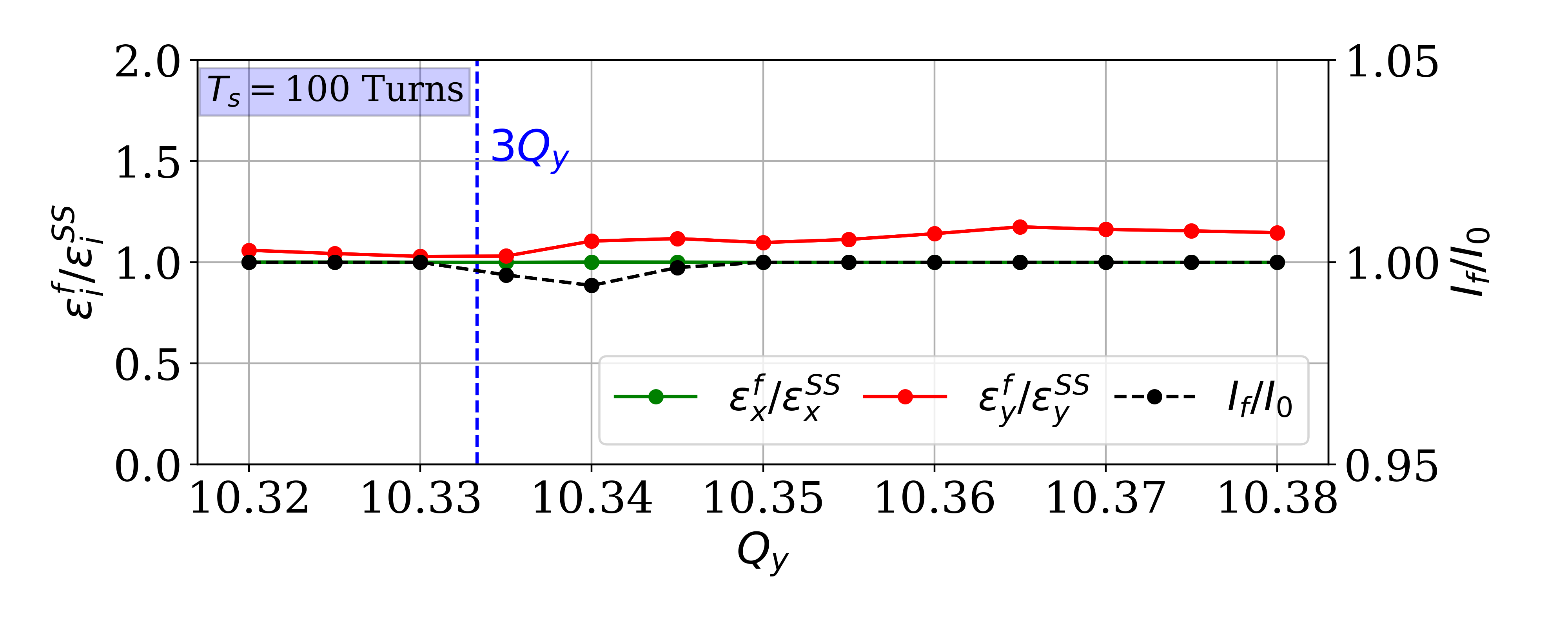} \\
        \small (c)
    \end{tabular}
    \begin{tabular}{@{}c@{}c@{}c@{}}
        \includegraphics[trim=20 10 30 0, clip, width=1.\columnwidth]{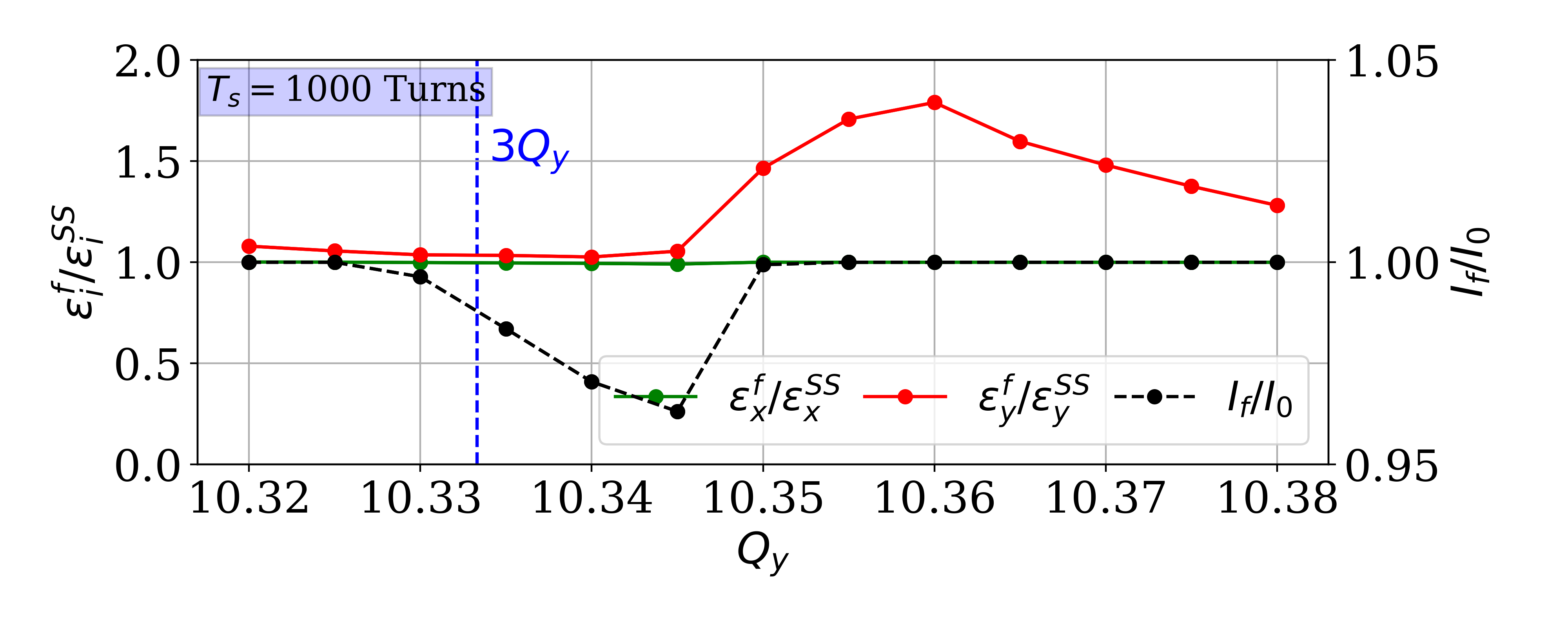} \\
        \small (d)
    \end{tabular}
    \caption{Ratios (final over initial) of the horizontal emittance~(green), vertical emittance~(red) 
            and intensity~(black) for a vertical tunescan around the $3Q_y$ 
            resonance for: a)~$T_s=25$ turns, b)~$T_s=50$ turns, c)~$T_s=100$ 
            turns and d)~$T_s=1000$ turns.} \label{fig:ts_scan}
\end{figure}

A full vertical tunescan around the $3Q_y$ resonance was performed for four chosen synchrotron periods; $25$, $50$, $100$ (nominal) and $1000$ turns, for a range of vertical tunes $Q_y=10.32\;...\;10.38$, for a constant horizontal tune $Q_x=48.39$ and for a duration of 14100~turns. All four cases are shown in Fig.~\ref{fig:ts_scan}, respectively. The case with the slower synchrotron motion shows the largest vertical emittance blow-up of the order of $80\%$ and $4\%$ of particle losses, followed by the nominal case with a blow-up of $15\%$ and $1\%$ of particle losses. In both cases, the losses are reduced when the emittance is blown-up, since the strength of the SC force is also reduced. For the two cases with faster synchrotron motion, the effect of the strongly excited resonance becomes negligible. For the nominal synchrotron motion of the CLIC DRs the beam degradation due to SC induced tune modulation close to a resonance is largely suppressed.

\section{\label{sec:5}Interplay of SC and IBS}
To study the interplay of IBS and SC in the vicinity of the excited $3Q_y=31$ resonance, simulations were conducted for the nominal synchrotron motion case with a period of $T_s=100$ turns including both SC and IBS effects. The focus of these studies is on the impact of the interplay of the two effects when the beam has reached the steady state emittances, i.e.~when the IBS effect is very strong. For this reason, the initial beam parameters were set as the ones of the steady state, but the simulation duration is still the 14100 turns of the full cycle. 

\begin{figure}[!htb]
    \centering
    \includegraphics[trim=10 0 0 0, clip, width=1.\columnwidth]{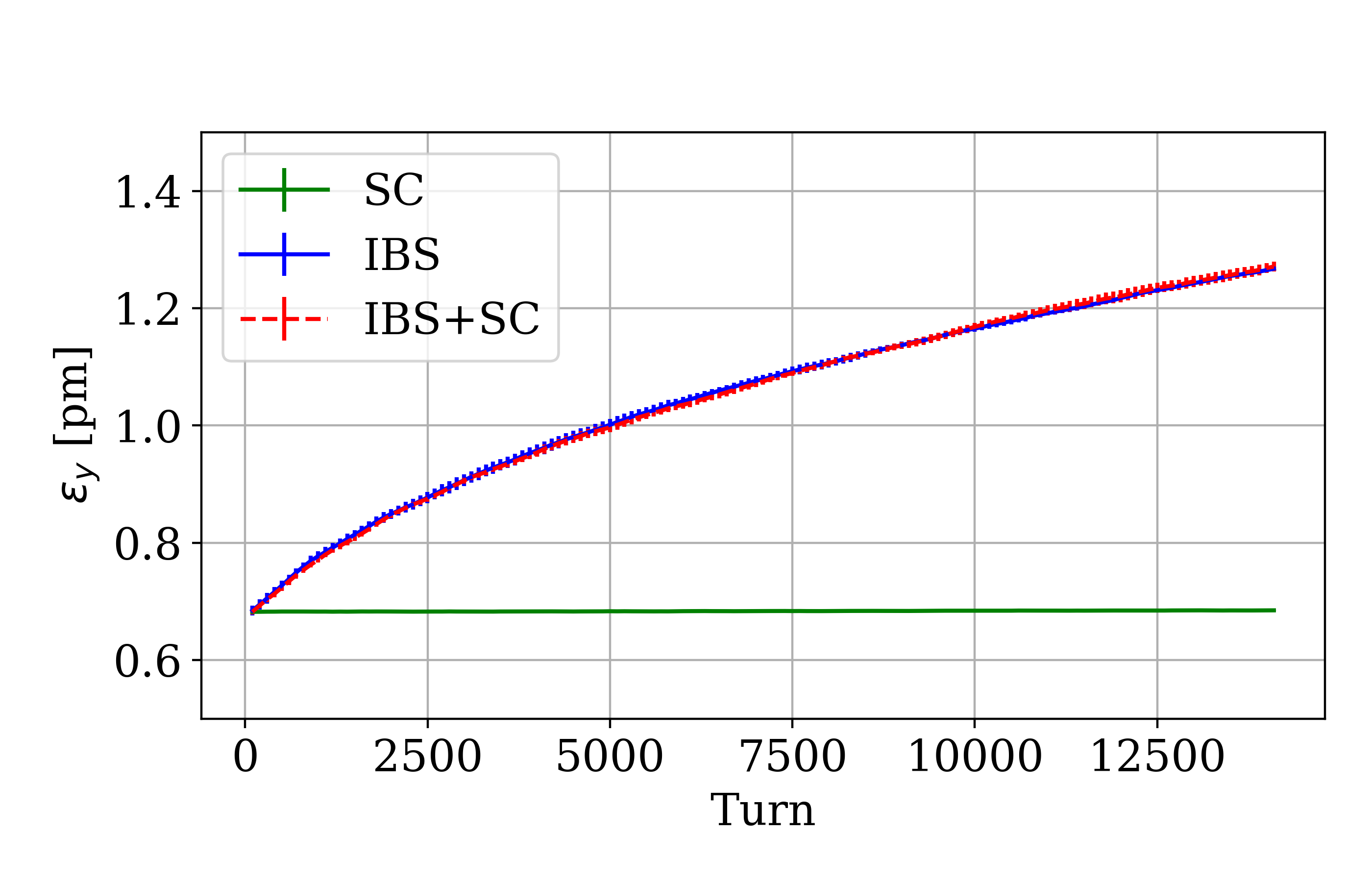}
    \caption{Vertical emittance evolution comparison with SC (green), IBS (blue)
            and combined SC with IBS (red) for $Q_x=48.39,\;Q_y=10.352$, nominal
            synchrotron period $T_s=100$ turns and the $3Q_y$ resonance not
            excited.}
    \label{fig:nom_k0}
\end{figure}

As a first test, simulations were performed without the skew sextupolar error, i.e.~the $3Q_y$ resonance not excited, to identify if there is any interplay between the two effects. The working point was set to $Q_x=48.39$, $Q_y=10.35$, which  is the one with the higher sensitivity to the resonance as shown before. Figure~\ref{fig:nom_k0} shows the vertical emittance evolution when only SC~(green), only IBS~(blue) and both IBS with SC~(red) act on the beam. The error-bars indicate the standard deviation over three different simulation runs, which appear to be negligible in this case. As expected, when only SC is considered in the simulations, the vertical emittance remains constant while in the case of the combined simulation with IBS and SC, the emittance evolution is dominated by IBS as the growth is the same as the one expected from the pure IBS effect. 

\begin{figure}
    \centering
    \begin{tabular}{@{}c@{}}
        \includegraphics[trim=5 5 20 0, clip, width=1.\columnwidth]{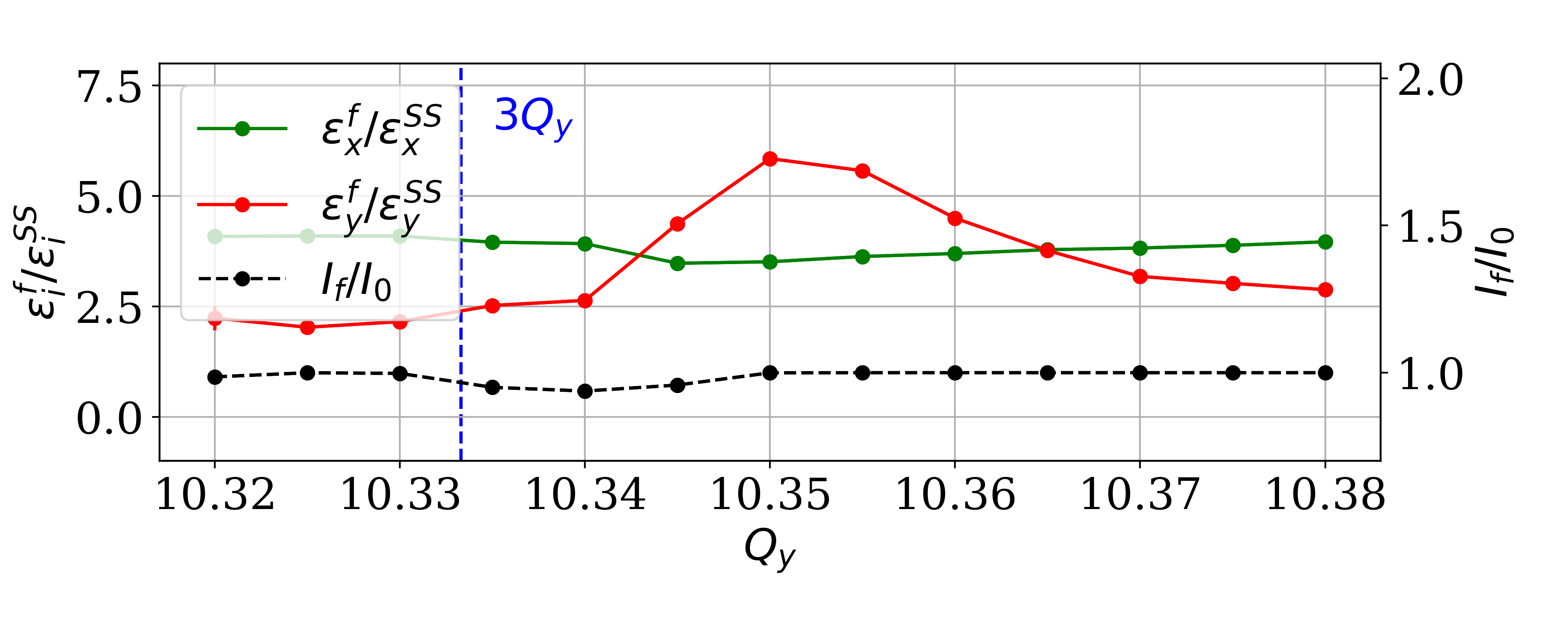} \\
        \small (a)
    \end{tabular}
    \begin{tabular}{@{}c@{}}
        \includegraphics[trim=5 5 20 0, clip, width=1.\columnwidth]{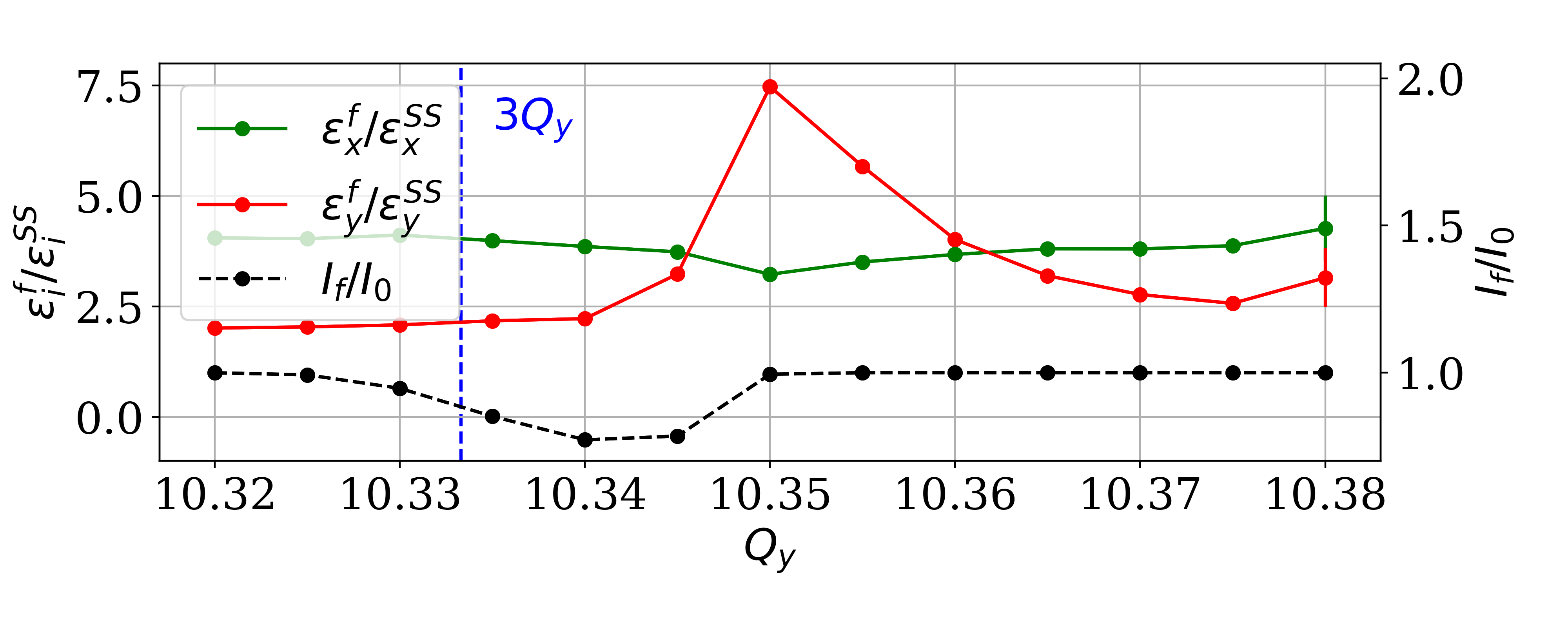} \\
        \small (b)
    \end{tabular}
    \caption{Ratios~(final over initial) of the horizontal~(green), vertical~(red) transeverse emittances and intensity~(black) for a vertical tunescan around the excited $3Q_y$ resonance, for $T_s=100$ turns~(a) and $T_s=1000$ turns~(b), with $Q_x=48.39$, combining IBS and SC.} 
    \label{fig:ibs_sc_tunescan}
\end{figure}

A full vertical tunescan around the $3Q_y$ resonance was performed for the nominal synchrotron motion case $T_s=100$~turns, for a range of vertical tunes $Q_y=10.32\;...\;10.38$ and for a constant horizontal tune $Q_x=48.39$. Here the $3Q_y$ resonance is excited by a skew sextupolar error with a normalized integrated strength $k_{\text{2s}L}=100~\text{m}^{-2}$. Figure~\ref{fig:ibs_sc_tunescan}a shows the relative transverse emittances and intensity when both IBS and SC act on the particle distribution. The sensitivity to the $3Q_y$ is strongly enhanced when both effects are present, producing a significantly larger vertical emittance blow-up and more losses for the working points affected by the resonance, while IBS dominates the emittance growth for the working points sufficiently far away from the resonance.

For comparison, the same tunescan was performed for the slow synchrotron period case $T_s=1000$ turns. The results are shown in Fig.~\ref{fig:ibs_sc_tunescan}b where a similar behavior as the previous case is observed, with enhanced vertical emittance growth and particle losses.

Comparing the two different synchrotron period cases with $T_s=100$ turns and $T_s=1000$ turns, two main observations should be pointed out. First, for both cases the interplay of IBS and SC leads to a significant increase of the sensitivity to the $3Q_y$ resonance in terms of vertical emittance blow-up. Second, the slow synchrotron motion case suffers from significantly more particle losses in the vicinity of the resonance.

\section{\label{sec:6}Interplay of SC, IBS and SR}
\begin{figure}[!htb]
    \centering
    \includegraphics[trim=35 15 10 0, clip, width=1.\columnwidth]{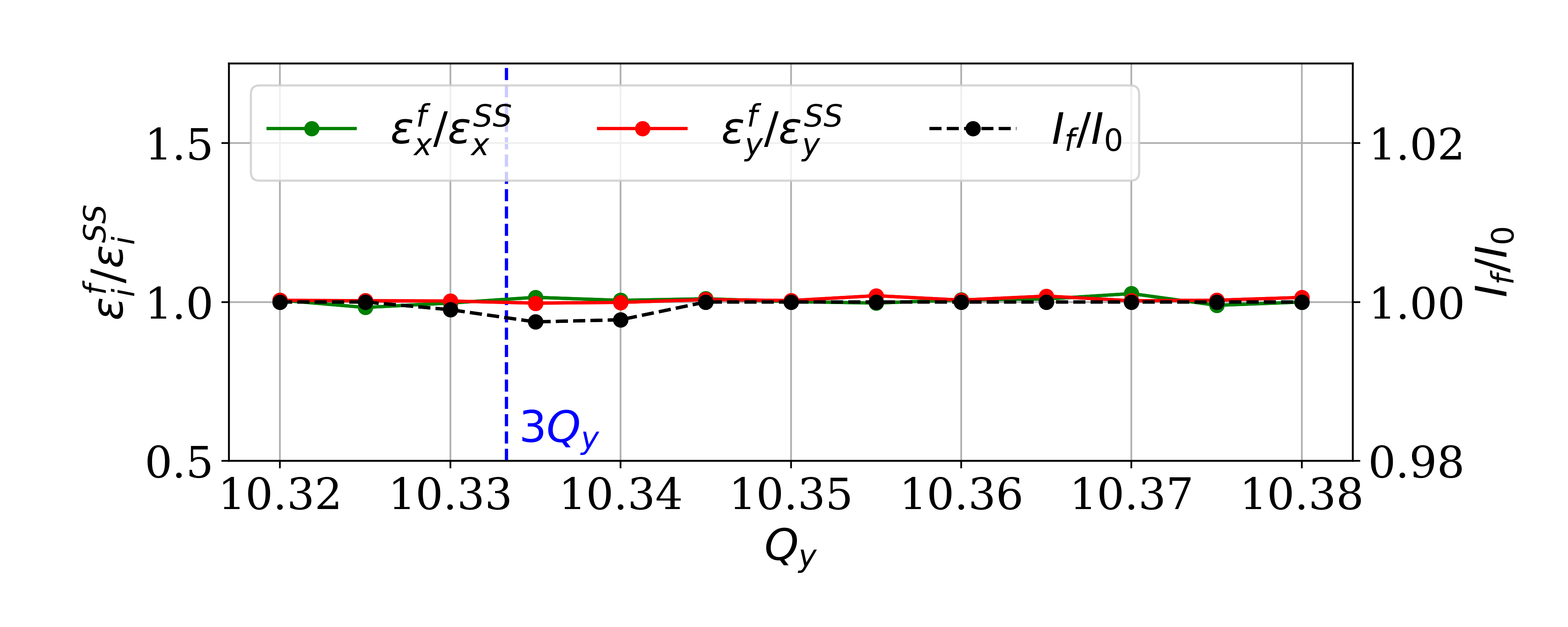}
    \caption{Ratios~(final over initial) of the horizontal~(green), vertical~(red) transeverse emittances and intensity~(black) for a vertical tunescan around the excited $3Q_y$ resonance with $Q_x=48.39$ and $T_s=100$ turns, combining SC, IBS and SR.}
    \label{fig:zoom_ibs_sc_sr}
\end{figure}

The final step is including the SR effect into the simulations to study the interplay of the three effects in the vicinity of an excited resonance. A tunescan was performed in a range of vertical tunes from 10.32 to 10.38, for the nominal synchrotron motion. Similar as before, the simulations are performed for a beam at the equilibrium state. Figure~\ref{fig:zoom_ibs_sc_sr} shows the ratio of the final over initial horizontal (green) and vertical (red) emittance and intensity (black). The strong damping from SR strongly mitigates the impact of the other two effects, leaving a maximum residual vertical emittance growth of the order of $5\%$. Particle losses are significantly less than $1\%$ and are observed only for the working points that were affected the most by the resonance, as shown in Sections~\ref{sec:4} and~\ref{sec:5}.

\begin{figure*}[!htb]
    \centering
    \includegraphics[width=1.\textwidth]{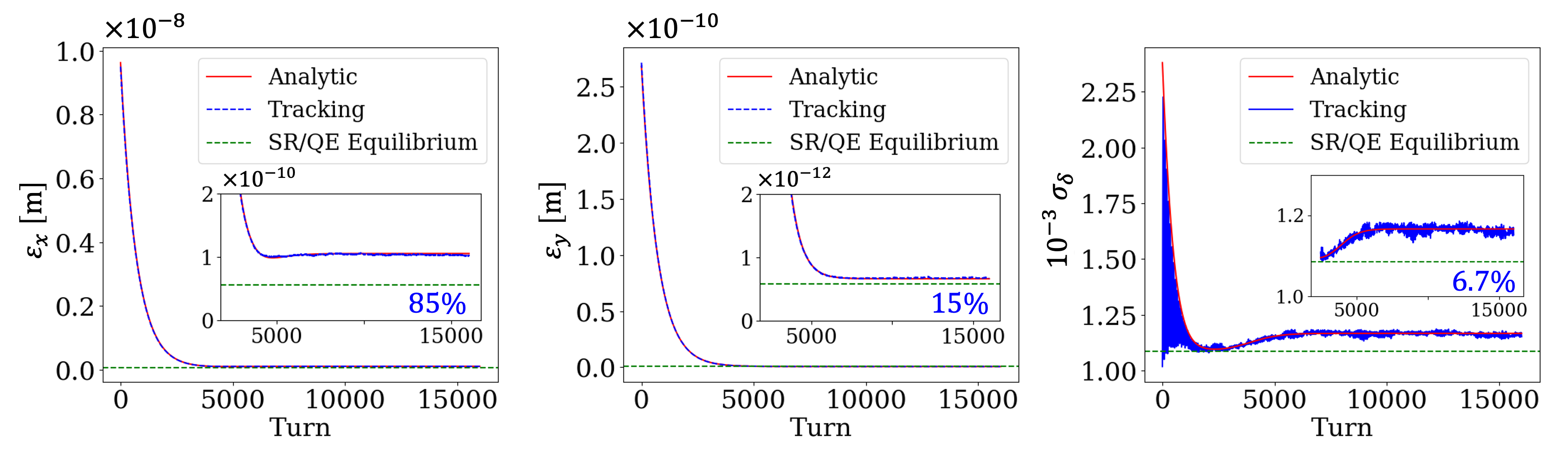}
    \caption{Evolution of the horizontal emittance (left), vertical emittance
            (center) and momentum spread (right) in the CLIC DR. SR and IBS 
            are taken into account for the analytic calculations. For the
            tracking SC is also included. Zoomed plots embedded in the main
            plots with the relative difference to the SR/QE Equilibriums.}
    \label{fig:clic_dr_cycle}
\end{figure*}

\subsection{\label{sec:6a}Full Cycle Simulations}
Due to the different regimes that the beam experiences through the operational cycle and the interaction with different resonances during this process, as was shown in Section~\ref{sec:4}, it is important to investigate the impact of the interplay of the SC, IBS and SR effects during the full cycle evolution. As a first study case, the interplay of SC, IBS and SR along the full nominal CLIC DR cycle was studied, starting with the injected beam parameters and tunes as previously shown in Tables~\ref{tab:params} and~\ref{tab:ineqams}. In these simulations, no lattice imperfections were taken into account, i.e. the ideal lattice was used, including sextupoles for chromaticity correction. The results of the tracking simulations were compared with analytical calculations, including only the effects of IBS and SR. 

Figure~\ref{fig:clic_dr_cycle} shows the comparison of the tracking simulation including the effects of SC, IBS and SR in blue and the analytical calculations of IBS and SR in red. The equilibrium emittances emerging from SR and QE, as calculated from the radiation integrals, are shown in the dashed green lines. As described earlier, the beam is injected with large transverse emittances. At that point, SC and IBS are very weak, and mostly radiation damping is acting on the beam. Due to the large amplitudes, detuning with amplitude due to the strong sextupoles might also affect the beam, leading to resonance crossing. As the beam becomes smaller, the effects of IBS and SC become stronger leading to larger emittance values than the equilibrium defined by SR and QE, known as steady state of the beam. The tracking simulation with SC, IBS and SR are in excellent agreement with the analytical predictions with only IBS and SR, leading to the conclusion that for the ideal lattice, i.e.~without imperfections, SC does not change the final steady state of the operational scenario despite the very small emittances and the large vertical tune spread that SC might induce.

\begin{figure}
    \centering
    \begin{tabular}{@{}c@{}c@{}}
        \includegraphics[trim=15 18 10 0, clip, width=1.\columnwidth]{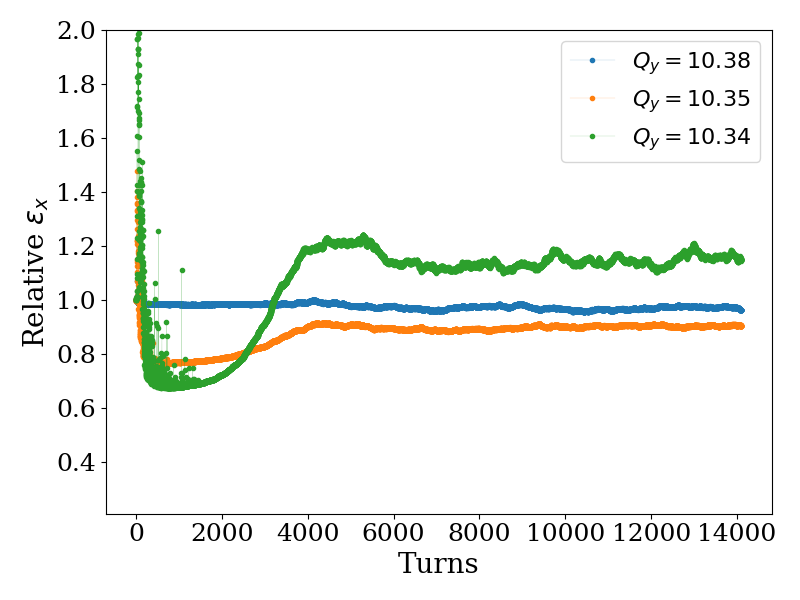} \\[\abovecaptionskip]
        \small (a)
    \end{tabular}
    \vspace{\floatsep}
    \begin{tabular}{@{}c@{}c@{}}
        \includegraphics[trim=15 18 10 0, clip, width=1.\columnwidth]{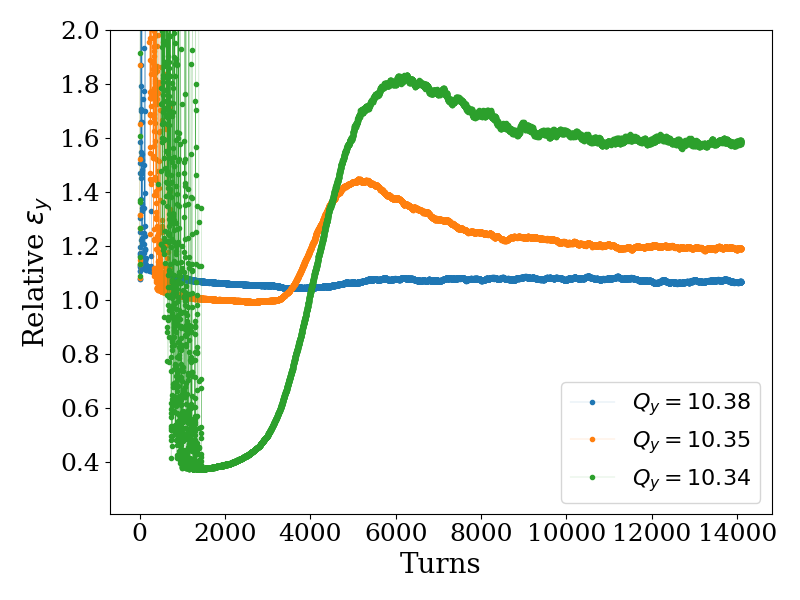} \\[\abovecaptionskip]
        \small (b)
    \end{tabular}
    \vspace{\floatsep}
    \begin{tabular}{@{}c@{}c@{}}
        \includegraphics[trim=15 18 10 15, clip, width=1.\columnwidth]{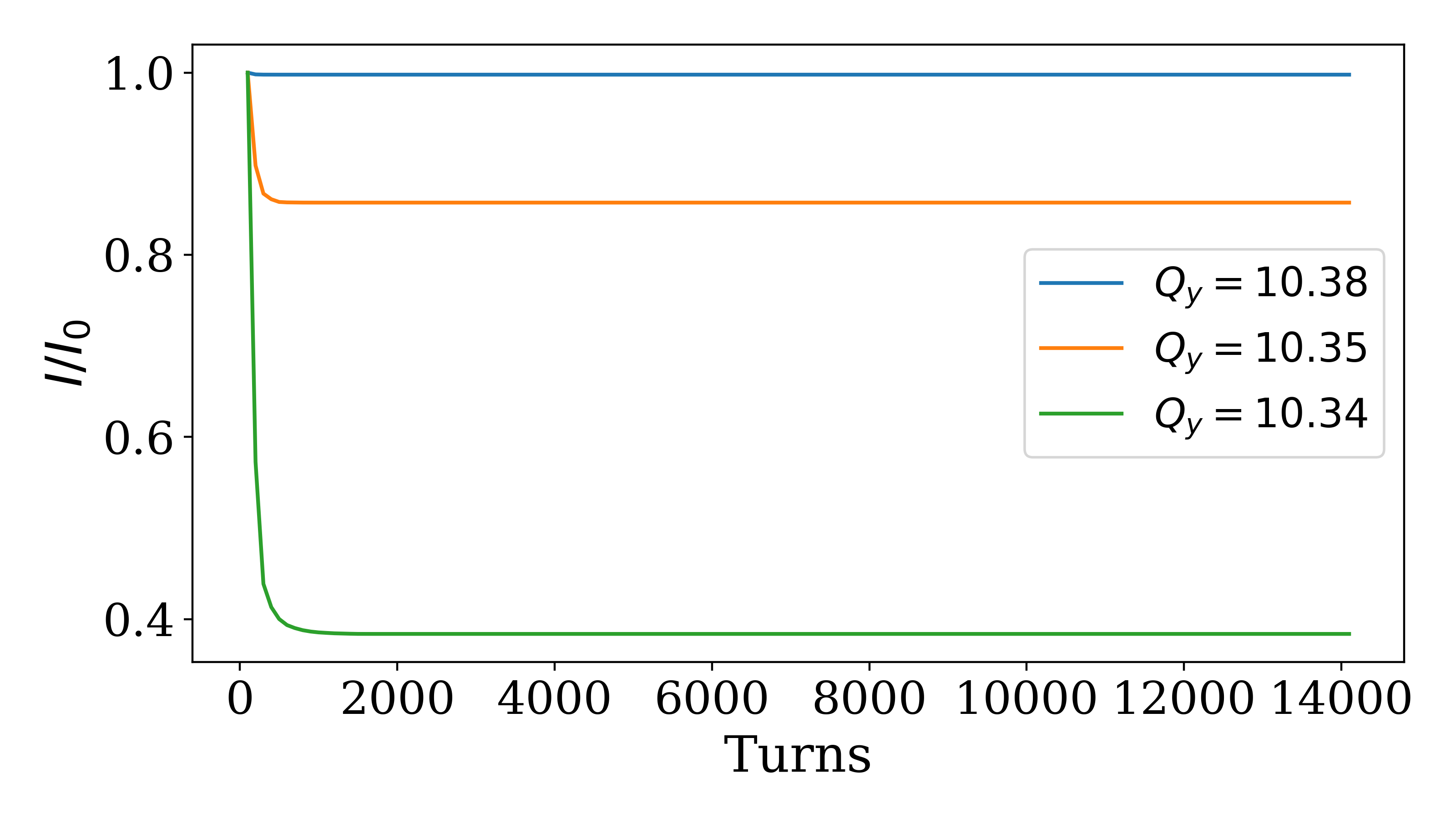} \\[\abovecaptionskip]
        \small (c)
    \end{tabular}
    \caption{Evolution of the horizontal emittance~(a), vertical emittance~(b) and intensity~(c) starting from injection including the skew sextupolar error with normalized integratedstrength $k_\text{2s}L=100~\text{m}^{-2}$. The plots show relative quantities, which are normalized to the beam evolution obtained from the analytical IBS/SR calculations considering no losses and no errors. Three different working points are shown.}\label{fig:rel_xyi}
\end{figure}

To investigate how the $3Q_y$ resonance can affect the beam during the full cycle of the CLIC DRs, simulations including the skew sextupole errors with normalized integrated strength $k_\text{2s}L=100\;\text{m}^{-2}$, were performed. The beam was again initialized to the injected beam parameters and the duration of the simulations is 14100 turns. The horizontal tune is set to the nominal $Q_x=48.357$. In particular, to understand the dependence of the interplay effect on the working point along the cycle, the simulations were performed for three different vertical tunes: $Q_y=10.38$, $Q_y=10.35$ and $Q_y=10.34$. The results of these simulations are summarized in Figs.~\ref{fig:rel_xyi}a and~\ref{fig:rel_xyi}b for the horizontal and vertical planes, respectively. In order to highlight the working point dependence, both the horizontal and vertical emittances are normalized to the ones from analytical calculations of IBS and SR, where no errors and no losses are observed.

Comparing the results from the three different working points, it becomes clear that even in the presence of strong SR, the working point choice is very important to avoid beam degradation. For the vertical tune $Q_y=10.38$, particles are away from the $3Q_y$ resonance and no beam degradation is observed. On the other hand, for the other two working points that are closer to the $3Q_y$ resonance, strong particle losses and significant emittance blow-up are observed. In particular, for $Q_y=10.34$, $60\%$ of the particles are lost soon after injection, explaining the fact that the emittance in the simulations is smaller than the one from the analytical calculations during the first part of the cycle (up to around 4000 turns). As the beam is shrinking fast due to the SR effect, the losses are eliminated and the interplay between the IBS and SC effects in the close vicinity of the strongly excited resonance, leads to a final vertical emittance that is $70\%$ higher than the expected one. The horizontal emittance increase observed between 2000 and 4000 turns, following a similar pattern as in the emittance evolution observed in the vertical plane, indicates some interaction of the beam with the coupling resonance, leading to emittance exchange in the two planes. The large fluctuation of the emittance values observed at the beginning of the cycle is associated to the strong particle losses during the first parts of the cycle, where the beam is still large. Figure~\ref{fig:rel_xyi}c shows the particle losses along the cycle for the three working points. 

\begin{figure}[!htb]
    \centering
    \includegraphics[trim=15 0 10 0, clip, width=1.\columnwidth]{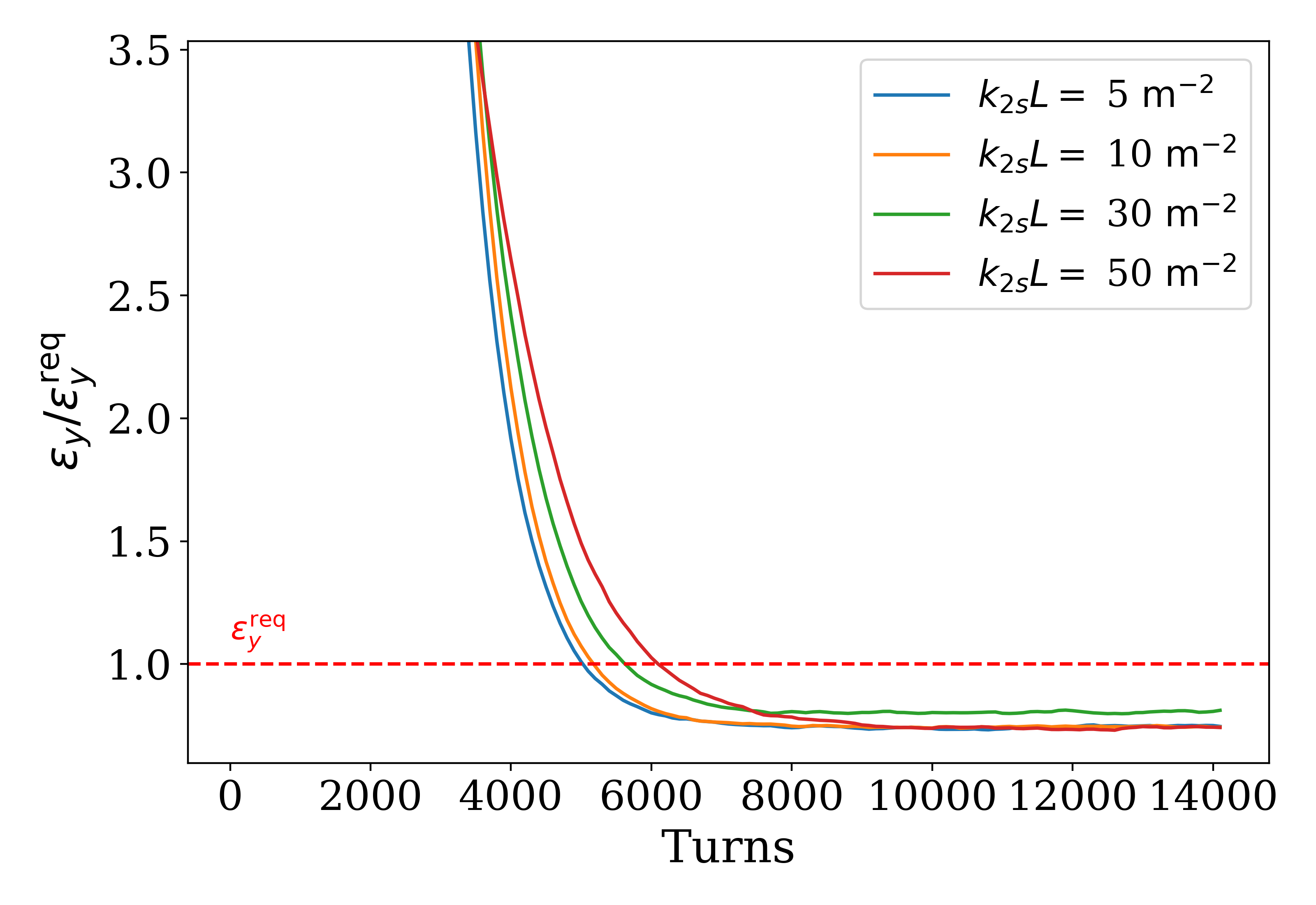}
    \caption{Vertical emittance evolution for different skew sextupole normalized integrated strength, for the working point $(Q_x, Q_y)~=~(48.357, 10.34)$. The emittance value is normalized to the required beam emittance.}
    \label{fig:eytol}
\end{figure}

\begin{figure}[!htb]
    \centering
    \includegraphics[trim=15 0 10 0, clip, width=1.\columnwidth]{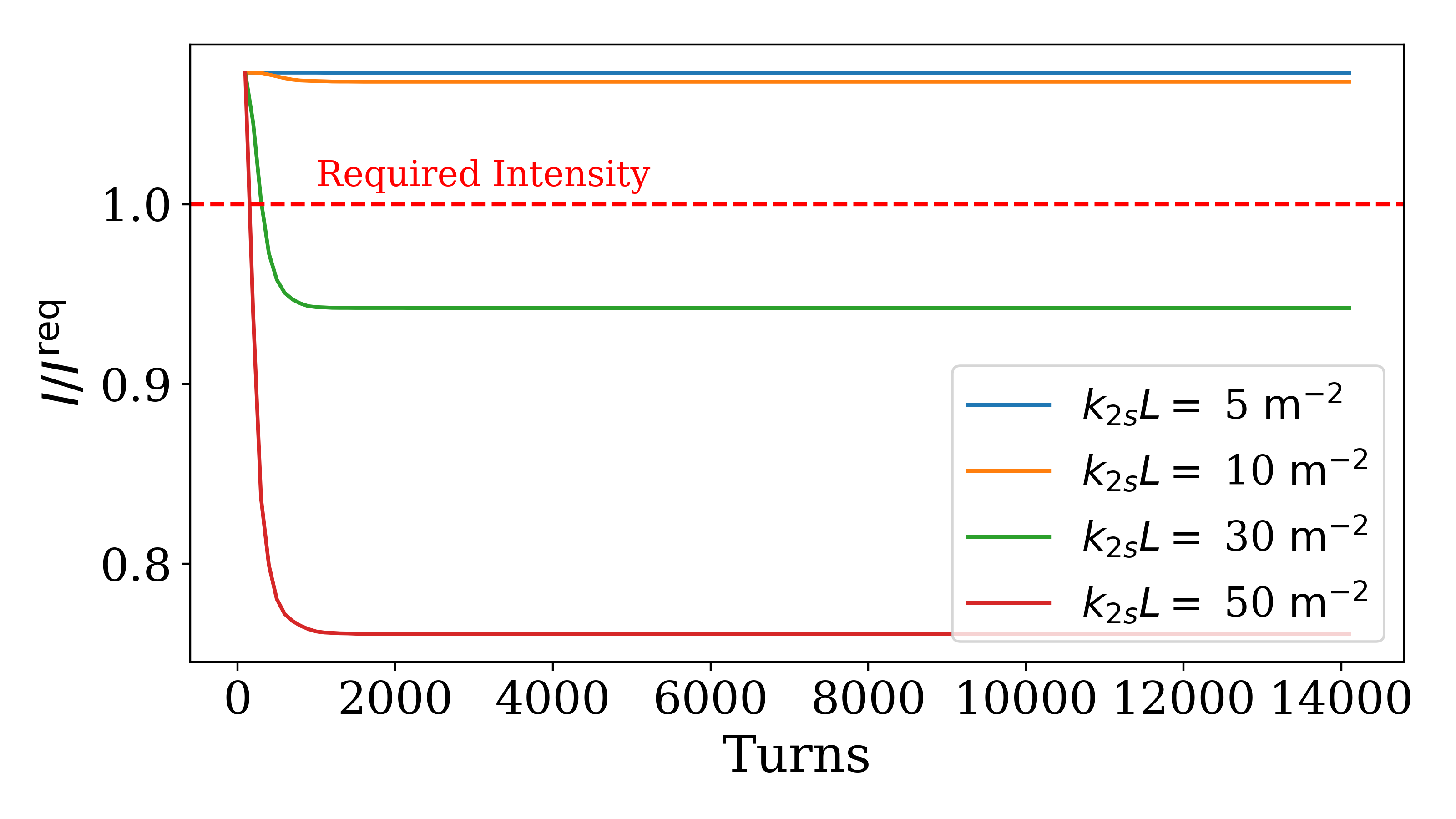}
    \caption{Intensity evolution for different skew sextupole normalized integrated strength, for the working point $(Q_x, Q_y)~=~(48.357, 10.34)$. The intensity values are normalized to the required intensity.}
    \label{fig:Itol}
\end{figure}

Our studies demonstrated that the working point choice is very important for minimising the impact of SC and IBS, even in the presence of strong SR. As a next step, we investigate the sensitivity to the skew sextupolar resonance and define a threshold strength at which the extraction requirements are not met in the case the chosen operational working point is affected by this resonance (i.e. $(Q_x, Q_y)~=~(48.357, 10.34)$). To this end, simulations for the full cycle were repeated for different strengths of the skew sextupolar error and the results are summarized in Fig.~\ref{fig:eytol} and Fig.~\ref{fig:Itol}, for the vertical emittance evolution and particle losses along the cycle, respectively. 

In terms of the vertical emittance, the extraction requirement (red dashed line) is met for all skew sextupole errors examined. On the other hand, losses are observed even for weak skew sextupoles with a threshold normalized integrated strength between $k_\text{2s}L=10~\text{m}^{-2}$ and $30~\text{m}^{-2}$. However, it should be emphasized that these skew sextupole errors are quite large and far from realistically expected field errors. For comparison, a normal sextupole used for the chromaticity correction in the CLIC DR lattice has a strength of $k_\text{2}L=30~\text{m}^{-2}$.

\begin{figure}
    \centering
    \begin{tabular}{@{}c@{}}
        \includegraphics[trim=15 0 10 0, clip, width=1.\columnwidth]{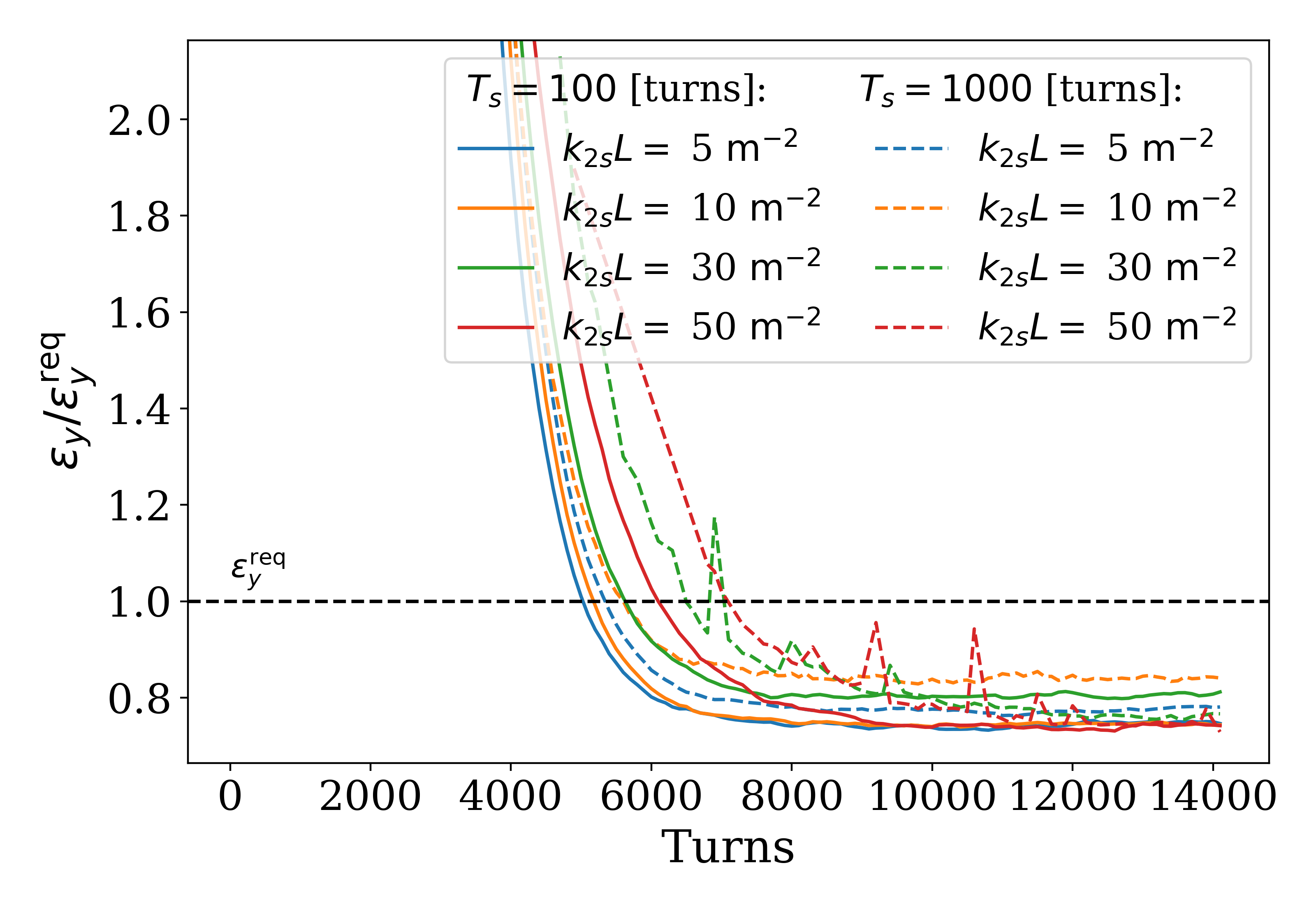} \\
        \small (a)
    \end{tabular}
    \vspace{1mm}
    \begin{tabular}{@{}c@{}}
    \includegraphics[trim=15 15 10 0, clip, width=1.\columnwidth]{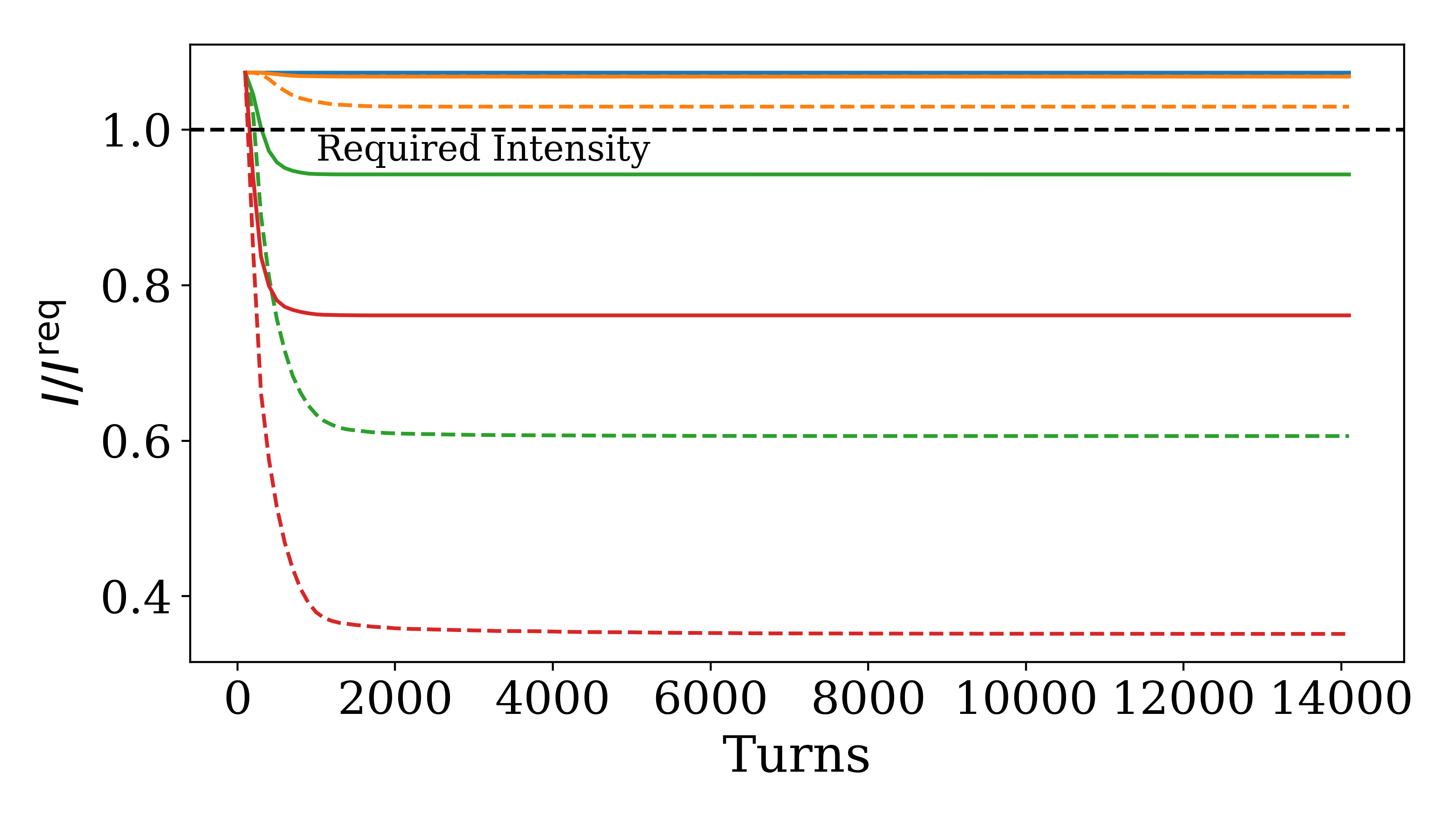} \\
        \small (b)
    \end{tabular}
    \caption{Evolution of the relative vertical emittance (a) and 
            relative intensity (b) over the extraction requirements, 
            starting from injection with SC, IBS and SR and for 
            different normalized integrated strengths $k_\text{2s}L$. 
            Comparison between the nominal synchrotron motion (full 
            lines) and slow synchrotron motion (dashed lines)}
    \label{fig:compeyItol}
\end{figure}

In Section~\ref{sec:4}, it was shown that for slower synchrotron motion, the beam response to the resonance is enhanced. To investigate the impact of the synchrotron period on the performance of the CLIC DR, similar tracking simulations were performed for the working point $(Q_x, Q_y)~=~(48.357, 10.34)$, for a synchrotron period of $T_s=1000$ turns. Figure~\ref{fig:compeyItol} shows the comparison of the vertical emittance~(a) and intensity~(b) ratios, for the nominal synchrotron motion case (full lines) and the slow synchrotron motion case (dashed lines) for normalized integrated strengths of $k_\text{2s}L=5~\text{m}^{-2}$ (blue) and $k_\text{2s}L=10\;\text{m}^{-2}$ (orange). Significantly higher losses are observed in the case of slower synchrotron motion, as already observed in section~\ref{sec:5}. Thus, the synchrotron motion in the CLIC DRs helps mitigating the impact of strong resonances on beam quality, in terms of particle losses and beam emittances.

\begin{figure}[!htb]
    \centering
    \includegraphics[trim=15 0 10 0, clip, width=1.\columnwidth]{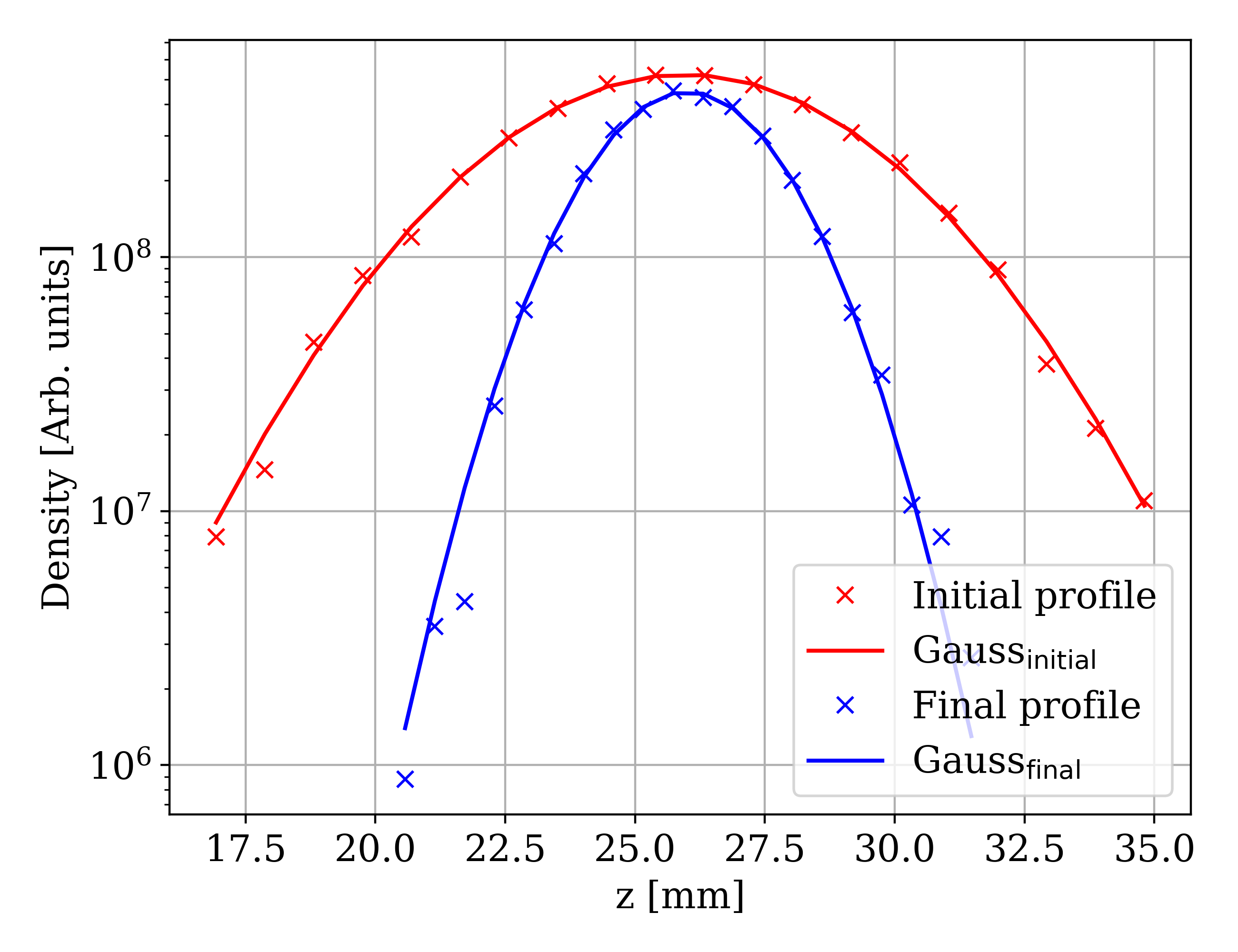}
    \caption{Initial (red) and final (blue) beam profiles, indicated with crosses and compared with Gaussian fits, indicated with solid lines. Nominal synchrotron period for the working point $(Q_x, Q_y)~=~(48.357, 10.34)$ and $k_\text{2s}L=50\;\text{m}^{-2}$}
    \label{fig:zprof}
\end{figure}

 Finally, the profile shape of the beam distribution also need to be investigated, as non-Gaussian profiles can impact the luminosity. Excited resonances and IBS can eventually change the shape of the beam profiles by populating the tails more, especially in the longitudinal plane. On the other hand, SR shrinks the profiles back to Gaussian or in some cases even to profiles with less than Gaussian tails~\cite{Papadopoulou:2018uvl,Bruce:ibs_kick}. To verify if in the case of the CLIC DRs the profiles remain Gaussian, the initial and the final longitudinal beam profiles are compared with Gaussian profiles for the nominal synchrotron period case and the skew sextupolar error with normalized integrated strength of $k_\text{2s}L=50\;\text{m}^{-2}$, which was the worst behaving case. Figure~\ref{fig:zprof} shows a comparison between the initial (red) and final (blue) beam profiles as produced from the simulations (crosses) and are compared with Gaussian fits (solid lines). As indicated, during the full cycle of the CLIC DRs, in the presence of SC, IBS and SR with the $3Q_y$ resonance excited, the longitudinal beam shape remains Gaussian.

\section{\label{sec:7}Impact of vertical dispersion}
In $e^-/e^+$ rings, the equilibrium vertical emittance is dominated by the coupling with the horizontal plane and the residual vertical dispersion, which can be induced by quadrupole tilt errors and dipole misalignments. It is therefore defined through the vertical emittance tuning of the machine~\cite{Ghasem}. In the previous sections the presented results did not take into account the vertical dispersion, the impact of which will be addressed here.

\begin{figure}[!t]
    \centering
    \includegraphics*[width=1.\columnwidth]{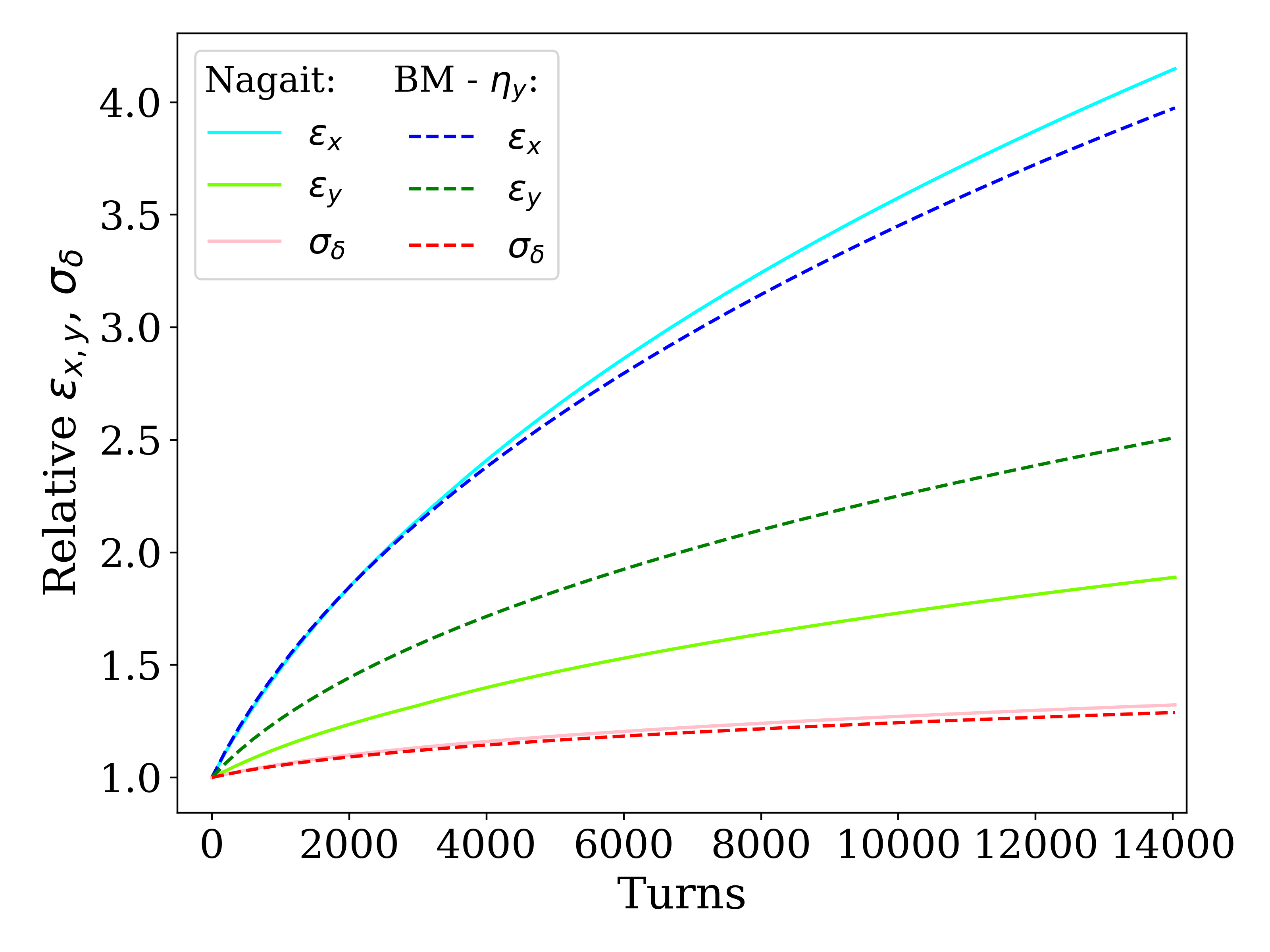}
    \caption{Comparison of the relative horizontal emittance (blue), relative
            vertical emittance (green) and relative momentum spread (red) with
            vertical dispersion (dark colors) and without (light colors). The
            analytical model of Bjorken and Mtingwa is used.}
    \label{fig:ibs_dy}
\end{figure}

Benchmarking simulations for IBS calculations were performed for a lattice with quadrupolar tilt errors, creating vertical dipole magnetic components and thus, vertical dispersion. The applied errors are arbitrary but still consistent with the CLIC requirements, that is an output normalised vertical emittance of below 5~nm. In this case, the original model of BM was used, which takes into account vertical dispersion. Figure~\ref{fig:ibs_dy} shows the results in comparison to the ones of the ideal lattice, previously shown in Fig.~\ref{fig:ibs_bench} using the model of Nagaitsev. Vertical dispersion does not contribute significantly in the IBS calculations for the horizontal and longitudinal planes, while for the vertical plane, the output emittance is 34$\%$ higher, as by neglecting vertical dispersion, the coupling between the longitudinal and vertical planes is neglected.

\begin{figure}[!htb]
    \centering
    \includegraphics[trim=10 0 0 0, clip, width=1.\columnwidth]{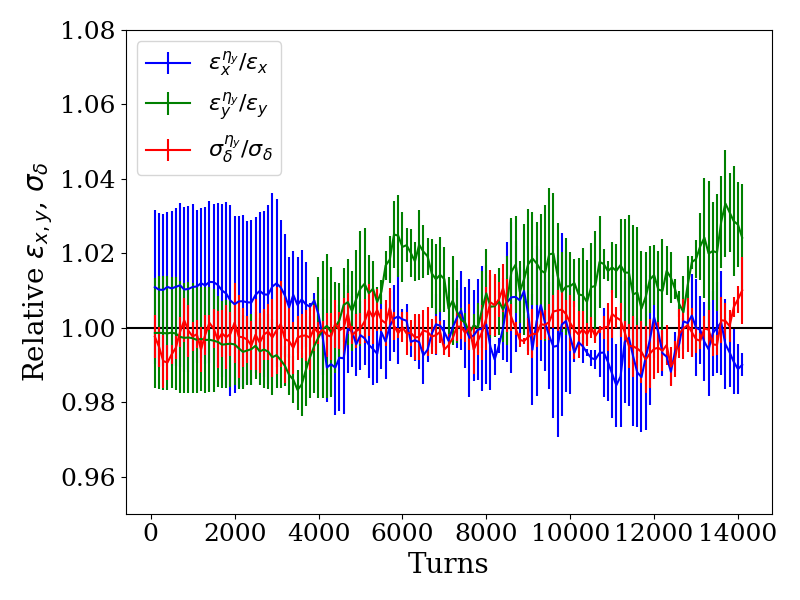}
    \caption{Ratios of the evolution of the horizontal emittance (blue),
            vertical emittance (green) and momentum spread (red), for a
            case including vertical dispersion over a case the case without
            vertical dispersion. No errors included.}
    \label{fig:Vdisprat}
\end{figure}

To understand the impact of vertical dispersion in the studies during the full cycle of the CLIC DRs, a tracking simulation similar to Section~\ref{sec:6a} was performed taking into account the vertical dispersion in the IBS calculations. In this study, vertical dispersion is produced by arbitrary quadrupolar errors such that the resulting equilibrium parameters from SR and QE are compliant with the target extraction values. Simulation results are presented in Fig.~\ref{fig:Vdisprat}, averaged over three runs, where the relative horizontal emittance (blue), vertical emittance (green) and momentum spread (red) are shown, in comparison with the case without vertical dispersion of Fig.~\ref{fig:clic_dr_cycle}. Results show that in the presence of SR, the induced vertical dispersion does not affect the horizontal and longitudinal planes, while it produces an extra $2\%$ increase in the vertical steady state emittance. Evidently, the effect of vertical dispersion in the final steady state is insignificant and can be neglected. This allowed the use of Nagaitsev's IBS model for more efficient computations~\cite{Nagaitsev} compared to the model of BM. In cases where additional errors are present, the impact of the resulting vertical dispersion could become significant and should be further studied.

\section{\label{sec:8}Conclusions}
Studies of the incoherent effects of SC, IBS and SR and their interplay were performed in the context of the CLIC DRs. To this end, two modules for simulating IBS and the SR effects were implemented in PyORBIT. Both modules were successfully benchmarked against analytical calculations and showed excellent agreement.

Simulation studies were performed to investigate the impact of an excited resonance (here the $3Q_y=31$ resonance was chosen) on the achievable beam parameters of the CLIC DRs. It was shown that at the steady state emittances, the beam degradation due to resonance crossing induced by SC alone is relatively weak. This is due to the synchrotron period of the CLIC DRs of about 100~turns which results in resonance side-bands rather than periodic resonance crossing, as demonstrated by the detailed investigation of the SC driven beam dynamics near the $3Q_y=31$ resonance. However, particle diffusion and thus emittance growth is strongly enhanced when IBS is also considered. Eventually, the strong damping from SR counteracts this effect and the beam degradation including SC, IBS and SR effects is of the order of a few percent only even for the case of the strong vertical sextupole resonance studied here. 

For the first time, combined macro-particle tracking simulations with SC, IBS and SR were performed for the operational scenario of the CLIC DRs, including all parts of the cycle, i.e.,~starting from injection with large emittances that are damped due to the strong radiation, leading to enhanced SC at the end of the cycle. The nominal working point was used and no errors in the lattice were considered. This is similar to using the full errors of the ring with the corresponding corrections that decrease the error spread back to the systematic errors. The fulfilment of the CLIC target parameters for ultra-high brightness beams is now demonstrated through macro-particle tracking simulations. In the absence of errors, SC does not play any role in the final steady state of the beam, which is then given by the combined effect of SR and IBS.

On the other hand, when resonance excitation is present due to errors, it was demonstrated that the choice of the working point is critical. Even in the case of the CLIC DRs with the strong radiation damping and the synchrotron period of about 100~turns, significant beam degradation was observed when the tune is set relatively close to the excited resonance. The beam quality degradation was shown for the example of the $3Q_y$ resonance, where both the vertical emittance and intensity requirements could not be met. Therefore, the design study of such machines need to include an assessment of resonance excitation and the tune spread evolution along the cycle for optimizing the working point. Finally, simulations with slower synchrotron motion showcased how the nominal synchrotron motion of the CLIC DRs makes the beam less sensitive to resonances.

As a last step, vertical dispersion was accounted for in the IBS calculations. It does not significantly affect the results obtained for the full cycle due to the strong SR.

\appendix
\begin{figure*}[!hbt]
    \centering
    \includegraphics[width=1.\textwidth]{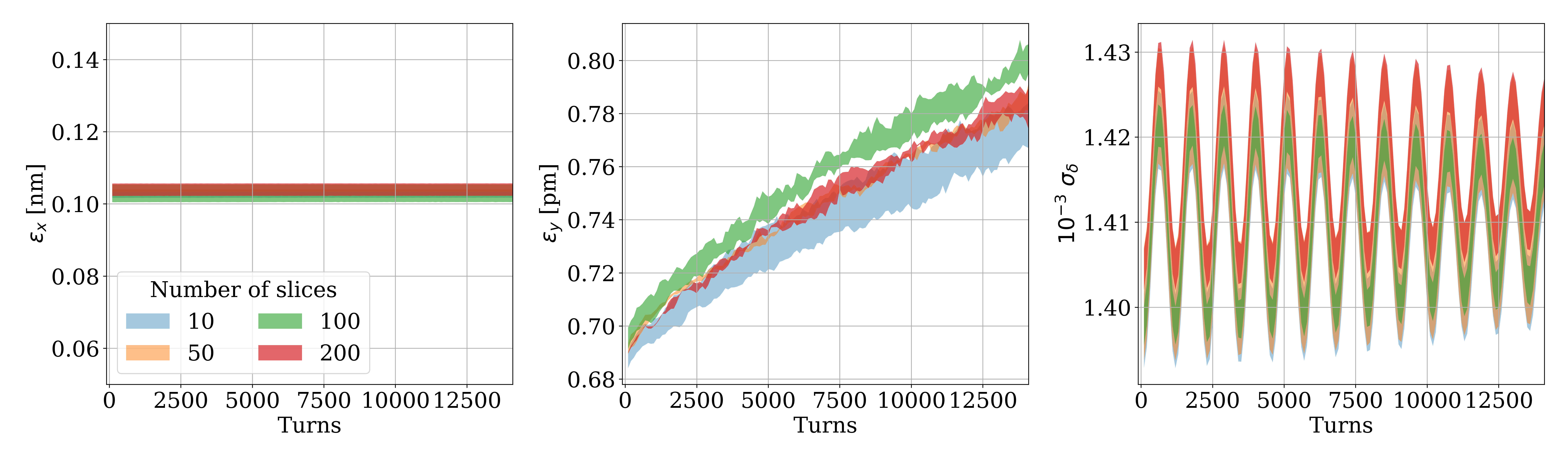}
    \caption{Evolution of the horizontal emittance~(left), vertical emittance~(center) and momentum spread~(right) in the CLIC DR, for the full CLIC DRs cycle, for four different numbers of longitudinal profile slices for the SC potential, while the $3Q_y$ resonance is strongly excited by a skew sextupolar error with normalized integrated strength $k_\text{2s}L=100~\text{m}^{-2}$. The simulations are performed with 10000 macro-particles for the working point $(Q_x, Q_y)~=~(48.39, 10.35)$, which is expected to be affected by the resonance. The spread in the evolution indicates the standard deviation over three different runs.}
    \label{fig:sls_conv1}
\end{figure*}

\begin{figure*}[!hbt]
    \centering
    \includegraphics[width=1.\textwidth]{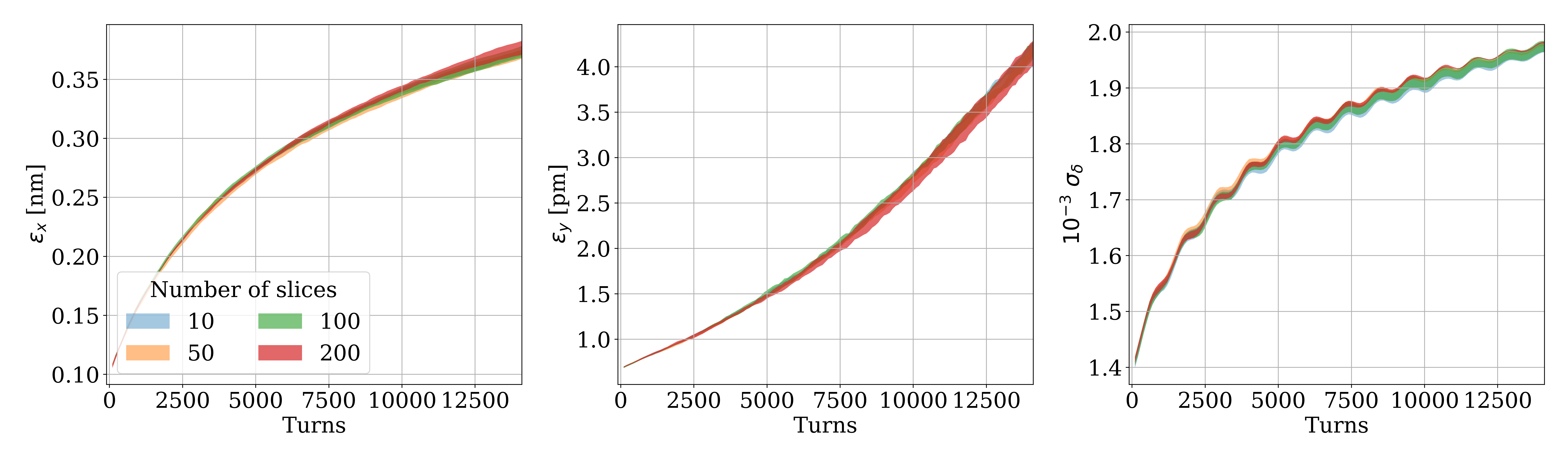}
    \caption{Evolution of the horizontal emittance~(left), vertical emittance~(center) and momentum spread~(right) in the CLIC DR, for the full CLIC DRs cycle, for four different numbers of longitudinal profile slices for the SC potential, in the presence of SC and IBS while the $3Q_y$ resonance is strongly excited by a skew sextupolar error with normalized integrated strength $k_\text{2s}L=100~\text{m}^{-2}$. The simulations are performed for the working point $(Q_x, Q_y)~=~(48.39, 10.35)$, which is expected to be affected by the resonance, with 10000 macro-particles. The spread in the evolution indicates the standard deviation over three different runs.}
    \label{fig:sls_conv2}
\end{figure*}

\begin{figure*}[!hbt]
    \centering
    \includegraphics[width=1.\textwidth]{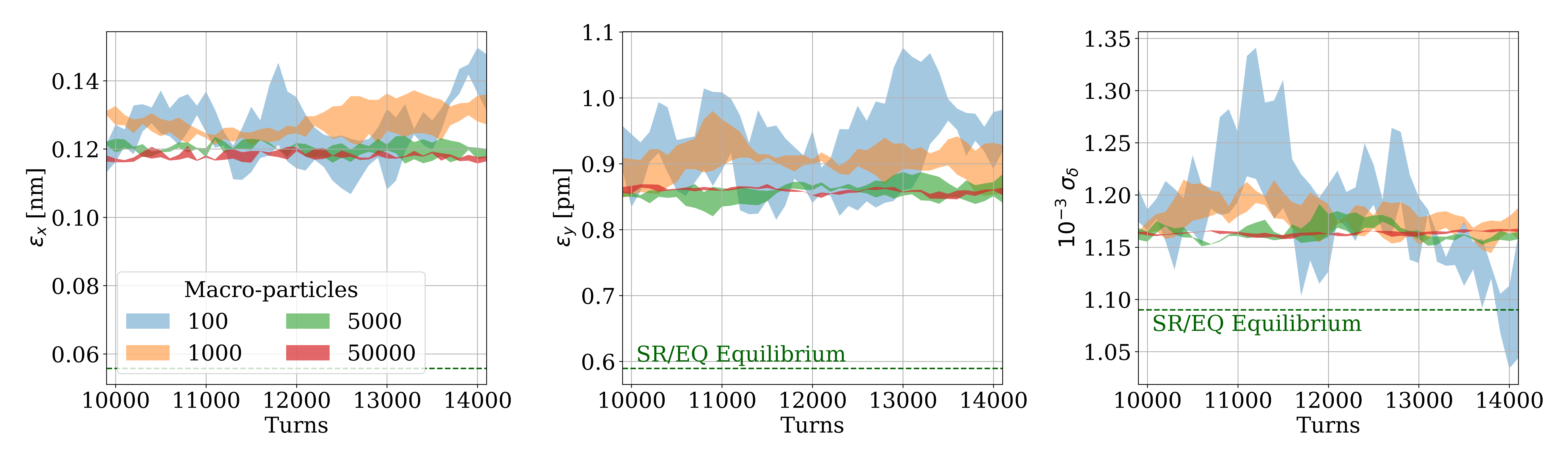}
    \caption{Evolution of the horizontal emittance~(left), vertical emittance~(center) and momentum spread~(right) in the CLIC DR, for the last 4000~turns of the cycle, for four different numbers of macro-particles, in the presence of SC, IBS and SR, while the $3Q_y$ resonance is strongly excited by a skew sextupolar error with normalized integrated strength $k_\text{2s}L=100~\text{m}^{-2}$. The simulations are performed for the working point $(Q_x, Q_y)~=~(48.39, 10.352)$, which is expected to be affected by the resonance. The spread in the evolution indicates the standard deviation over three different runs.}
    \label{fig:mps_conv}
\end{figure*}
\section{Convergence studies}\label{app:conv}

Choosing the correct number of slices for the SC potential and the number of macro-particles in tracking simulations, while trying to keep the computing resources to a minimum without sacrificing accuracy in the results, is crucial. To this end, a convergence study has been performed for three different cases. In these study cases, different number of slices or different number of macro-particles were scanned, in order to decide the optimal parameters.

The configuration of the first two convergence studies is the following: the initial beam parameters are the steady state parameters and the distribution consists of 10000 macro-particles. The $3Q_y$ resonance is strongly excited by a skew sextupolar error with normalized integrated strength $k_\text{2s}L=100~\text{m}^{-2}$, as presented in Sec.~\ref{sec:4} and the working point is set to $(Q_x, Q_y)~=~(48.39, 10.35)$ which is expected to be affected by the $3Q_y$ resonance.

The first case includes only SC with different number of longitudinal profile slices used in the calculation of the SC potential. Results are summarized in Fig.~\ref{fig:sls_conv1} where the evolution of the transverse emittances and the momentum spread are shown for the duration of the full cycle of the CLIC DRs. Four different number of slices are chosen; 10, 50, 100 and 200 slices and three different runs were performed for each case to calculate the standard deviation. Differences in the results arise mostly due to the initial conditions of the simulations, with no significant difference among the different number of slices.

Following exactly the same configuration, a second study case was performed with the addition of IBS in combination with SC. The evolution of the transverse emittances and the momentum spread for the same number of slices as the previous case are presented in Fig.~\ref{fig:sls_conv2}, where no significant differences were observed for different number of slices. Hence, the maximum number number of 200 slices was chosen for all simulations.

The third convergence study case follows a different configuration. The initial beam parameters are injection parameters, shown in Table~\ref{tab:ineqams}, while SC, IBS and SR are included in the simulations. Again, the $3Q_y$ resonance is strongly excited while the working point is set to $(Q_x, Q_y)~=~(48.39, 10.352)$. The SC potential is calculated using 200 slices and four different number of macro-particles are chosen; 100, 1000, 5000 and 50000 macro-particles. Figure~\ref{fig:mps_conv} summarizes the results of the transverse emittances and momentum spread evolution for the last 4000 turns of the cycle that the steady state is reached. The spread is again defined by the standard deviation over three different simulations. In the case of very few macro-particles, e.g., 100 or 1000, the final beam parameters deviate due to the high noise levels in the simulations. Increasing the number of macro-particles to 5000 or above produces accurate results, with less than $1\%$ of relative difference in all three planes when comparing to the case of 50000 macro-particles. Thus, a particle distribution that consists of 5000 macro-particles is a good compromise that minimizes the run time of the simulations while maintaining accurate results.

\nocite{*}

\bibliography{main}

\providecommand{\noopsort}[1]{}\providecommand{\singleletter}[1]{#1}%
\begin{thebibliography}{67}%
\makeatletter
\providecommand \@ifxundefined [1]{%
 \@ifx{#1\undefined}
}%
\providecommand \@ifnum [1]{%
 \ifnum #1\expandafter \@firstoftwo
 \else \expandafter \@secondoftwo
 \fi
}%
\providecommand \@ifx [1]{%
 \ifx #1\expandafter \@firstoftwo
 \else \expandafter \@secondoftwo
 \fi
}%
\providecommand \natexlab [1]{#1}%
\providecommand \enquote  [1]{``#1''}%
\providecommand \bibnamefont  [1]{#1}%
\providecommand \bibfnamefont [1]{#1}%
\providecommand \citenamefont [1]{#1}%
\providecommand \href@noop [0]{\@secondoftwo}%
\providecommand \href [0]{\begingroup \@sanitize@url \@href}%
\providecommand \@href[1]{\@@startlink{#1}\@@href}%
\providecommand \@@href[1]{\endgroup#1\@@endlink}%
\providecommand \@sanitize@url [0]{\catcode `\\12\catcode `\$12\catcode `\&12\catcode `\#12\catcode `\^12\catcode `\_12\catcode `\%12\relax}%
\providecommand \@@startlink[1]{}%
\providecommand \@@endlink[0]{}%
\providecommand \url  [0]{\begingroup\@sanitize@url \@url }%
\providecommand \@url [1]{\endgroup\@href {#1}{\urlprefix }}%
\providecommand \urlprefix  [0]{URL }%
\providecommand \Eprint [0]{\href }%
\providecommand \doibase [0]{https://doi.org/}%
\providecommand \selectlanguage [0]{\@gobble}%
\providecommand \bibinfo  [0]{\@secondoftwo}%
\providecommand \bibfield  [0]{\@secondoftwo}%
\providecommand \translation [1]{[#1]}%
\providecommand \BibitemOpen [0]{}%
\providecommand \bibitemStop [0]{}%
\providecommand \bibitemNoStop [0]{.\EOS\space}%
\providecommand \EOS [0]{\spacefactor3000\relax}%
\providecommand \BibitemShut  [1]{\csname bibitem#1\endcsname}%
\let\auto@bib@innerbib\@empty
\bibitem [{\citenamefont {Mertens}(2011)}]{MSc_Mertens}%
  \BibitemOpen
  \bibfield  {author} {\bibinfo {author} {\bibfnamefont {T.}~\bibnamefont {Mertens}},\ }\emph {\bibinfo {title} {{Intrabeam scattering in the LHC}}},\ \href {https://cds.cern.ch/record/1364596} {Master's thesis},\ \bibinfo  {school} {{University of Porto}} (\bibinfo {year} {2011})\BibitemShut {NoStop}%
\bibitem [{\citenamefont {Wei}\ \emph {et~al.}(2005)\citenamefont {Wei}, \citenamefont {Fedotov}, \citenamefont {Fischer}, \citenamefont {Malitsky}, \citenamefont {Parzen},\ and\ \citenamefont {Qiang}}]{JWei_RHIC}%
  \BibitemOpen
  \bibfield  {author} {\bibinfo {author} {\bibfnamefont {J.}~\bibnamefont {Wei}}, \bibinfo {author} {\bibfnamefont {A.}~\bibnamefont {Fedotov}}, \bibinfo {author} {\bibfnamefont {W.}~\bibnamefont {Fischer}}, \bibinfo {author} {\bibfnamefont {N.}~\bibnamefont {Malitsky}}, \bibinfo {author} {\bibfnamefont {G.}~\bibnamefont {Parzen}},\ and\ \bibinfo {author} {\bibfnamefont {J.}~\bibnamefont {Qiang}},\ }\bibfield  {title} {\bibinfo {title} {Intra‐beam scattering theory and rhic experiments},\ }\href {https://doi.org/10.1063/1.1949570} {\bibfield  {journal} {\bibinfo  {journal} {AIP Conference Proceedings}\ }\textbf {\bibinfo {volume} {773}},\ \bibinfo {pages} {389} (\bibinfo {year} {2005})}\BibitemShut {NoStop}%
\bibitem [{\citenamefont {Fischer}\ \emph {et~al.}(2001)\citenamefont {Fischer}, \citenamefont {Bai}, \citenamefont {Blaskiewicz}, \citenamefont {Brennan}, \citenamefont {Cameron}, \citenamefont {Connolly}, \citenamefont {Lehrach}, \citenamefont {Parzen}, \citenamefont {Tepikian}, \citenamefont {Zeno},\ and\ \citenamefont {Van~Zeijts}}]{Fischer_RHIC}%
  \BibitemOpen
  \bibfield  {author} {\bibinfo {author} {\bibfnamefont {W.}~\bibnamefont {Fischer}}, \bibinfo {author} {\bibfnamefont {M.}~\bibnamefont {Bai}}, \bibinfo {author} {\bibfnamefont {M.}~\bibnamefont {Blaskiewicz}}, \bibinfo {author} {\bibfnamefont {J.~M.}\ \bibnamefont {Brennan}}, \bibinfo {author} {\bibfnamefont {P.}~\bibnamefont {Cameron}}, \bibinfo {author} {\bibfnamefont {R.}~\bibnamefont {Connolly}}, \bibinfo {author} {\bibfnamefont {A.}~\bibnamefont {Lehrach}}, \bibinfo {author} {\bibfnamefont {G.}~\bibnamefont {Parzen}}, \bibinfo {author} {\bibfnamefont {S.}~\bibnamefont {Tepikian}}, \bibinfo {author} {\bibfnamefont {K.}~\bibnamefont {Zeno}},\ and\ \bibinfo {author} {\bibfnamefont {J.}~\bibnamefont {Van~Zeijts}},\ }\bibfield  {title} {\bibinfo {title} {{Measurements of Intrabeam scattering growth times with gold beam below transition in RHIC}},\ }\href@noop {} {\bibfield  {journal} {\bibinfo  {journal} {Conf. Proc. C}\ }\textbf {\bibinfo {volume} {0106181}},\ \bibinfo {pages} {2857} (\bibinfo {year}
  {2001})}\BibitemShut {NoStop}%
\bibitem [{\citenamefont {Bruce}\ \emph {et~al.}(2010)\citenamefont {Bruce}, \citenamefont {Jowett}, \citenamefont {Blaskiewicz},\ and\ \citenamefont {Fischer}}]{Bruce:ibs_kick}%
  \BibitemOpen
  \bibfield  {author} {\bibinfo {author} {\bibfnamefont {R.}~\bibnamefont {Bruce}}, \bibinfo {author} {\bibfnamefont {J.~M.}\ \bibnamefont {Jowett}}, \bibinfo {author} {\bibfnamefont {M.}~\bibnamefont {Blaskiewicz}},\ and\ \bibinfo {author} {\bibfnamefont {W.}~\bibnamefont {Fischer}},\ }\bibfield  {title} {\bibinfo {title} {Time evolution of the luminosity of colliding heavy-ion beams in bnl relativistic heavy ion collider and cern large hadron collider},\ }\href {https://doi.org/10.1103/PhysRevSTAB.13.091001} {\bibfield  {journal} {\bibinfo  {journal} {Phys. Rev. ST Accel. Beams}\ }\textbf {\bibinfo {volume} {13}},\ \bibinfo {pages} {091001} (\bibinfo {year} {2010})}\BibitemShut {NoStop}%
\bibitem [{\citenamefont {Lebedev}(2005)}]{Lebedev_1}%
  \BibitemOpen
  \bibfield  {author} {\bibinfo {author} {\bibfnamefont {V.}~\bibnamefont {Lebedev}},\ }\bibfield  {title} {\bibinfo {title} {{Single and Multiple Intrabeam Scattering in Hadron Colliders}},\ }\href {https://doi.org/10.1063/1.1949581} {\bibfield  {journal} {\bibinfo  {journal} {AIP Conference Proceedings}\ }\textbf {\bibinfo {volume} {773}},\ \bibinfo {pages} {440} (\bibinfo {year} {2005})},\ \Eprint {https://arxiv.org/abs/https://pubs.aip.org/aip/acp/article-pdf/773/1/440/11391455/440\_1\_online.pdf} {https://pubs.aip.org/aip/acp/article-pdf/773/1/440/11391455/440\_1\_online.pdf} \BibitemShut {NoStop}%
\bibitem [{\citenamefont {Papadopoulou}\ \emph {et~al.}(2020)\citenamefont {Papadopoulou}, \citenamefont {Antoniou}, \citenamefont {Argyropoulos}, \citenamefont {Hostettler}, \citenamefont {Papaphilippou},\ and\ \citenamefont {Trad}}]{Papadopoulou:2018uvl}%
  \BibitemOpen
  \bibfield  {author} {\bibinfo {author} {\bibfnamefont {S.}~\bibnamefont {Papadopoulou}}, \bibinfo {author} {\bibfnamefont {F.}~\bibnamefont {Antoniou}}, \bibinfo {author} {\bibfnamefont {T.}~\bibnamefont {Argyropoulos}}, \bibinfo {author} {\bibfnamefont {M.}~\bibnamefont {Hostettler}}, \bibinfo {author} {\bibfnamefont {Y.}~\bibnamefont {Papaphilippou}},\ and\ \bibinfo {author} {\bibfnamefont {G.}~\bibnamefont {Trad}},\ }\bibfield  {title} {\bibinfo {title} {Impact of non-gaussian beam profiles in the performance of hadron colliders},\ }\href {https://doi.org/10.1103/PhysRevAccelBeams.23.101004} {\bibfield  {journal} {\bibinfo  {journal} {Phys. Rev. Accel. Beams}\ }\textbf {\bibinfo {volume} {23}},\ \bibinfo {pages} {101004} (\bibinfo {year} {2020})}\BibitemShut {NoStop}%
\bibitem [{\citenamefont {Tom\'as}\ \emph {et~al.}(2020)\citenamefont {Tom\'as}, \citenamefont {Keintzel},\ and\ \citenamefont {Papadopoulou}}]{Rogelio}%
  \BibitemOpen
  \bibfield  {author} {\bibinfo {author} {\bibfnamefont {R.}~\bibnamefont {Tom\'as}}, \bibinfo {author} {\bibfnamefont {J.}~\bibnamefont {Keintzel}},\ and\ \bibinfo {author} {\bibfnamefont {S.}~\bibnamefont {Papadopoulou}},\ }\bibfield  {title} {\bibinfo {title} {{Emittance growth from luminosity burn-off in future hadron colliders}},\ }\href {https://doi.org/10.1103/PhysRevAccelBeams.23.031002} {\bibfield  {journal} {\bibinfo  {journal} {Phys. Rev. Accel. Beams}\ }\textbf {\bibinfo {volume} {23}},\ \bibinfo {pages} {031002} (\bibinfo {year} {2020})}\BibitemShut {NoStop}%
\bibitem [{\citenamefont {Antoniou}\ \emph {et~al.}(2018)\citenamefont {Antoniou}, \citenamefont {Alekou}, \citenamefont {Bartosik}, \citenamefont {Bohl}, \citenamefont {Calaga}, \citenamefont {Carver}, \citenamefont {Repond},\ and\ \citenamefont {Vandoni}}]{Antoniou:SPS}%
  \BibitemOpen
  \bibfield  {author} {\bibinfo {author} {\bibfnamefont {F.}~\bibnamefont {Antoniou}}, \bibinfo {author} {\bibfnamefont {A.}~\bibnamefont {Alekou}}, \bibinfo {author} {\bibfnamefont {H.}~\bibnamefont {Bartosik}}, \bibinfo {author} {\bibfnamefont {T.}~\bibnamefont {Bohl}}, \bibinfo {author} {\bibfnamefont {R.}~\bibnamefont {Calaga}}, \bibinfo {author} {\bibfnamefont {L.}~\bibnamefont {Carver}}, \bibinfo {author} {\bibfnamefont {J.}~\bibnamefont {Repond}},\ and\ \bibinfo {author} {\bibfnamefont {G.}~\bibnamefont {Vandoni}},\ }\bibfield  {title} {\bibinfo {title} {{Emittance growth in coast in the SPS at CERN}},\ }\href {https://doi.org/10.1088/1742-6596/1067/2/022008} {\bibfield  {journal} {\bibinfo  {journal} {Journal of Physics: Conference Series}\ }\textbf {\bibinfo {volume} {1067}},\ \bibinfo {pages} {022008} (\bibinfo {year} {2018})}\BibitemShut {NoStop}%
\bibitem [{\citenamefont {Evans}\ and\ \citenamefont {Zotter}(1980)}]{Evans:1980}%
  \BibitemOpen
  \bibfield  {author} {\bibinfo {author} {\bibfnamefont {L.~R.}\ \bibnamefont {Evans}}\ and\ \bibinfo {author} {\bibfnamefont {B.~W.}\ \bibnamefont {Zotter}},\ }\href {https://cds.cern.ch/record/126036} {\emph {\bibinfo {title} {{Intrabeam scattering in the SPS}}}},\ \bibinfo {type} {Tech. Rep.}\ (\bibinfo  {institution} {CERN},\ \bibinfo {address} {Geneva},\ \bibinfo {year} {1980})\BibitemShut {NoStop}%
\bibitem [{\citenamefont {Papadopoulou}(2019)}]{PhD_Papadopoulou}%
  \BibitemOpen
  \bibfield  {author} {\bibinfo {author} {\bibfnamefont {P.~S.}\ \bibnamefont {Papadopoulou}},\ }\emph {\bibinfo {title} {{Bunch characteristics evolution for lepton and hadron rings under the influence of the Intra-beam scattering effect}}},\ \href@noop {} {Ph.D. thesis},\ \bibinfo  {school} {Crete U.} (\bibinfo {year} {2019})\BibitemShut {NoStop}%
\bibitem [{\citenamefont {Lebedev}\ \emph {et~al.}(2020)\citenamefont {Lebedev}, \citenamefont {Lobach}, \citenamefont {Romanov},\ and\ \citenamefont {Valishev}}]{Lebedev_2}%
  \BibitemOpen
  \bibfield  {author} {\bibinfo {author} {\bibfnamefont {V.~A.}\ \bibnamefont {Lebedev}}, \bibinfo {author} {\bibfnamefont {I.}~\bibnamefont {Lobach}}, \bibinfo {author} {\bibfnamefont {A.}~\bibnamefont {Romanov}},\ and\ \bibinfo {author} {\bibfnamefont {A.}~\bibnamefont {Valishev}},\ }\href {https://doi.org/10.2172/1764146} {\emph {\bibinfo {title} {{Report on Single and Multiple Intrabeam Scattering Measurements in IOTA Ring in Fermilab}}}},\ \bibinfo {type} {Tech. Rep.}\ (\bibinfo  {institution} {Fermilab},\ \bibinfo {year} {2020})\BibitemShut {NoStop}%
\bibitem [{\citenamefont {Antoniou}(2012)}]{PhD_Antoniou}%
  \BibitemOpen
  \bibfield  {author} {\bibinfo {author} {\bibfnamefont {F.}~\bibnamefont {Antoniou}},\ }\emph {\bibinfo {title} {{Optics design of Intrabeam Scattering dominated damping rings}}},\ \href {http://cds.cern.ch/record/1666863} {Ph.D. thesis},\ \bibinfo  {school} {{}} (\bibinfo {year} {2012}),\ \bibinfo {note} {presented 08 Jan 2013}\BibitemShut {NoStop}%
\bibitem [{\citenamefont {Antoniou}\ \emph {et~al.}(2012)\citenamefont {Antoniou}, \citenamefont {Papaphilippou}, \citenamefont {Aiba}, \citenamefont {Boege}, \citenamefont {Milas}, \citenamefont {Streun},\ and\ \citenamefont {Demma}}]{Antoniou:SLS}%
  \BibitemOpen
  \bibfield  {author} {\bibinfo {author} {\bibfnamefont {F.}~\bibnamefont {Antoniou}}, \bibinfo {author} {\bibfnamefont {Y.}~\bibnamefont {Papaphilippou}}, \bibinfo {author} {\bibfnamefont {M.}~\bibnamefont {Aiba}}, \bibinfo {author} {\bibfnamefont {M.}~\bibnamefont {Boege}}, \bibinfo {author} {\bibfnamefont {N.}~\bibnamefont {Milas}}, \bibinfo {author} {\bibfnamefont {A.}~\bibnamefont {Streun}},\ and\ \bibinfo {author} {\bibfnamefont {T.}~\bibnamefont {Demma}},\ }\bibfield  {title} {\bibinfo {title} {{Intrabeam scattering studies at the Swiss light source}},\ }\href {https://cds.cern.ch/record/1464110} {\bibfield  {journal} {\bibinfo  {journal} {Conf. Proc.}\ }\textbf {\bibinfo {volume} {C1205201}},\ \bibinfo {pages} {TUPPR057. 3 p} (\bibinfo {year} {2012})}\BibitemShut {NoStop}%
\bibitem [{\citenamefont {Bane}(2012)}]{Bane_2}%
  \BibitemOpen
  \bibfield  {author} {\bibinfo {author} {\bibfnamefont {K.~L.~F.}\ \bibnamefont {Bane}},\ }\href {https://www.osti.gov/biblio/1037597} {\emph {\bibinfo {title} {Intra-Beam Scattering, Impedance, and Instabilities in Ultimate Storage Rings}}},\ \bibinfo {type} {Tech. Rep.}\ (\bibinfo  {institution} {SLAC National Accelerator Lab., Menlo Park, CA (United States)},\ \bibinfo {year} {2012})\BibitemShut {NoStop}%
\bibitem [{\citenamefont {Bane}\ \emph {et~al.}(2002)\citenamefont {Bane}, \citenamefont {Hayano}, \citenamefont {Kubo}, \citenamefont {Naito}, \citenamefont {Okugi},\ and\ \citenamefont {Urakawa}}]{Bane_KEK}%
  \BibitemOpen
  \bibfield  {author} {\bibinfo {author} {\bibfnamefont {K.~L.~F.}\ \bibnamefont {Bane}}, \bibinfo {author} {\bibfnamefont {H.}~\bibnamefont {Hayano}}, \bibinfo {author} {\bibfnamefont {K.}~\bibnamefont {Kubo}}, \bibinfo {author} {\bibfnamefont {T.}~\bibnamefont {Naito}}, \bibinfo {author} {\bibfnamefont {T.}~\bibnamefont {Okugi}},\ and\ \bibinfo {author} {\bibfnamefont {J.}~\bibnamefont {Urakawa}},\ }\bibfield  {title} {\bibinfo {title} {Intrabeam scattering analysis of measurements at {KEK}'s accelerator test facility damping ring},\ }\bibfield  {journal} {\bibinfo  {journal} {Physical Review Special Topics - Accelerators and Beams}\ }\textbf {\bibinfo {volume} {5}},\ \href {https://doi.org/10.1103/physrevstab.5.084403} {10.1103/physrevstab.5.084403} (\bibinfo {year} {2002})\BibitemShut {NoStop}%
\bibitem [{\citenamefont {Kubo}\ \emph {et~al.}(2005)\citenamefont {Kubo}, \citenamefont {Mtingwa},\ and\ \citenamefont {Wolski}}]{Kubo_KEK}%
  \BibitemOpen
  \bibfield  {author} {\bibinfo {author} {\bibfnamefont {K.}~\bibnamefont {Kubo}}, \bibinfo {author} {\bibfnamefont {S.~K.}\ \bibnamefont {Mtingwa}},\ and\ \bibinfo {author} {\bibfnamefont {A.}~\bibnamefont {Wolski}},\ }\bibfield  {title} {\bibinfo {title} {Intrabeam scattering formulas for high energy beams},\ }\href {https://doi.org/10.1103/PhysRevSTAB.8.081001} {\bibfield  {journal} {\bibinfo  {journal} {Phys. Rev. ST Accel. Beams}\ }\textbf {\bibinfo {volume} {8}},\ \bibinfo {pages} {081001} (\bibinfo {year} {2005})}\BibitemShut {NoStop}%
\bibitem [{\citenamefont {Huang}(2002)}]{Huang}%
  \BibitemOpen
  \bibfield  {author} {\bibinfo {author} {\bibfnamefont {Z.}~\bibnamefont {Huang}},\ }\href@noop {} {\emph {\bibinfo {title} {{Intrabeam scattering in an X-ray FEL driver}}}},\ \bibinfo {type} {Tech. Rep.}\ (\bibinfo  {institution} {SLAC National Accelerator Lab., Menlo Park, CA (United States)},\ \bibinfo {year} {2002})\BibitemShut {NoStop}%
\bibitem [{\citenamefont {Xiao}\ and\ \citenamefont {Borland}(2010)}]{Xiao_2010}%
  \BibitemOpen
  \bibfield  {author} {\bibinfo {author} {\bibfnamefont {A.}~\bibnamefont {Xiao}}\ and\ \bibinfo {author} {\bibfnamefont {M.}~\bibnamefont {Borland}},\ }\bibfield  {title} {\bibinfo {title} {{Intrabeam Scattering Effect Calculated for a Non-Gaussian Distributed Linac Beam}},\ }in\ \href@noop {} {\emph {\bibinfo {booktitle} {{Particle Accelerator Conference (PAC 09)}}}}\ (\bibinfo {year} {2010})\ p.\ \bibinfo {pages} {TH5PFP038}\BibitemShut {NoStop}%
\bibitem [{\citenamefont {Mitri}\ \emph {et~al.}(2020)\citenamefont {Mitri}, \citenamefont {Perosa}, \citenamefont {Brynes}, \citenamefont {Setija}, \citenamefont {Spampinati}, \citenamefont {Williams}, \citenamefont {Wolski}, \citenamefont {Allaria}, \citenamefont {Brussaard}, \citenamefont {Giannessi}, \citenamefont {Penco}, \citenamefont {Rebernik},\ and\ \citenamefont {Trovò}}]{DiMitri_2020}%
  \BibitemOpen
  \bibfield  {author} {\bibinfo {author} {\bibfnamefont {S.~D.}\ \bibnamefont {Mitri}}, \bibinfo {author} {\bibfnamefont {G.}~\bibnamefont {Perosa}}, \bibinfo {author} {\bibfnamefont {A.}~\bibnamefont {Brynes}}, \bibinfo {author} {\bibfnamefont {I.}~\bibnamefont {Setija}}, \bibinfo {author} {\bibfnamefont {S.}~\bibnamefont {Spampinati}}, \bibinfo {author} {\bibfnamefont {P.~H.}\ \bibnamefont {Williams}}, \bibinfo {author} {\bibfnamefont {A.}~\bibnamefont {Wolski}}, \bibinfo {author} {\bibfnamefont {E.}~\bibnamefont {Allaria}}, \bibinfo {author} {\bibfnamefont {S.}~\bibnamefont {Brussaard}}, \bibinfo {author} {\bibfnamefont {L.}~\bibnamefont {Giannessi}}, \bibinfo {author} {\bibfnamefont {G.}~\bibnamefont {Penco}}, \bibinfo {author} {\bibfnamefont {P.~R.}\ \bibnamefont {Rebernik}},\ and\ \bibinfo {author} {\bibfnamefont {M.}~\bibnamefont {Trovò}},\ }\bibfield  {title} {\bibinfo {title} {Experimental evidence of intrabeam scattering in a free-electron laser driver},\ }\href
  {https://doi.org/10.1088/1367-2630/aba572} {\bibfield  {journal} {\bibinfo  {journal} {New Journal of Physics}\ }\textbf {\bibinfo {volume} {22}},\ \bibinfo {pages} {083053} (\bibinfo {year} {2020})}\BibitemShut {NoStop}%
\bibitem [{\citenamefont {Dalena}\ \emph {et~al.}(2022)\citenamefont {Dalena}, \citenamefont {Chance}, \citenamefont {Antoniou}, \citenamefont {Etisken}, \citenamefont {Mashal}, \citenamefont {Raubenheimer}, \citenamefont {Zampetakis},\ and\ \citenamefont {Zimmermann}}]{Dalena_2022}%
  \BibitemOpen
  \bibfield  {author} {\bibinfo {author} {\bibfnamefont {B.}~\bibnamefont {Dalena}}, \bibinfo {author} {\bibfnamefont {A.}~\bibnamefont {Chance}}, \bibinfo {author} {\bibfnamefont {F.}~\bibnamefont {Antoniou}}, \bibinfo {author} {\bibfnamefont {O.}~\bibnamefont {Etisken}}, \bibinfo {author} {\bibfnamefont {A.}~\bibnamefont {Mashal}}, \bibinfo {author} {\bibfnamefont {T.}~\bibnamefont {Raubenheimer}}, \bibinfo {author} {\bibfnamefont {M.}~\bibnamefont {Zampetakis}},\ and\ \bibinfo {author} {\bibfnamefont {F.}~\bibnamefont {Zimmermann}},\ }\bibfield  {title} {\bibinfo {title} {{Status of the High Energy Booster of the lepton option of the future circular collider}},\ }in\ \href {https://doi.org/10.22323/1.414.0042} {\emph {\bibinfo {booktitle} {Proceedings of 41st International Conference on High Energy physics {\textemdash} PoS(ICHEP2022)}}},\ Vol.\ \bibinfo {volume} {414}\ (\bibinfo {year} {2022})\ p.\ \bibinfo {pages} {042}\BibitemShut {NoStop}%
\bibitem [{\citenamefont {Etisken}\ \emph {et~al.}(2017)\citenamefont {Etisken}, \citenamefont {Papaphilippou},\ and\ \citenamefont {Ciftci}}]{Etisken_2017}%
  \BibitemOpen
  \bibfield  {author} {\bibinfo {author} {\bibfnamefont {O.}~\bibnamefont {Etisken}}, \bibinfo {author} {\bibfnamefont {Y.}~\bibnamefont {Papaphilippou}},\ and\ \bibinfo {author} {\bibfnamefont {A.~K.}\ \bibnamefont {Ciftci}},\ }\bibfield  {title} {\bibinfo {title} {{Conceptual design of a pre-booster ring for FCC e+e- injector}},\ }\href {https://doi.org/10.1088/1742-6596/874/1/012014} {\bibfield  {journal} {\bibinfo  {journal} {Journal of Physics: Conference Series}\ }\textbf {\bibinfo {volume} {874}},\ \bibinfo {pages} {012014} (\bibinfo {year} {2017})}\BibitemShut {NoStop}%
\bibitem [{\citenamefont {Franchetti}\ \emph {et~al.}(2003)\citenamefont {Franchetti}, \citenamefont {Hofmann}, \citenamefont {Giovannozzi}, \citenamefont {Martini},\ and\ \citenamefont {Metral}}]{Franchetti:2003aa}%
  \BibitemOpen
  \bibfield  {author} {\bibinfo {author} {\bibfnamefont {G.}~\bibnamefont {Franchetti}}, \bibinfo {author} {\bibfnamefont {I.}~\bibnamefont {Hofmann}}, \bibinfo {author} {\bibfnamefont {M.}~\bibnamefont {Giovannozzi}}, \bibinfo {author} {\bibfnamefont {M.}~\bibnamefont {Martini}},\ and\ \bibinfo {author} {\bibfnamefont {E.}~\bibnamefont {Metral}},\ }\bibfield  {title} {\bibinfo {title} {{Space charge and octupole driven resonance trapping observed at the CERN Proton Synchrotron}},\ }\href {https://doi.org/10.1103/PhysRevSTAB.6.124201} {\bibfield  {journal} {\bibinfo  {journal} {Phys. Rev. ST Accel. Beams}\ }\textbf {\bibinfo {volume} {6}},\ \bibinfo {pages} {124201} (\bibinfo {year} {2003})}\BibitemShut {NoStop}%
\bibitem [{\citenamefont {M\'etral}\ \emph {et~al.}(2006)\citenamefont {M\'etral}, \citenamefont {Franchetti}, \citenamefont {Giovannozzi}, \citenamefont {Hofmann}, \citenamefont {Martini},\ and\ \citenamefont {Steerenberg}}]{Metral:2006qw}%
  \BibitemOpen
  \bibfield  {author} {\bibinfo {author} {\bibfnamefont {E.}~\bibnamefont {M\'etral}}, \bibinfo {author} {\bibfnamefont {G.}~\bibnamefont {Franchetti}}, \bibinfo {author} {\bibfnamefont {M.}~\bibnamefont {Giovannozzi}}, \bibinfo {author} {\bibfnamefont {I.}~\bibnamefont {Hofmann}}, \bibinfo {author} {\bibfnamefont {M.}~\bibnamefont {Martini}},\ and\ \bibinfo {author} {\bibfnamefont {R.}~\bibnamefont {Steerenberg}},\ }\bibfield  {title} {\bibinfo {title} {Observation of octupole driven resonance phenomena with space charge at the cern proton synchrotron},\ }\href {https://doi.org/https://doi.org/10.1016/j.nima.2006.01.029} {\bibfield  {journal} {\bibinfo  {journal} {Nuclear Instruments and Methods in Physics Research Section A: Accelerators, Spectrometers, Detectors and Associated Equipment}\ }\textbf {\bibinfo {volume} {561}},\ \bibinfo {pages} {257} (\bibinfo {year} {2006})},\ \bibinfo {note} {proceedings of the Workshop on High Intensity Beam Dynamics}\BibitemShut {NoStop}%
\bibitem [{\citenamefont {Franchetti}\ \emph {et~al.}(2010)\citenamefont {Franchetti}, \citenamefont {Chorniy}, \citenamefont {Hofmann}, \citenamefont {Bayer}, \citenamefont {Becker}, \citenamefont {Forck}, \citenamefont {Giacomini}, \citenamefont {Kirk}, \citenamefont {Mohite}, \citenamefont {Omet}, \citenamefont {Parfenova},\ and\ \citenamefont {Sch\"utt}}]{Franchetti:2010zz}%
  \BibitemOpen
  \bibfield  {author} {\bibinfo {author} {\bibfnamefont {G.}~\bibnamefont {Franchetti}}, \bibinfo {author} {\bibfnamefont {O.}~\bibnamefont {Chorniy}}, \bibinfo {author} {\bibfnamefont {I.}~\bibnamefont {Hofmann}}, \bibinfo {author} {\bibfnamefont {W.}~\bibnamefont {Bayer}}, \bibinfo {author} {\bibfnamefont {F.}~\bibnamefont {Becker}}, \bibinfo {author} {\bibfnamefont {P.}~\bibnamefont {Forck}}, \bibinfo {author} {\bibfnamefont {T.}~\bibnamefont {Giacomini}}, \bibinfo {author} {\bibfnamefont {M.}~\bibnamefont {Kirk}}, \bibinfo {author} {\bibfnamefont {T.}~\bibnamefont {Mohite}}, \bibinfo {author} {\bibfnamefont {C.}~\bibnamefont {Omet}}, \bibinfo {author} {\bibfnamefont {A.}~\bibnamefont {Parfenova}},\ and\ \bibinfo {author} {\bibfnamefont {P.}~\bibnamefont {Sch\"utt}},\ }\bibfield  {title} {\bibinfo {title} {Experiment on space charge driven nonlinear resonance crossing in an ion synchrotron},\ }\href {https://doi.org/10.1103/PhysRevSTAB.13.114203} {\bibfield  {journal} {\bibinfo  {journal} {Phys. Rev. ST
  Accel. Beams}\ }\textbf {\bibinfo {volume} {13}},\ \bibinfo {pages} {114203} (\bibinfo {year} {2010})}\BibitemShut {NoStop}%
\bibitem [{\citenamefont {Franchetti}\ \emph {et~al.}(2017)\citenamefont {Franchetti}, \citenamefont {Gilardoni}, \citenamefont {Huschauer}, \citenamefont {Schmidt},\ and\ \citenamefont {Wasef}}]{Franchetti:2017aa}%
  \BibitemOpen
  \bibfield  {author} {\bibinfo {author} {\bibfnamefont {G.}~\bibnamefont {Franchetti}}, \bibinfo {author} {\bibfnamefont {S.}~\bibnamefont {Gilardoni}}, \bibinfo {author} {\bibfnamefont {A.}~\bibnamefont {Huschauer}}, \bibinfo {author} {\bibfnamefont {F.}~\bibnamefont {Schmidt}},\ and\ \bibinfo {author} {\bibfnamefont {R.}~\bibnamefont {Wasef}},\ }\bibfield  {title} {\bibinfo {title} {Space charge effects on the third order coupled resonance},\ }\href {https://doi.org/10.1103/PhysRevAccelBeams.20.081006} {\bibfield  {journal} {\bibinfo  {journal} {Phys. Rev. Accel. Beams}\ }\textbf {\bibinfo {volume} {20}},\ \bibinfo {pages} {081006} (\bibinfo {year} {2017})}\BibitemShut {NoStop}%
\bibitem [{\citenamefont {Ainsworth}\ \emph {et~al.}(2019)\citenamefont {Ainsworth}, \citenamefont {Adamson}, \citenamefont {Amundson}, \citenamefont {Kourbanis}, \citenamefont {Lu},\ and\ \citenamefont {Stern}}]{Fermi_res}%
  \BibitemOpen
  \bibfield  {author} {\bibinfo {author} {\bibfnamefont {R.}~\bibnamefont {Ainsworth}}, \bibinfo {author} {\bibfnamefont {P.}~\bibnamefont {Adamson}}, \bibinfo {author} {\bibfnamefont {J.}~\bibnamefont {Amundson}}, \bibinfo {author} {\bibfnamefont {I.}~\bibnamefont {Kourbanis}}, \bibinfo {author} {\bibfnamefont {Q.}~\bibnamefont {Lu}},\ and\ \bibinfo {author} {\bibfnamefont {E.}~\bibnamefont {Stern}},\ }\bibfield  {title} {\bibinfo {title} {High intensity space charge effects on slip stacked beam in the fermilab recycler},\ }\href {https://doi.org/10.1103/PhysRevAccelBeams.22.020404} {\bibfield  {journal} {\bibinfo  {journal} {Phys. Rev. Accel. Beams}\ }\textbf {\bibinfo {volume} {22}},\ \bibinfo {pages} {020404} (\bibinfo {year} {2019})}\BibitemShut {NoStop}%
\bibitem [{\citenamefont {Shiltsev}(2021)}]{Fermi_IOTA}%
  \BibitemOpen
  \bibfield  {author} {\bibinfo {author} {\bibfnamefont {V.~D.}\ \bibnamefont {Shiltsev}},\ }\bibfield  {title} {\bibinfo {title} {Space-charge and other effects in fermilab booster and iota rings’ ionization profile monitors},\ }\bibfield  {journal} {\bibinfo  {journal} {JACoW}\ }\textbf {\bibinfo {volume} {IBIC2021}},\ \href {https://doi.org/10.18429/JACoW-IBIC2021-TUPP05} {10.18429/JACoW-IBIC2021-TUPP05} (\bibinfo {year} {2021})\BibitemShut {NoStop}%
\bibitem [{\citenamefont {Alexahin}\ and\ \citenamefont {Kapin}(2021)}]{Fermi_boost}%
  \BibitemOpen
  \bibfield  {author} {\bibinfo {author} {\bibfnamefont {Y.}~\bibnamefont {Alexahin}}\ and\ \bibinfo {author} {\bibfnamefont {V.}~\bibnamefont {Kapin}},\ }\bibfield  {title} {\bibinfo {title} {On possibility of space-charge compensation in the fermilab booster with multiple electron columns},\ }\href {https://doi.org/10.1088/1748-0221/16/03/P03049} {\bibfield  {journal} {\bibinfo  {journal} {Journal of Instrumentation}\ }\textbf {\bibinfo {volume} {16}}\bibinfo  {number} { (03)},\ \bibinfo {pages} {P03049}}\BibitemShut {NoStop}%
\bibitem [{\citenamefont {Chung}\ \emph {et~al.}(2013)\citenamefont {Chung}, \citenamefont {Shiltsev},\ and\ \citenamefont {Prost}}]{Fermi_highI}%
  \BibitemOpen
\bibfield  {number} {  }\bibfield  {author} {\bibinfo {author} {\bibfnamefont {M.}~\bibnamefont {Chung}}, \bibinfo {author} {\bibfnamefont {V.}~\bibnamefont {Shiltsev}},\ and\ \bibinfo {author} {\bibfnamefont {L.}~\bibnamefont {Prost}},\ }\bibfield  {title} {\bibinfo {title} {{Space-Charge Compensation for High-Intensity Linear and Circular Accelerators at Fermilab}},\ }in\ \href@noop {} {\emph {\bibinfo {booktitle} {{1st North American Particle Accelerator Conference}}}}\ (\bibinfo {year} {2013})\BibitemShut {NoStop}%
\bibitem [{\citenamefont {Dell}\ and\ \citenamefont {Peggs}(1995)}]{SC_RHIC}%
  \BibitemOpen
  \bibfield  {author} {\bibinfo {author} {\bibfnamefont {G.}~\bibnamefont {Dell}}\ and\ \bibinfo {author} {\bibfnamefont {S.}~\bibnamefont {Peggs}},\ }\bibfield  {title} {\bibinfo {title} {Simulation of the space charge effect in rhic},\ }in\ \href {https://doi.org/10.1109/PAC.1995.505541} {\emph {\bibinfo {booktitle} {Proceedings Particle Accelerator Conference}}},\ Vol.~\bibinfo {volume} {4}\ (\bibinfo {year} {1995})\ pp.\ \bibinfo {pages} {2327--2329 vol.4}\BibitemShut {NoStop}%
\bibitem [{\citenamefont {Litvinenko}\ and\ \citenamefont {Wang}(2014)}]{SC_eRHIC}%
  \BibitemOpen
  \bibfield  {author} {\bibinfo {author} {\bibfnamefont {V.~N.}\ \bibnamefont {Litvinenko}}\ and\ \bibinfo {author} {\bibfnamefont {G.}~\bibnamefont {Wang}},\ }\bibfield  {title} {\bibinfo {title} {Compensating tune spread induced by space charge in bunched beams},\ }\href {https://doi.org/10.1103/PhysRevSTAB.17.114401} {\bibfield  {journal} {\bibinfo  {journal} {Phys. Rev. ST Accel. Beams}\ }\textbf {\bibinfo {volume} {17}},\ \bibinfo {pages} {114401} (\bibinfo {year} {2014})}\BibitemShut {NoStop}%
\bibitem [{\citenamefont {Asvesta}\ \emph {et~al.}(2020)\citenamefont {Asvesta}, \citenamefont {Bartosik}, \citenamefont {Gilardoni}, \citenamefont {Huschauer}, \citenamefont {Machida}, \citenamefont {Papaphilippou},\ and\ \citenamefont {Wasef}}]{Asvesta:ps}%
  \BibitemOpen
  \bibfield  {author} {\bibinfo {author} {\bibfnamefont {F.}~\bibnamefont {Asvesta}}, \bibinfo {author} {\bibfnamefont {H.}~\bibnamefont {Bartosik}}, \bibinfo {author} {\bibfnamefont {S.}~\bibnamefont {Gilardoni}}, \bibinfo {author} {\bibfnamefont {A.}~\bibnamefont {Huschauer}}, \bibinfo {author} {\bibfnamefont {S.}~\bibnamefont {Machida}}, \bibinfo {author} {\bibfnamefont {Y.}~\bibnamefont {Papaphilippou}},\ and\ \bibinfo {author} {\bibfnamefont {R.}~\bibnamefont {Wasef}},\ }\bibfield  {title} {\bibinfo {title} {Identification and characterization of high order incoherent space charge driven structure resonances in the cern proton synchrotron},\ }\href {https://doi.org/10.1103/PhysRevAccelBeams.23.091001} {\bibfield  {journal} {\bibinfo  {journal} {Phys. Rev. Accel. Beams}\ }\textbf {\bibinfo {volume} {23}},\ \bibinfo {pages} {091001} (\bibinfo {year} {2020})}\BibitemShut {NoStop}%
\bibitem [{\citenamefont {Sa\'a~Hern\'andez}\ \emph {et~al.}(2018)\citenamefont {Sa\'a~Hern\'andez}, \citenamefont {Bartosik}, \citenamefont {Biancacci}, \citenamefont {Hirlaender}, \citenamefont {Huschauer},\ and\ \citenamefont {Moreno~Garcia}}]{SaaHernandez:2018zqv}%
  \BibitemOpen
  \bibfield  {author} {\bibinfo {author} {\bibfnamefont {A.}~\bibnamefont {Sa\'a~Hern\'andez}}, \bibinfo {author} {\bibfnamefont {H.}~\bibnamefont {Bartosik}}, \bibinfo {author} {\bibfnamefont {N.}~\bibnamefont {Biancacci}}, \bibinfo {author} {\bibfnamefont {S.}~\bibnamefont {Hirlaender}}, \bibinfo {author} {\bibfnamefont {A.}~\bibnamefont {Huschauer}},\ and\ \bibinfo {author} {\bibfnamefont {D.}~\bibnamefont {Moreno~Garcia}},\ }\bibfield  {title} {\bibinfo {title} {{Space Charge Studies on LEIR}},\ }\href {https://doi.org/10.18429/JACoW-IPAC2018-THPAF055} {\bibfield  {journal} {\bibinfo  {journal} {J. Phys. Conf. Ser.}\ }\textbf {\bibinfo {volume} {1067}},\ \bibinfo {pages} {062020} (\bibinfo {year} {2018})}\BibitemShut {NoStop}%
\bibitem [{\citenamefont {Venturini}\ \emph {et~al.}(2006)\citenamefont {Venturini}, \citenamefont {Oide},\ and\ \citenamefont {Wolski}}]{Venturini:2006rp}%
  \BibitemOpen
  \bibfield  {author} {\bibinfo {author} {\bibfnamefont {M.}~\bibnamefont {Venturini}}, \bibinfo {author} {\bibfnamefont {K.}~\bibnamefont {Oide}},\ and\ \bibinfo {author} {\bibfnamefont {A.}~\bibnamefont {Wolski}},\ }\bibfield  {title} {\bibinfo {title} {{Space charge and equilibrium emittances in damping rings}},\ }\href@noop {} {\bibfield  {journal} {\bibinfo  {journal} {Conf. Proc. C}\ }\textbf {\bibinfo {volume} {060626}},\ \bibinfo {pages} {882} (\bibinfo {year} {2006})}\BibitemShut {NoStop}%
\bibitem [{\citenamefont {Venturini}\ and\ \citenamefont {Oide}(2006)}]{Venturini:2006ilc}%
  \BibitemOpen
  \bibfield  {author} {\bibinfo {author} {\bibfnamefont {M.}~\bibnamefont {Venturini}}\ and\ \bibinfo {author} {\bibfnamefont {K.}~\bibnamefont {Oide}},\ }\bibfield  {title} {\bibinfo {title} {{Direct space charge effects on the ILC damping rings: Task force report}},\ }\bibfield  {journal} {\bibinfo  {journal} {{}}\ }\href {https://doi.org/10.2172/889308} {10.2172/889308} (\bibinfo {year} {2006})\BibitemShut {NoStop}%
\bibitem [{\citenamefont {Venturini}(2007)}]{Venturini:2007ler}%
  \BibitemOpen
  \bibfield  {author} {\bibinfo {author} {\bibfnamefont {M.}~\bibnamefont {Venturini}},\ }\bibfield  {title} {\bibinfo {title} {Space-charge effects in the super b-factory ler},\ }\bibfield  {journal} {\bibinfo  {journal} {{}}\ }\href {https://doi.org/10.2172/923014} {10.2172/923014} (\bibinfo {year} {2007})\BibitemShut {NoStop}%
\bibitem [{\citenamefont {Xiao}\ \emph {et~al.}(2007)\citenamefont {Xiao}, \citenamefont {Borland}, \citenamefont {Emery}, \citenamefont {Wang},\ and\ \citenamefont {Ng}}]{Xiao:2007ilc}%
  \BibitemOpen
  \bibfield  {author} {\bibinfo {author} {\bibfnamefont {A.}~\bibnamefont {Xiao}}, \bibinfo {author} {\bibfnamefont {M.}~\bibnamefont {Borland}}, \bibinfo {author} {\bibfnamefont {L.}~\bibnamefont {Emery}}, \bibinfo {author} {\bibfnamefont {Y.}~\bibnamefont {Wang}},\ and\ \bibinfo {author} {\bibfnamefont {K.~Y.}\ \bibnamefont {Ng}},\ }\bibfield  {title} {\bibinfo {title} {{Direct Space Charge Calculation in Elegant and Its Application to the ILC Damping Ring}},\ }\href {https://doi.org/10.2172/921985} {\bibfield  {journal} {\bibinfo  {journal} {Conf. Proc. C}\ }\textbf {\bibinfo {volume} {070625}},\ \bibinfo {pages} {3456} (\bibinfo {year} {2007})}\BibitemShut {NoStop}%
\bibitem [{\citenamefont {Aicheler}\ \emph {et~al.}(2012)\citenamefont {Aicheler}, \citenamefont {Burrows}, \citenamefont {Draper}, \citenamefont {Garvey}, \citenamefont {Lebrun}, \citenamefont {Peach}, \citenamefont {Phinney}, \citenamefont {Schmickler}, \citenamefont {Schulte},\ and\ \citenamefont {Toge}}]{CLIC_CDR}%
  \BibitemOpen
  \bibfield  {author} {\bibinfo {author} {\bibfnamefont {M.}~\bibnamefont {Aicheler}}, \bibinfo {author} {\bibfnamefont {P.}~\bibnamefont {Burrows}}, \bibinfo {author} {\bibfnamefont {M.}~\bibnamefont {Draper}}, \bibinfo {author} {\bibfnamefont {T.}~\bibnamefont {Garvey}}, \bibinfo {author} {\bibfnamefont {P.}~\bibnamefont {Lebrun}}, \bibinfo {author} {\bibfnamefont {K.}~\bibnamefont {Peach}}, \bibinfo {author} {\bibfnamefont {N.}~\bibnamefont {Phinney}}, \bibinfo {author} {\bibfnamefont {H.}~\bibnamefont {Schmickler}}, \bibinfo {author} {\bibfnamefont {D.}~\bibnamefont {Schulte}},\ and\ \bibinfo {author} {\bibfnamefont {N.}~\bibnamefont {Toge}},\ }\href {https://doi.org/10.5170/CERN-2012-007} {\emph {\bibinfo {title} {{A Multi-TeV Linear Collider Based on CLIC Technology: CLIC Conceptual Design Report}}}},\ CERN Yellow Reports: Monographs\ (\bibinfo  {publisher} {CERN},\ \bibinfo {address} {Geneva},\ \bibinfo {year} {2012})\BibitemShut {NoStop}%
\bibitem [{\citenamefont {Ghasem}\ \emph {et~al.}(2016)\citenamefont {Ghasem}, \citenamefont {Alabau-Gonzalvo}, \citenamefont {Antoniou}, \citenamefont {Papadopoulou},\ and\ \citenamefont {Papaphilippou}}]{Ghasem}%
  \BibitemOpen
  \bibfield  {author} {\bibinfo {author} {\bibfnamefont {H.}~\bibnamefont {Ghasem}}, \bibinfo {author} {\bibfnamefont {J.}~\bibnamefont {Alabau-Gonzalvo}}, \bibinfo {author} {\bibfnamefont {F.}~\bibnamefont {Antoniou}}, \bibinfo {author} {\bibfnamefont {S.}~\bibnamefont {Papadopoulou}},\ and\ \bibinfo {author} {\bibfnamefont {Y.}~\bibnamefont {Papaphilippou}},\ }\bibfield  {title} {\bibinfo {title} {{NONLINEAR OPTIMIZATION OF CLIC DRS NEW DESIGN WITH VARIABLE BENDS AND HIGH FIELD WIGGLERS}},\ }in\ \href {https://doi.org/10.18429/JACoW-IPAC2016-WEPMW003} {\emph {\bibinfo {booktitle} {{7th International Particle Accelerator Conference}}}}\ (\bibinfo {year} {2016})\ p.\ \bibinfo {pages} {WEPMW003}\BibitemShut {NoStop}%
\bibitem [{\citenamefont {Shishlo}\ \emph {et~al.}(2015)\citenamefont {Shishlo}, \citenamefont {Cousineau}, \citenamefont {Holmes},\ and\ \citenamefont {Gorlov}}]{pyorbit}%
  \BibitemOpen
  \bibfield  {author} {\bibinfo {author} {\bibfnamefont {A.}~\bibnamefont {Shishlo}}, \bibinfo {author} {\bibfnamefont {S.}~\bibnamefont {Cousineau}}, \bibinfo {author} {\bibfnamefont {J.}~\bibnamefont {Holmes}},\ and\ \bibinfo {author} {\bibfnamefont {T.}~\bibnamefont {Gorlov}},\ }\bibfield  {title} {\bibinfo {title} {The particle accelerator simulation code pyorbit},\ }\href {https://doi.org/https://doi.org/10.1016/j.procs.2015.05.312} {\bibfield  {journal} {\bibinfo  {journal} {Procedia Computer Science}\ }\textbf {\bibinfo {volume} {51}},\ \bibinfo {pages} {1272} (\bibinfo {year} {2015})},\ \bibinfo {note} {international Conference On Computational Science, ICCS 2015}\BibitemShut {NoStop}%
\bibitem [{\citenamefont {Papaphilippou}\ \emph {et~al.}(2012)\citenamefont {Papaphilippou}, \citenamefont {Antoniou}, \citenamefont {Barnes}, \citenamefont {Calatroni}, \citenamefont {Chiggiato} \emph {et~al.}}]{Papaphilippou:1464093}%
  \BibitemOpen
  \bibfield  {author} {\bibinfo {author} {\bibfnamefont {Y.}~\bibnamefont {Papaphilippou}}, \bibinfo {author} {\bibfnamefont {F.}~\bibnamefont {Antoniou}}, \bibinfo {author} {\bibfnamefont {M.}~\bibnamefont {Barnes}}, \bibinfo {author} {\bibfnamefont {S.}~\bibnamefont {Calatroni}}, \bibinfo {author} {\bibfnamefont {P.}~\bibnamefont {Chiggiato}}, \emph {et~al.},\ }\bibfield  {title} {\bibinfo {title} {{Conceptual Design of the CLIC Damping Rings}},\ }\href {https://cds.cern.ch/record/1464093} {\bibfield  {journal} {\bibinfo  {journal} {Conf. Proc.}\ }\textbf {\bibinfo {volume} {C1205201}},\ \bibinfo {pages} {TUPPC086. 3 p} (\bibinfo {year} {2012})}\BibitemShut {NoStop}%
\bibitem [{\citenamefont {Papadopoulou}\ \emph {et~al.}(2019)\citenamefont {Papadopoulou}, \citenamefont {Antoniou},\ and\ \citenamefont {Papaphilippou}}]{Papadopoulou_2019}%
  \BibitemOpen
  \bibfield  {author} {\bibinfo {author} {\bibfnamefont {S.}~\bibnamefont {Papadopoulou}}, \bibinfo {author} {\bibfnamefont {F.}~\bibnamefont {Antoniou}},\ and\ \bibinfo {author} {\bibfnamefont {Y.}~\bibnamefont {Papaphilippou}},\ }\bibfield  {title} {\bibinfo {title} {Emittance reduction with variable bending magnet strengths: Analytical optics considerations and application to the compact linear collider damping ring design},\ }\href {https://doi.org/10.1103/PhysRevAccelBeams.22.091601} {\bibfield  {journal} {\bibinfo  {journal} {Phys. Rev. Accel. Beams}\ }\textbf {\bibinfo {volume} {22}},\ \bibinfo {pages} {091601} (\bibinfo {year} {2019})}\BibitemShut {NoStop}%
\bibitem [{\citenamefont {Schmidt}\ \emph {et~al.}(2002)\citenamefont {Schmidt}, \citenamefont {Forest},\ and\ \citenamefont {McIntosh}}]{PTC}%
  \BibitemOpen
  \bibfield  {author} {\bibinfo {author} {\bibfnamefont {F.}~\bibnamefont {Schmidt}}, \bibinfo {author} {\bibfnamefont {E.}~\bibnamefont {Forest}},\ and\ \bibinfo {author} {\bibfnamefont {E.}~\bibnamefont {McIntosh}},\ }\href {http://cds.cern.ch/record/573082} {\emph {\bibinfo {title} {{Introduction to the polymorphic tracking code: Fibre bundles, polymorphic Taylor types and ``Exact tracking''}}}},\ \bibinfo {type} {Tech. Rep.}\ (\bibinfo  {institution} {CERN},\ \bibinfo {address} {Geneva},\ \bibinfo {year} {2002})\BibitemShut {NoStop}%
\bibitem [{\citenamefont {Bassetti}\ and\ \citenamefont {Erskine}(1980)}]{Bassetti}%
  \BibitemOpen
  \bibfield  {author} {\bibinfo {author} {\bibfnamefont {M.}~\bibnamefont {Bassetti}}\ and\ \bibinfo {author} {\bibfnamefont {G.~A.}\ \bibnamefont {Erskine}},\ }\href {https://cds.cern.ch/record/122227} {\emph {\bibinfo {title} {{Closed expression for the electrical field of a two-dimensional Gaussian charge}}}},\ \bibinfo {type} {Tech. Rep.}\ (\bibinfo  {institution} {CERN},\ \bibinfo {address} {Geneva, Switzerland},\ \bibinfo {year} {1980})\BibitemShut {NoStop}%
\bibitem [{\citenamefont {CERN}(2021)}]{xsuite}%
  \BibitemOpen
  \bibfield  {author} {\bibinfo {author} {\bibnamefont {CERN}},\ }\href@noop {} {\bibinfo {title} {{Xsuite tracking code}}},\ \bibinfo {howpublished} {\url{https://xsuite.readthedocs.io/en/latest/}} (\bibinfo {year} {2021})\BibitemShut {NoStop}%
\bibitem [{\citenamefont {Timko}\ \emph {et~al.}(2023)\citenamefont {Timko}, \citenamefont {Albright}, \citenamefont {Argyropoulos}, \citenamefont {Damerau}, \citenamefont {Iliakis}, \citenamefont {Intelisano}, \citenamefont {Karlsen-Baeck}, \citenamefont {Karpov}, \citenamefont {Lasheen}, \citenamefont {Medina}, \citenamefont {Quartullo}, \citenamefont {Repond}, \citenamefont {Vanel}, \citenamefont {M\"uller}, \citenamefont {Schwarz}, \citenamefont {Tsapatsaris},\ and\ \citenamefont {Typaldos}}]{blond}%
  \BibitemOpen
  \bibfield  {author} {\bibinfo {author} {\bibfnamefont {H.}~\bibnamefont {Timko}}, \bibinfo {author} {\bibfnamefont {S.}~\bibnamefont {Albright}}, \bibinfo {author} {\bibfnamefont {T.}~\bibnamefont {Argyropoulos}}, \bibinfo {author} {\bibfnamefont {H.}~\bibnamefont {Damerau}}, \bibinfo {author} {\bibfnamefont {K.}~\bibnamefont {Iliakis}}, \bibinfo {author} {\bibfnamefont {L.}~\bibnamefont {Intelisano}}, \bibinfo {author} {\bibfnamefont {B.~E.}\ \bibnamefont {Karlsen-Baeck}}, \bibinfo {author} {\bibfnamefont {I.}~\bibnamefont {Karpov}}, \bibinfo {author} {\bibfnamefont {A.}~\bibnamefont {Lasheen}}, \bibinfo {author} {\bibfnamefont {L.}~\bibnamefont {Medina}}, \bibinfo {author} {\bibfnamefont {D.}~\bibnamefont {Quartullo}}, \bibinfo {author} {\bibfnamefont {J.}~\bibnamefont {Repond}}, \bibinfo {author} {\bibfnamefont {A.~L.}\ \bibnamefont {Vanel}}, \bibinfo {author} {\bibfnamefont {J.~E.}\ \bibnamefont {M\"uller}}, \bibinfo {author} {\bibfnamefont {M.}~\bibnamefont {Schwarz}}, \bibinfo {author} {\bibfnamefont
  {P.}~\bibnamefont {Tsapatsaris}},\ and\ \bibinfo {author} {\bibfnamefont {G.}~\bibnamefont {Typaldos}},\ }\bibfield  {title} {\bibinfo {title} {Beam longitudinal dynamics simulation studies},\ }\href {https://doi.org/10.1103/PhysRevAccelBeams.26.114602} {\bibfield  {journal} {\bibinfo  {journal} {Phys. Rev. Accel. Beams}\ }\textbf {\bibinfo {volume} {26}},\ \bibinfo {pages} {114602} (\bibinfo {year} {2023})}\BibitemShut {NoStop}%
\bibitem [{\citenamefont {{CERN}}(2016)}]{pyHT}%
  \BibitemOpen
  \bibfield  {author} {\bibinfo {author} {\bibnamefont {{CERN}}},\ }\href@noop {} {\bibinfo {title} {{PyHEADTAIL} code repository}},\ \bibinfo {howpublished} {\url{https://github.com/PyCOMPLETE/}} (\bibinfo {year} {2016})\BibitemShut {NoStop}%
\bibitem [{\citenamefont {Helm}\ \emph {et~al.}(1973)\citenamefont {Helm}, \citenamefont {Lee}, \citenamefont {Morton},\ and\ \citenamefont {Sands}}]{SRint}%
  \BibitemOpen
  \bibfield  {author} {\bibinfo {author} {\bibfnamefont {R.~H.}\ \bibnamefont {Helm}}, \bibinfo {author} {\bibfnamefont {M.~J.}\ \bibnamefont {Lee}}, \bibinfo {author} {\bibfnamefont {P.~L.}\ \bibnamefont {Morton}},\ and\ \bibinfo {author} {\bibfnamefont {M.}~\bibnamefont {Sands}},\ }\bibfield  {title} {\bibinfo {title} {Evaluation of synchrotron radiation integrals},\ }\href {https://doi.org/10.1109/TNS.1973.4327284} {\bibfield  {journal} {\bibinfo  {journal} {IEEE Transactions on Nuclear Science}\ }\textbf {\bibinfo {volume} {20}},\ \bibinfo {pages} {900} (\bibinfo {year} {1973})}\BibitemShut {NoStop}%
\bibitem [{\citenamefont {Sands}(1969)}]{Sands:SR}%
  \BibitemOpen
  \bibfield  {author} {\bibinfo {author} {\bibfnamefont {M.}~\bibnamefont {Sands}},\ }\bibfield  {title} {\bibinfo {title} {{The Physics of Electron Storage Rings: An Introduction}},\ }\href@noop {} {\bibfield  {journal} {\bibinfo  {journal} {Conf. Proc. C}\ }\textbf {\bibinfo {volume} {6906161}},\ \bibinfo {pages} {257} (\bibinfo {year} {1969})}\BibitemShut {NoStop}%
\bibitem [{\citenamefont {Piwinski}(1974)}]{Piwinski:0}%
  \BibitemOpen
  \bibfield  {author} {\bibinfo {author} {\bibfnamefont {A.}~\bibnamefont {Piwinski}},\ }\bibfield  {title} {\bibinfo {title} {{Intra-beam-Scattering}},\ }in\ \href {https://doi.org/10.5170/CERN-1992-001.226} {\emph {\bibinfo {booktitle} {{9th International Conference on High-Energy Accelerators}}}}\ (\bibinfo {year} {1974})\ pp.\ \bibinfo {pages} {405--409}\BibitemShut {NoStop}%
\bibitem [{\citenamefont {Bjorken}\ and\ \citenamefont {Mtingwa}(1982)}]{Bjorken:0}%
  \BibitemOpen
  \bibfield  {author} {\bibinfo {author} {\bibfnamefont {J.~D.}\ \bibnamefont {Bjorken}}\ and\ \bibinfo {author} {\bibfnamefont {S.~K.}\ \bibnamefont {Mtingwa}},\ }\bibfield  {title} {\bibinfo {title} {{Intrabeam scattering}},\ }\href {https://cds.cern.ch/record/140304} {\bibfield  {journal} {\bibinfo  {journal} {Part. Accel.}\ }\textbf {\bibinfo {volume} {13}},\ \bibinfo {pages} {115} (\bibinfo {year} {1982})}\BibitemShut {NoStop}%
\bibitem [{\citenamefont {Bane}(2002)}]{Bane_HE}%
  \BibitemOpen
  \bibfield  {author} {\bibinfo {author} {\bibfnamefont {K.~L.~F.}\ \bibnamefont {Bane}},\ }\href {https://doi.org/10.2172/799081} {\emph {\bibinfo {title} {A Simplified Model of Intrabeam Scattering}}},\ \bibinfo {type} {Tech. Rep.}\ (\bibinfo  {institution} {SLAC National Accelerator Lab., Menlo Park, CA (United States)},\ \bibinfo {year} {2002})\BibitemShut {NoStop}%
\bibitem [{\citenamefont {Martini}\ \emph {et~al.}(2016)\citenamefont {Martini}, \citenamefont {Antoniou},\ and\ \citenamefont {Papaphilippou}}]{Martini:2016}%
  \BibitemOpen
  \bibfield  {author} {\bibinfo {author} {\bibfnamefont {M.}~\bibnamefont {Martini}}, \bibinfo {author} {\bibfnamefont {F.}~\bibnamefont {Antoniou}},\ and\ \bibinfo {author} {\bibfnamefont {Y.}~\bibnamefont {Papaphilippou}},\ }\bibfield  {title} {\bibinfo {title} {{Intrabeam Scattering}},\ }\href {https://cds.cern.ch/record/2266030} {\bibfield  {journal} {\bibinfo  {journal} {ICFA Beam Dyn. Newsl.}\ }\textbf {\bibinfo {volume} {69}},\ \bibinfo {pages} {38} (\bibinfo {year} {2016})}\BibitemShut {NoStop}%
\bibitem [{\citenamefont {Martini}(2017)}]{Martini:2017}%
  \BibitemOpen
  \bibfield  {author} {\bibinfo {author} {\bibfnamefont {M.}~\bibnamefont {Martini}},\ }\bibfield  {title} {\bibinfo {title} {{Intrabeam Scattering: Anatomy of the Theory}},\ }\href {https://doi.org/10.23730/CYRSP-2017-003.291} {\bibfield  {journal} {\bibinfo  {journal} {CERN Yellow Rep. School Proc.}\ }\textbf {\bibinfo {volume} {3}},\ \bibinfo {pages} {291} (\bibinfo {year} {2017})}\BibitemShut {NoStop}%
\bibitem [{\citenamefont {Raubenheimer}(1991)}]{PhD_Raubenheimer}%
  \BibitemOpen
  \bibfield  {author} {\bibinfo {author} {\bibfnamefont {T.~O.}\ \bibnamefont {Raubenheimer}},\ }\href@noop {} {\emph {\bibinfo {title} {{The Generation and acceleration of low emittance flat beams for future linear colliders}}}},\ \bibinfo {type} {Tech. Rep.}\ (\bibinfo  {institution} {Stanford Linear Accelerator Center},\ \bibinfo {year} {1991})\BibitemShut {NoStop}%
\bibitem [{\citenamefont {Vivoli}\ and\ \citenamefont {Martini}(2010)}]{vivoli:1}%
  \BibitemOpen
  \bibfield  {author} {\bibinfo {author} {\bibfnamefont {A.}~\bibnamefont {Vivoli}}\ and\ \bibinfo {author} {\bibfnamefont {M.}~\bibnamefont {Martini}},\ }\bibfield  {title} {\bibinfo {title} {{Intra-Beam Scattering in the CLIC Damping Rings}},\ }\href@noop {} {\bibfield  {journal} {\bibinfo  {journal} {Conf. Proc. C}\ }\textbf {\bibinfo {volume} {100523}},\ \bibinfo {pages} {WEPE090} (\bibinfo {year} {2010})}\BibitemShut {NoStop}%
\bibitem [{\citenamefont {Biagini}\ \emph {et~al.}(2012)\citenamefont {Biagini}, \citenamefont {Boscolo}, \citenamefont {Demma}, \citenamefont {Chao}, \citenamefont {Bane},\ and\ \citenamefont {Pivi}}]{osti_1}%
  \BibitemOpen
  \bibfield  {author} {\bibinfo {author} {\bibfnamefont {M.}~\bibnamefont {Biagini}}, \bibinfo {author} {\bibfnamefont {M.}~\bibnamefont {Boscolo}}, \bibinfo {author} {\bibfnamefont {T.}~\bibnamefont {Demma}}, \bibinfo {author} {\bibfnamefont {A.}~\bibnamefont {Chao}}, \bibinfo {author} {\bibfnamefont {K.}~\bibnamefont {Bane}},\ and\ \bibinfo {author} {\bibfnamefont {M.}~\bibnamefont {Pivi}},\ }\bibfield  {title} {\bibinfo {title} {Multiparticle simulation of intrabeam scattering for superb},\ }\href@noop {} {\bibfield  {journal} {\bibinfo  {journal} {IPAC 2011 - 2nd International Particle Accelerator Conference}\ } (\bibinfo {year} {2012})}\BibitemShut {NoStop}%
\bibitem [{\citenamefont {Pivi}(2007)}]{Pivi:1}%
  \BibitemOpen
  \bibfield  {author} {\bibinfo {author} {\bibfnamefont {M.~T.~F.}\ \bibnamefont {Pivi}},\ }\bibfield  {title} {\bibinfo {title} {{CMAD: A new self-consistent parallel code to simulate the electron cloud build-up and instabilities}},\ }\href {https://doi.org/10.1109/PAC.2007.4440517} {\bibfield  {journal} {\bibinfo  {journal} {Conf. Proc. C}\ }\textbf {\bibinfo {volume} {070625}},\ \bibinfo {pages} {3636} (\bibinfo {year} {2007})}\BibitemShut {NoStop}%
\bibitem [{\citenamefont {Sonnad}\ \emph {et~al.}(2012)\citenamefont {Sonnad}, \citenamefont {Antoniou}, \citenamefont {Papaphilippou}, \citenamefont {Li}, \citenamefont {Boscolo}, \citenamefont {Demma}, \citenamefont {Chao}, \citenamefont {Rivetta},\ and\ \citenamefont {Pivi}}]{Sonnad:1}%
  \BibitemOpen
  \bibfield  {author} {\bibinfo {author} {\bibfnamefont {K.~G.}\ \bibnamefont {Sonnad}}, \bibinfo {author} {\bibfnamefont {F.}~\bibnamefont {Antoniou}}, \bibinfo {author} {\bibfnamefont {Y.}~\bibnamefont {Papaphilippou}}, \bibinfo {author} {\bibfnamefont {K.~S.~B.}\ \bibnamefont {Li}}, \bibinfo {author} {\bibfnamefont {M.}~\bibnamefont {Boscolo}}, \bibinfo {author} {\bibfnamefont {T.}~\bibnamefont {Demma}}, \bibinfo {author} {\bibfnamefont {A.}~\bibnamefont {Chao}}, \bibinfo {author} {\bibfnamefont {C.~H.}\ \bibnamefont {Rivetta}},\ and\ \bibinfo {author} {\bibfnamefont {M.~T.~F.}\ \bibnamefont {Pivi}},\ }\bibfield  {title} {\bibinfo {title} {{Multi-Particle Simulation Codes Implementation to Include Models of a Novel Single-bunch Feedback System and Intra-beam Scattering}},\ }\href@noop {} {\bibfield  {journal} {\bibinfo  {journal} {Conf. Proc. C}\ }\textbf {\bibinfo {volume} {1205201}},\ \bibinfo {pages} {3147} (\bibinfo {year} {2012})}\BibitemShut {NoStop}%
\bibitem [{\citenamefont {Zenkevich}\ \emph {et~al.}(2006)\citenamefont {Zenkevich}, \citenamefont {Boine-Frankenheim},\ and\ \citenamefont {Bolshakov}}]{Zenkevich:0}%
  \BibitemOpen
  \bibfield  {author} {\bibinfo {author} {\bibfnamefont {P.}~\bibnamefont {Zenkevich}}, \bibinfo {author} {\bibfnamefont {O.}~\bibnamefont {Boine-Frankenheim}},\ and\ \bibinfo {author} {\bibfnamefont {A.}~\bibnamefont {Bolshakov}},\ }\bibfield  {title} {\bibinfo {title} {A new algorithm for the kinetic analysis of intra-beam scattering in storage rings},\ }\href {https://doi.org/https://doi.org/10.1016/j.nima.2006.01.013} {\bibfield  {journal} {\bibinfo  {journal} {Nuclear Instruments and Methods in Physics Research Section A: Accelerators, Spectrometers, Detectors and Associated Equipment}\ }\textbf {\bibinfo {volume} {561}},\ \bibinfo {pages} {284} (\bibinfo {year} {2006})},\ \bibinfo {note} {proceedings of the Workshop on High Intensity Beam Dynamics}\BibitemShut {NoStop}%
\bibitem [{\citenamefont {Zenkevich}\ \emph {et~al.}(2005)\citenamefont {Zenkevich}, \citenamefont {Bolshakov},\ and\ \citenamefont {Boine-Frankenheim}}]{Zenkevich:1}%
  \BibitemOpen
  \bibfield  {author} {\bibinfo {author} {\bibfnamefont {P.}~\bibnamefont {Zenkevich}}, \bibinfo {author} {\bibfnamefont {A.}~\bibnamefont {Bolshakov}},\ and\ \bibinfo {author} {\bibfnamefont {O.}~\bibnamefont {Boine-Frankenheim}},\ }\bibfield  {title} {\bibinfo {title} {{Kinetic effects in multiple intra-beam scattering}},\ }\href {https://doi.org/10.1063/1.1949577} {\bibfield  {journal} {\bibinfo  {journal} {AIP Conf. Proc.}\ }\textbf {\bibinfo {volume} {773}},\ \bibinfo {pages} {425} (\bibinfo {year} {2005})}\BibitemShut {NoStop}%
\bibitem [{\citenamefont {Nagaitsev}(2005)}]{Nagaitsev}%
  \BibitemOpen
  \bibfield  {author} {\bibinfo {author} {\bibfnamefont {S.}~\bibnamefont {Nagaitsev}},\ }\bibfield  {title} {\bibinfo {title} {Intrabeam scattering formulas for fast numerical evaluation},\ }\href {https://doi.org/10.1103/PhysRevSTAB.8.064403} {\bibfield  {journal} {\bibinfo  {journal} {Phys. Rev. ST Accel. Beams}\ }\textbf {\bibinfo {volume} {8}},\ \bibinfo {pages} {064403} (\bibinfo {year} {2005})}\BibitemShut {NoStop}%
\bibitem [{\citenamefont {Papaphilippou}(2014)}]{Papafilippou:naff}%
  \BibitemOpen
  \bibfield  {author} {\bibinfo {author} {\bibfnamefont {Y.}~\bibnamefont {Papaphilippou}},\ }\bibfield  {title} {\bibinfo {title} {Detecting chaos in particle accelerators through the frequency map analysis method},\ }\href {https://doi.org/10.1063/1.4884495} {\bibfield  {journal} {\bibinfo  {journal} {Chaos: An Interdisciplinary Journal of Nonlinear Science}\ }\textbf {\bibinfo {volume} {24}},\ \bibinfo {pages} {024412} (\bibinfo {year} {2014})},\ \Eprint {https://arxiv.org/abs/https://doi.org/10.1063/1.4884495} {https://doi.org/10.1063/1.4884495} \BibitemShut {NoStop}%
\bibitem [{\citenamefont {CERN}(2002)}]{madx}%
  \BibitemOpen
  \bibfield  {author} {\bibinfo {author} {\bibnamefont {CERN}},\ }\href {https://mad.web.cern.ch/mad/} {\bibinfo {title} {{Methodical Accelerator Design (MAD-X)} website}} (\bibinfo {year} {2002})\BibitemShut {NoStop}%
\bibitem [{\citenamefont {Satogata}\ \emph {et~al.}(1992)\citenamefont {Satogata}, \citenamefont {Chen}, \citenamefont {Cole}, \citenamefont {Finley}, \citenamefont {Gerasimov}, \citenamefont {Goderre}, \citenamefont {Harrison}, \citenamefont {Johnson}, \citenamefont {Kourbanis}, \citenamefont {Manz}, \citenamefont {Merminga}, \citenamefont {Michelotti}, \citenamefont {Peggs}, \citenamefont {Pilat}, \citenamefont {Pruss}, \citenamefont {Saltmarsh}, \citenamefont {Saritepe}, \citenamefont {Talman}, \citenamefont {Trahern},\ and\ \citenamefont {Tsironis}}]{Satogata}%
  \BibitemOpen
  \bibfield  {author} {\bibinfo {author} {\bibfnamefont {T.}~\bibnamefont {Satogata}}, \bibinfo {author} {\bibfnamefont {T.}~\bibnamefont {Chen}}, \bibinfo {author} {\bibfnamefont {B.}~\bibnamefont {Cole}}, \bibinfo {author} {\bibfnamefont {D.}~\bibnamefont {Finley}}, \bibinfo {author} {\bibfnamefont {A.}~\bibnamefont {Gerasimov}}, \bibinfo {author} {\bibfnamefont {G.}~\bibnamefont {Goderre}}, \bibinfo {author} {\bibfnamefont {M.}~\bibnamefont {Harrison}}, \bibinfo {author} {\bibfnamefont {R.}~\bibnamefont {Johnson}}, \bibinfo {author} {\bibfnamefont {I.}~\bibnamefont {Kourbanis}}, \bibinfo {author} {\bibfnamefont {C.}~\bibnamefont {Manz}}, \bibinfo {author} {\bibfnamefont {N.}~\bibnamefont {Merminga}}, \bibinfo {author} {\bibfnamefont {L.}~\bibnamefont {Michelotti}}, \bibinfo {author} {\bibfnamefont {S.}~\bibnamefont {Peggs}}, \bibinfo {author} {\bibfnamefont {F.}~\bibnamefont {Pilat}}, \bibinfo {author} {\bibfnamefont {S.}~\bibnamefont {Pruss}}, \bibinfo {author} {\bibfnamefont {C.}~\bibnamefont
  {Saltmarsh}}, \bibinfo {author} {\bibfnamefont {S.}~\bibnamefont {Saritepe}}, \bibinfo {author} {\bibfnamefont {R.}~\bibnamefont {Talman}}, \bibinfo {author} {\bibfnamefont {C.~G.}\ \bibnamefont {Trahern}},\ and\ \bibinfo {author} {\bibfnamefont {G.}~\bibnamefont {Tsironis}},\ }\bibfield  {title} {\bibinfo {title} {Driven response of a trapped particle beam},\ }\href {https://doi.org/10.1103/PhysRevLett.68.1838} {\bibfield  {journal} {\bibinfo  {journal} {Phys. Rev. Lett.}\ }\textbf {\bibinfo {volume} {68}},\ \bibinfo {pages} {1838} (\bibinfo {year} {1992})}\BibitemShut {NoStop}%
\bibitem [{\citenamefont {Kostoglou}\ \emph {et~al.}(2020)\citenamefont {Kostoglou}, \citenamefont {Bartosik}, \citenamefont {Papaphilippou}, \citenamefont {Sterbini},\ and\ \citenamefont {Triantafyllou}}]{Kostoglou}%
  \BibitemOpen
  \bibfield  {author} {\bibinfo {author} {\bibfnamefont {S.}~\bibnamefont {Kostoglou}}, \bibinfo {author} {\bibfnamefont {H.}~\bibnamefont {Bartosik}}, \bibinfo {author} {\bibfnamefont {Y.}~\bibnamefont {Papaphilippou}}, \bibinfo {author} {\bibfnamefont {G.}~\bibnamefont {Sterbini}},\ and\ \bibinfo {author} {\bibfnamefont {N.}~\bibnamefont {Triantafyllou}},\ }\bibfield  {title} {\bibinfo {title} {Tune modulation effects for colliding beams in the high luminosity large hadron collider},\ }\href {https://doi.org/10.1103/PhysRevAccelBeams.23.121001} {\bibfield  {journal} {\bibinfo  {journal} {Phys. Rev. Accel. Beams}\ }\textbf {\bibinfo {volume} {23}},\ \bibinfo {pages} {121001} (\bibinfo {year} {2020})}\BibitemShut {NoStop}%
\bibitem [{\citenamefont {Franchetti}\ and\ \citenamefont {Hofmann}(2006)}]{Franchetti:2006aa}%
  \BibitemOpen
  \bibfield  {author} {\bibinfo {author} {\bibfnamefont {G.}~\bibnamefont {Franchetti}}\ and\ \bibinfo {author} {\bibfnamefont {I.}~\bibnamefont {Hofmann}},\ }\bibfield  {title} {\bibinfo {title} {Particle trapping by nonlinear resonances and space charge},\ }\href {https://doi.org/https://doi.org/10.1016/j.nima.2006.01.031} {\bibfield  {journal} {\bibinfo  {journal} {Nuclear Instruments and Methods in Physics Research Section A: Accelerators, Spectrometers, Detectors and Associated Equipment}\ }\textbf {\bibinfo {volume} {561}},\ \bibinfo {pages} {195} (\bibinfo {year} {2006})},\ \bibinfo {note} {proceedings of the Workshop on High Intensity Beam Dynamics}\BibitemShut {NoStop}%
\end{thebibliography}%

\end{document}